\documentclass[12pt,preprint]{aastex}

\usepackage{float}
\usepackage{CJKutf8}
\usepackage{graphicx}
\usepackage{epsfig}
\usepackage[perpage,symbol*]{footmisc}
\usepackage{natbib}
\usepackage{dcolumn}
\usepackage{bm}
\usepackage[figureright]{rotating}

\shorttitle{TMRT observations of 26 pulsars at 8.6~GHz}
\shortauthors{R.S. Zhao et al.}

\begin{document}


\title{TMRT observations of 26 pulsars at 8.6~GHz}


\author{
Ru-Shuang~ZHAO~(\begin{CJK}{UTF8}{gbsn}赵汝双\end{CJK})\altaffilmark{1,3}, 
Xin-Ji~WU~(\begin{CJK}{UTF8}{gbsn}吴鑫基\end{CJK})\altaffilmark{2}, 
Zhen~YAN~(\begin{CJK}{UTF8}{gbsn}闫振\end{CJK})\altaffilmark{1,4}, 
Zhi-Qiang~SHEN~(\begin{CJK}{UTF8}{gbsn}沈志强\end{CJK})\altaffilmark{1,4}\thanks{zshen@shao.ac.cn},
R.~N.~Manchester\altaffilmark{5},
Guo-Jun~QIAO~(\begin{CJK}{UTF8}{gbsn}乔国俊\end{CJK})\altaffilmark{2}, 
Ren-Xin~XU~(\begin{CJK}{UTF8}{gbsn}徐仁新\end{CJK})\altaffilmark{2}, 
Ya-Jun~WU~(\begin{CJK}{UTF8}{gbsn}吴亚军\end{CJK})\altaffilmark{1,4}, 
Rong-Bing~ZHAO~(\begin{CJK}{UTF8}{gbsn}赵融冰\end{CJK})\altaffilmark{1,4}, 
Bin~LI~(\begin{CJK}{UTF8}{gbsn}李斌\end{CJK})\altaffilmark{1,4}, 
Yuan-Jie~DU~(\begin{CJK}{UTF8}{gbsn}杜源杰\end{CJK})\altaffilmark{6}, 
Ke-Jia~LEE~(\begin{CJK}{UTF8}{gbsn}李柯伽\end{CJK})\altaffilmark{7}, 
Long-Fei~HAO~(\begin{CJK}{UTF8}{gbsn}郝龙飞\end{CJK})\altaffilmark{8}, 
Qing-Hui~LIU~(\begin{CJK}{UTF8}{gbsn}刘庆会\end{CJK})\altaffilmark{1,4}, 
Ji-Guang~LU~(\begin{CJK}{UTF8}{gbsn}卢吉光\end{CJK})\altaffilmark{2,10}, 
Lun-Hua~SHANG~(\begin{CJK}{UTF8}{gbsn}尚伦华\end{CJK})\altaffilmark{9,11}, 
Jin-Qing~WANG~(\begin{CJK}{UTF8}{gbsn}王锦清\end{CJK})\altaffilmark{1,4}, 
Min~WANG~(\begin{CJK}{UTF8}{gbsn}汪敏\end{CJK})\altaffilmark{8}, 
Jin~YUAN~(\begin{CJK}{UTF8}{gbsn}袁瑾\end{CJK})\altaffilmark{1,4}, 
Qi-Jun~ZHI~(\begin{CJK}{UTF8}{gbsn}支启军\end{CJK})\altaffilmark{9}, 
and Wei-Ye~ZHONG~(\begin{CJK}{UTF8}{gbsn}仲伟业\end{CJK})\altaffilmark{1,4}
}

\altaffiltext{1}{Shanghai Astronomical Observatory, Chinese Academy of Sciences, Shanghai 200030, China}
\altaffiltext{2}{School of Physics, Peking University, Beijing 100871, China}
\altaffiltext{3}{University of Chinese Academy of Sciences, Beijing 100049, China}
\altaffiltext{4}{Key Laboratory of Radio Astronomy, Chinese Academy of Sciences, Nanjing 210008, China}
\altaffiltext{5}{CSIRO Astronomy and Space Science, P.O. Box 76, Epping, NSW 1710, Australia}
\altaffiltext{6}{ Qian Xuesen Laboratory of Space Technology, NO. 104, Youyi Road, Haidian District, Beijing 100094, China}
\altaffiltext{7}{Kavil Institute for Astronomy and Astrophysics, Peking University, Beijing 100871, China}
\altaffiltext{8}{Yunnan Astronomical Observatory, Chinese Academy of Sciences, Kunming 650011, China}
\altaffiltext{9}{School of Physics and Electronic Science, Guizhou Normal University, Guiyang 550001, China}
\altaffiltext{10}{State Key Laboratory of nuclear Science and Technology, Peking University, Beijing 100871, China}
\altaffiltext{11}{National Astronomical Observatories, Chinese Academy of Sciences, Beijing 100012, China}


\begin{abstract}
Integrated pulse profiles at 8.6~GHz obtained with the Shanghai Tian
Ma Radio Telescope (TMRT) are presented for a sample of 26
pulsars. Mean flux densities and pulse width parameters of these
pulsars are estimated. For eleven pulsars these are the first
high-frequency observations and for a further four, our observations
have a better signal-to-noise ratio than previous
observations. For one (PSRs J0742$-$2822) the 8.6~GHz profiles
  differs from previously observed profiles. A comparison of 19
profiles with those at other frequencies shows that in nine cases the
separation between the outmost leading and trailing components
decreases with frequency, roughly in agreement with
radius-to-frequency mapping, whereas in the other ten the separation
is nearly constant. Different spectral indices of profile components
lead to the variation of integrated pulse profile shapes with
frequency. In seven pulsars with multi-component profiles,
the spectral indices of the central components are steeper than those
of the outer components. For the 12 pulsars with
  multi-component profiles in the high-frequency sample, we estimate
  the core width using gaussian fitting and discuss the width-period
  relationsip.
\end{abstract}


\keywords{pulsars: general}



\section{Introduction}\label{sec:intro}
Integrated pulse profiles can be thought of as `fingerprints' for two
reasons. One is that for a given pulsar, even though the individual
pulses show different shapes, the integrated pulse profile at a given
frequency is usually very stable. The other is that, for different
pulsars, the integrated pulse profiles are morphologically different
\citep{lm88,mgs+81}. At different frequencies, the shape of the
integrated pulse profile often changes substantially
\citep{srw75}. These variations might be related to the different ways
in which the line of sight cuts across the radiation beam and
intrinsic asymmetry and irregularity within the individual pulsar
beams. Therefore, studying integrated pulse profiles at multiple
frequencies helps to investigate both the geometry and the mechanism
of pulsar radiation.

There have been many studies of pulse profiles at different
frequencies during the history of pulsar research
\citep[e.g.,][]{ran83,lm88,ran92,mr02a,jkmg08,hr10}. However, 
because of the steep spectrum of pulsar emission, with a typical 
spectral index $\alpha = -1.8$ ($S_\nu \propto \nu^{\alpha}$)
\citep{sie73,mgj+94,mkk+00}, limited samples have been observed at
high frequencies. Using the Effelsberg 100-m radio telescope,
\citet{mgs+81} observed 15 northern hemisphere pulsars at 8.7~GHz,
\citet{kxj+97} observed 8 pulsars between 1.4 and 32~GHz including
8.5~GHz, \citet{mkw04} observed 4 millisecond pulsars at 8.35~GHz and
\citet{msk+13} observed 12 pulsars at 8.35~GHz. \citet{jkw06} observed
32 southern hemisphere pulsars at 8.4~GHz using Parkes 64-m telescope. In
total, only 48 pulsars have been observed in a frequency range of 8~GHz
to 9~GHz. Obviously, more high-frequency observations will help our
understanding the pulse emission mechanism. Thus, we initiated
observations of a sample of 26 pulsars with 1.4 GHz flux density of
$\gtrsim$ 4~mJy using the newly-built Shanghai Tian Ma Radio 
Telescope (TMRT). As a result, 26 pulsars were firmly detected at
8.6~GHz.

In this paper we present integrated pulse profiles at 8.6~GHz for
these 26 pulsars observed with the TMRT. In
\S\ref{sec:obs}, the TMRT observations and data reduction methods are
described. Results for individual pulsars are presented in
\S\ref{sec:results} and discussed in the context of other observations
in \S\ref{sec:discn}. Our results are summarised in \S\ref{sec:concl}.

\section{Observations and data reduction}\label{sec:obs}
The TMRT is a 65-m diameter shaped Cassegrain radio telescope with an
elevation-azimuth pedestal and an active reflector for gravity
compensation located in Shanghai, China. Observations of a sample of
26 pulsars were performed with the TMRT in July 2014, May 2015 and
January 2016 at a frequency of 8.6~GHz. The observations were made
with incoherent dedispersion and on-line folding by the FPGA-based
spectrometer DIBAS \citep{ysw+15}. The total recording bandwidth of
800 MHz (8.2$-$9.0~GHz) was subdivided into 512 frequency channels and
each pulsar period was divided into 1024 phase bins. Observation times
were typically 30~min or 60~min and were divided into 30~s
subintegrations. Observational data were written out in 8-bit {\sc
  psrfits} format and {\sc psrchive} programs \citep{hvm04} were used
for data editing and processing to produce the integrated pulse
profiles. Pulsar parameters for on-line folding were obtained from the
ATNF Pulsar Catalogue
\citep{mhth05}\footnote{http://www.atnf.csiro.au/research/pulsar/psrcat}.

Since no real-time flux calibration system was available, we estimated
the profile peak flux density in Jy using the radiometer equation:
\begin{equation}
  S_{\rm pk} = \sigma_b\;\mathcal{S} =\frac{\eta S_{\rm sys}
    \mathcal{S}}{\sqrt{(T/N_b)\Delta\nu}}
\end{equation}
where $\sigma_b$ is the off-pulse rms noise in Jy, $\mathcal{S}$
is the observed signal-to-noise ratio (S/N) of the integrated profile,
$\eta$ is a factor, taken to be 1.5, which allows for digitiser and other
inefficencies, $S_{\rm sys}$ is the system equivalent flux density in
Jy, $T$ is the on-source observation time in s, $\Delta\nu$ is the
recorded bandwidth in Hz and $N_b$ is the number of phase bins
in the integrated profile. The mean flux density in Jy can be
obtained from
\begin{equation}
S = \frac{\sum_{i=1}^{n_p}S_i}{N_b}
\end{equation}
where $n_p$ is the number of on-pulse bins and $S_i$ is
the flux density in on-pulse bin $i$. The uncertainty in $S$ is
\begin{equation}
\sigma_S = \frac{\sigma_b}{\sqrt{N_b}}.
\end{equation}
For our 8.6~GHz TMRT observations, the measured $S_{\rm sys}$ is
$\geq$ 50 Jy depending on elevation \citep{wzy+15}, $\Delta \nu =
8\times10^{8}$ Hz and $N_b$ = 1024.

Mean pulse profile widths at 10\% ($W_{10}$) and 50\% ($W_{50}$) of 
the peak amplitude are commonly tabulated. We estimate the uncertainty in $W_{50}$, 
by considering each component in the profile to have a gaussian shape. 
For a mean profile with just one component above the 50\% level, 
the uncertainty in $W_{50}$ is:
\begin{equation}
\sigma_{W_{50}}=\frac{W_{50}}{\sqrt{2}\;\ln(2)\;\mathcal{S}^{\prime}}
\end{equation}
where $\mathcal{S}^{\prime}$ is the S/N of the component.
For a multi-component profile, $W_{50}$ is measured at 50\% level of
the strongest component and its uncertainty is the quadrature sum of 
\begin{equation}
\sigma_{W_R}=\frac{W_R}{4R\;\ln(1/R)\;\mathcal{S}^{\prime}}
\end{equation}
for the two outermost components that reach the 50\% level, where $R$
is the fractional amplitude of each component at the level at which
$W_{50}$ is measured, $W_R$ is the component full width at that level
and $\mathcal{S}^{\prime}$ is the S/N of that component. 

\section{Observational results}\label{sec:results}

In this section we present the integrated pulse profiles for the 26
pulsars observed by TMRT and discuss their properties. Observational
and profile parameters are given in Table~\ref{tab:fluxpsr}. The first
four columns give the pulsar J2000 name, the pulsar period, the
observation date as a Modified Julian Day (MJD) and the on-source
time. The next three columns give the separation between the outermost
components with number of components in parenthesis and the observed
pulse widths at 50\% and 10\% of the profile peak, all in units of
degrees of pulse phase. Uncertainties are estimated for $W_{50}$ but
not for $W_{10}$ which is often just approximate because of the low
signal to noise ratio. For all the $W_{50}$ estimates, the S/N
  at 50\% of the peak amplitude is greater than 3. For the $W_{10}$
  measurements, the S/N at 10\% of the peak amplitude is greater than
  3 for 10 of the 26 pulsars. In Table 1, those with S/N between 1
  and 3 are marked by `?' and the value is rounded to the nearest
  degree. For the two (PSRs J0659+1414 and J1239+2453) with S/N
  $<1$, the $W_{10}$ measurement is omitted. The next two columns
give the measured 8.6~GHz mean and peak flux densities. For
comparison, the next three columns give pulse widths and flux
densities at frequencies close to 8~GHz as measured by others. For 11
pulsars (PSRs J0437$-$4715, J1239$+$2453, J1645$-$0317, J1705$-$1906,
J1745$-$3040, J1803$-$2137, J1807$-$0847, J1829$-$1751, J1848$-$0123,
J1948$+$3540 and J2048$-$1616), these are the first published profiles
for frequencies around 8~GHz. For 4 pulsars (PSRs J0742-2822,
J1709-1640, J1932+1059 and J2022+5154), these profiles have a much
better signal-to-noise ratio than previous observations \citep[cf.,
][]{kxj+97, jkw06}. Figure~\ref{fg:prf} gives 8.6~GHz pulse profiles
for the 26 pulsars observed by TMRT and Figure~\ref{fg:4freq} compares
the TMRT 8.6~GHz profiles with profiles at other frequencies obtained
from the EPN profile database.\footnote{http://www.epta.eu.org/epndb}
The relationship of components at the different frequencies is
indicated by dashed lines where the power-law frequency dependences of
component phase discussed in \S\ref{sec:w-nu} are assumed.
\begin{sidewaystable*}[p]
{\footnotesize
\caption{Parameters for the 26 pulsars observed with the TMRT}\label{tab:fluxpsr}

\begin{tabular}{l r r r r r l r r r r r r}
\hline
\hline
\\
\multicolumn{1}{c}{PSR J2000} & \multicolumn{1}{c}{$P$} & \multicolumn{1}{c}{MJD} & \multicolumn{1}{c}{$T$} & \multicolumn{1}{c}{$\Delta\phi(N)$} & \multicolumn{1}{c}{$W_{50}$} & \multicolumn{1}{c}{$W_{10}$} & \multicolumn{1}{c}{$S_{8.6} \pm \sigma_{\rm S}$} & \multicolumn{1}{c}{$S^{\rm pk}_{8.6} \pm \sigma_{\rm b}$}  & \multicolumn{1}{c}{$W_{50}^\dag$} & \multicolumn{1}{c}{$W_{10}^\dag$} & \multicolumn{2}{c}{$S^\dag$}\\
\multicolumn{1}{c}{Name} & \multicolumn{1}{c}{(ms)}  &  \multicolumn{1}{c}{(d)}  & \multicolumn{1}{c}{(min)} & \multicolumn{1}{c}{(deg)} & \multicolumn{1}{c}{(deg)} & \multicolumn{1}{c}{(deg)} & \multicolumn{1}{c}{(mJy)} & \multicolumn{1}{c}{(mJy)} & \multicolumn{1}{c}{(deg)} & \multicolumn{1}{c}{(deg)} & \multicolumn{1}{c}{$S_{8.35}$(mJy)} & \multicolumn{1}{c}{$S_{8.7}$(mJy)}\\
\hline
\\
J0437$-$4715 &	5.8   & 57156.203  & 30  & 143.0(4) & 6.3$\pm0.1$  & 13.9 & 5.24$\pm0.12$ & 162.4$\pm4.0$ & $-$ & $-$ & $-$ &$-$\\
J0659$+$1414 & 384.9  & 57156.533  & 30  & $-$(1)   & 12.7$\pm1.5$ & $-$ & 0.96$\pm0.11$ &  27.4$\pm3.4$ & 11.8$^1$  & 26$^1$  & 2.5$^1$ & $-$\\
J0738$-$4042 & 374.9  & 57156.370  & 30  & 7.7(2)   & 15.9$\pm0.6$ & 27$^?$ & 3.69$\pm0.09$ & 78.8$\pm3.0$  & 14.9$^1$  & 24$^1$  & 7.0$^1$ & $-$\\
J0742$-$2822 & 562.6  & 57156.305  & 30  & 5.6(2)   & 8.9$\pm0.2$  & 14.2 & 1.39$\pm0.07$ & 68.1$\pm2.2$  & 1.9$^1$  & 14$^1$  & 1.6$^1$   & $0.55^3$\\
J0837$-$4135 & 751.6  & 57398.667  & 120 &11.0(3)   & 2.9$\pm0.1$  & 13$^?$ & 0.39$\pm0.05$ & 30.9$\pm1.6$  & 2.5$^1$  & 13$^1$  & 1.3$^1$ & $-$\\
J0953$+$0755 & 253.1  & 57156.419  & 30  &$-$(1)    & 12.7$\pm0.6$ & 30$^?$ & 2.31$\pm0.08$ & 53.0$\pm2.4$  & 11.9$^1$  & 30$^1$  &  
4.3$^1$ & $0.58^3$\\
             &        &            &                &              &      &               &        &       &  &  & $1.09\pm0.22^4$ & $-$\\ % $0.58^3$,

J1136$+$1551 & 1188.0 & 56863.411  & 50  & 5.1(2)   & 1.4$\pm0.01$  & $\,\ 7.3$  & 0.86$\pm0.05$ & 151.2$\pm1.1$ & 1.4$^1$  & 8$^1$  &  
0.6$^1$ & $0.62^3$\\
             &        &            &                &              &      &               &        &       &  &  & $0.73\pm0.15^4$ &$-$\\ 
&        &            &                &              &      &               &        &       &  &  & $0.79\pm0.16^4$ &$-$\\                                              &        &            &                &              &      &               &        &       &  &  & $0.88\pm0.18^4$ &$-$\\
&        &            &                &              &      &               &        &       &  &  & $0.92\pm0.18^4$ &$-$\\
J1239$+$2453 & 1382.4 & 57156.463  & 30  & 8.3(2)   & 9.7$\pm0.2$  & $-$ & 0.31$\pm0.07$ & 21.9$\pm2.3$  & $-$ & $-$ & $-$ &$-$\\
J1644$-$4559 & 455.1  & 57156.668  & 30  &20.6(3)   & 5.7$\pm0.04$  & 11.9 & 10.20$\pm0.12$ & 531.3$\pm4.0$ & 5.1$^1$  & 16$^1$  & 2.2$^1$ & $-$\\
J1645$-$0317 & 387.6  & 57398.882  & 120  &12.3(3)  & 13.8$\pm0.8$ & 19$^?$ & 0.50$\pm0.03$ & 18.8$\pm1.1$  & $-$ & $-$ & $-$ & $-$\\
J1705$-$1906 & 298.9  & 57156.582  & 30  & 5.6(2)   & 11.2$\pm0.4$ & 16$^?$ & 1.17$\pm0.09$ & 42.0$\pm1.5$  & $-$ & $-$ & $-$ & $-$\\
J1709$-$1640 & 166.8  & 57156.828  & 30  & $-$(1)   &  2.1$\pm0.03$ & $\,\ 6.5$  & 1.41$\pm0.07$ & 163.8$\pm2.2$ & 3.7$^2$ & 9.7$^2$ & $-$ & 0.24$^3$ \\
J1709$-$4429 & 102.5  & 57156.689  & 30  & $-$(1)   & 15.9$\pm1.5$ & 31$^?$ & 1.75$\pm0.11$ & 37.5$\pm3.5$  & 16.7$^1$  & 32$^1$  & 5.5$^1$ & $-$\\
J1721$-$3532 & 280.4  & 57156.753  & 30  & $-$(1)   &  9.5$\pm0.2$ & 19.9 & 4.30$\pm0.08$ & 145.1$\pm2.5$ & 8.8$^1$  & 17$^1$  & 1.9$^1$ & $-$ \\
J1740$-$3015 & 226.5  & 57156.774  & 30  & 1.2(2)   &  3.3$\pm0.1$ &  $\,\ 6.0$ & 1.21$\pm0.07$ & 124.5$\pm2.5$ & 3.8$^1$  & 8$^1$  & 0.5$^1$ & $-$\\
J1745$-$3040 & 367.4  & 57156.795  & 60  &  9.5(3)  & 14.2$\pm0.3$ & 22.9 & 1.55$\pm0.05$ & 47.6$\pm1.6$  & $-$ & $-$ & $-$ & $-$\\
J1752$-$2806 & 529.2  & 57399.095  & 60  &  5.0(2)  &  2.8$\pm0.2$ & 11$^?$ & 0.30$\pm0.05$ &  23.6$\pm1.5$ & 2.5$^1$  & 12$^1$  & 0.9$^1$ & $-$\\
J1803$-$2137 & 133.7  & 57156.873  & 30  & 40.2(2)  & 23.4$\pm0.6$ & 81$^?$ & 4.42$\pm0.07$ & 58.7$\pm2.2$  & $-$ & $-$ & $-$ & $-$\\
J1807$-$0847 & 163.7  & 57156.894  & 30  & 16.4(3)  & 20.9$\pm0.6$ & 26$^?$ & 1.56$\pm0.08$ & 34.4$\pm2.5$  & $-$ & $-$ & $-$ & $-$\\
J1829$-$1751 & 307.1  & 57157.609  & 30  & 11.3(2)  &  2.7$\pm0.2$ & 18$^?$ & 0.69$\pm0.11$ & 55.4$\pm3.5$  & $-$ & $-$ & $-$ & $-$\\
J1848$-$0123 & 659.4  & 57157.630  & 60  & 12.8(3)  & 15.3$\pm0.3$ & 21$^?$ & 0.87$\pm0.05$ & 32.4$\pm1.5$  & $-$ & $-$ & $-$ & $-$\\
J1932$+$1059 & 653.1  & 57156.890  & 30  & 2.2(2)   &  6.1$\pm0.1$ & 14.9 & 3.62$\pm0.06$ & 169.6$\pm2.0$ & 6.6$^2$ & 14.5$^2$ & $-$ & $1.2^3 \,\ $\\
J1935$+$1616 & 358.7  & 57399.231  & 60  & 14.3(3)  &  9.7$\pm0.3$ & 18$^?$ & 0.43$\pm0.05$ &  20.9$\pm1.6$ & $-$ & $-$ & $-$ & 0.18$^3$\\
J1948$+$3540 & 717.3  & 57156.974  & 60  & 12.4(3)  & 14.7$\pm0.2$ & 17$^?$ & 0.30$\pm0.05$ &  21.3$\pm1.5$ & $-$ & $-$ & $-$ & $-$\\
J2022$+$5154 & 606.9  & 57157.017  & 30  & $-$(1)   &  6.7$\pm0.04$ & 14.1 & 6.31$\pm0.06$ & 293.0$\pm2.0$ & 6.6$^2$ & 13.8$^2$ & $1.73\pm0.35^4$ & 2.84$^3$ \\ 
J2048$-$1616 & 1961.0 & 57399.274  & 60  & 9.8(3)   & 10.9$\pm0.2$ & 13$^?$ & 0.30$\pm0.04$ & 31.8$\pm1.4$  & $-$ & $-$ & $-$ & $-$\\

\hline
\hline
\end{tabular}
\\
$^\dag$ Previously published results: 1: \citet{jkw06}, 2: \citet{kxj+97}, 3: \citet{mgs+81}, 4: \citet{msk+13}\\
$?$ $W_{10}$ estimate has low S/N.
}
\end{sidewaystable*}

{\bf PSR J0437$-$4715.} Since its discovery by \citet{jlh+93}, PSR
J0437$-$4715 has been extensively studied as the closest and brightest
binary millisecond pulsar. \citet{nms+97} found that there are at
least 10 pulse components in the profile and observed a
frequency-dependent lag of the total-intensity profile with respect to
the polarization profile. \citet{ymv+11} showed that multiple
overlapping components cover at least 85\% of the rotation period at
1.3~GHz. The integrated pulse profile at 8.6~GHz in
  Fig.~\ref{fg:prf} shows over-lapping central components and weaker
  emission on both sides to outlying components around $104\degr$ and
  $250\degr$. On the trailing side especially, there is a bridge of
  emission between the central and outlying
  component. Fig.~\ref{fg:4freq} shows that the trailing central
  component 2 has a steeper spectrum than the leading component 1 and
  that the component separation is essentially independent of
  frequency \citep[cf.,][]{dhm+15}.  At 8.6~GHz the center components
  are more dominant and the side components (around $140\degr$ and
  $230\degr$) are relatively weaker than the components at $104\degr$
  and $250\degr$.

{\bf PSR J0659$+$1414 (B0656$+$14).} This pulsar is a nearby
middle-aged pulsar that is a strong source of pulsed high-energy
emission \citep{waa+10}. In the radio band, \citet{wwsr06} found that
profile instabilities are caused by very bright and narrow pulses
which may be related to the pulses observed from RRATs (Rotating Radio
Transients). The TMRT 8.6~GHz pulse profile shown in Fig.~\ref{fg:prf}
has relatively low S/N but is similar to the 8.4~GHz of
\citet{jkw06}. This pulsar has a single component profile at all radio
frequencies \citep{lylg95,hoe99} and gets narrower with increasing
frequency.

{\bf PSR J0738$-$4042 (B0736$-$40).} Our 8.6~GHz profile for this
pulsar shown in Fig.~\ref{fg:prf} has two obvious components and a
third overlapping component on the leading edge. As
Fig.~\ref{fg:4freq} shows, this leading component is not as obvious in
the 8.4~GHz Parkes profile of \citet{jkw06}. Systematic variations in
the leading components of the pulse profile at frequencies around
1.4~GHz that are related to changes in spin-down rate have been found
for this pulsar \citep{krj+11,bkb+14}. It is likely that the
differences in the 8~GHz profiles are related to the changes seen at
lower frequencies.

{\bf PSR J0742$-$2822 (B0740$-$28).} At 8.6~GHz the profile
(Fig.~\ref{fg:prf}) has two clear components.  At lower frequencies,
the pulsar consists of as many as seven components \citep{kwj+94} but
the weaker ones are not visible in the TMRT profile. The width of
pulse profile is almost the same over a wide frequency range
\citep{cw14}. In observations at 1.4~GHz and 3.1~GHz \citep{ksj13},
this pulsar has an interesting mode-changing behaviour on timescales
of several hundred days that appears to be modified by a glitch.  In
the more common Mode I the trailing component is relatively weaker
than in the less common Mode II.  Fig.~\ref{fg:mode} shows that the
TMRT profile differs from the 8.4~GHz profile of \citet{jkw06} in the
same way as for the two modes observed at lower frequencies. Evidently
the Parkes data were taken when the pulsars was in Mode I, whereas the
pulsar was in Mode II during the TMRT observations.

{\bf PSR J0837$-$4135 (B0835$-$41).} This pulsar was discovered at
408~MHz by \citet{lvw68} and \citet{wmz+01} used the Nanshan radio
telescope at Xinjiang Astronomical Observatory to update the period
and period derivative. The 8.6~GHz profile for this pulsar
(Fig.~\ref{fg:prf}) has a strong central component with outlier
components on each side. As Fig.~\ref{fg:4freq} shows, the frequency
evolution of the profile shows classical behaviour with the outlier
components having flatter spectra so that at 1.4 and 3.1~GHz the
central component completely dominates the profile \citep{kj06}. 
The separation of the components in longitude is only weakly 
dependent on frequency.

{\bf PSR J0953$+$0755 (B0950$+$08).} This famous pulsar, one of the
original four pulsars discovered at Cambridge \citep{phbc68}, has been
extensively studied over a wide frequency range. The profile has a
weak interpulse and bridge emission with the separation of the main
component and interpulse close to $150\degr$ over a wide frequency
range \citep{hf86}. At low frequencies, the main pulse has two
distinct components \citep{phs+16} but at higher frequencies there is
just one peak with an extension on the leading side. The 8.6~GHz
profile shown in Fig.~\ref{fg:prf} has this form. Giant pulses and
microstructure have been detected at different frequencies
\citep{han71,cjd04,tsa+16}.

{\bf PSR J1136$+$1551 (B1133$+$16).} This is another of the original
four pulsars discovered at Cambridge \citep{phbc68} that has a classic
two-component profile and has been studied at many different
frequencies. At low radio frequencies, the two components are of
comparable amplitude \citep{bkk+16} but, as Fig.~\ref{fg:prf} shows,
at 8.6~GHz, the trailing component is much weaker, indicating a much
steeper spectrum. \citet{tho91a} has shown that the component
separation $\Delta\phi$ for most double profiles can be fitted by the
function of the form:
\begin{equation}
  \Delta\phi = A\nu^{-\beta} + \Delta\phi_{\rm min}.
\end{equation}
where $\nu$ is in MHz. 
For PSR J1136$+$1551, fitting to data at frequencies between 26~MHz and 10~GHz gives
$A=53\degr$, $\beta=0.50$ and $\Delta\phi_{\rm min}=4.4\degr$. For
8.6 GHz, this equation gives $\Delta\phi=5.0\degr$, in agreement with
the measured value $5.1\degr$ (Table~\ref{tab:fluxpsr}). 

{\bf PSR J1239$+$2453 (B1237$+$25).} The mean pulse profile for this
well-known bright pulsar has five distinct components at low radio
frequencies and exhibits dramatic mode changing affecting mainly the
central and trailing components \citep{sr05}. The polarisation
properties are consistent with a very small impact parameter, i.e.,
the magnetic axis passes within a fraction of one degree of the line
of sight to the pulsar, so essentially a full diameter of the beam is
traversed \citep{sr05}. Fig.~\ref{fg:4freq} shows that, at
  higher frequencies, the outer components more closely spaced. The
  component separation at 8.6~GHz is $8.3\degr$
  (Table~\ref{tab:fluxpsr}), which is close to the $8.6\degr$
  predicted by \citet{tho91a} from a fit to data at frequencies
  between 80 and 4900~MHz. Fig.~\ref{fg:4freq} also shows that, at
  4.8~GHz and 8.6~GHz, the trailing component is stronger than the
  leading component, whereas the opposite is true at frequencies
  around and below 1~GHz. Clearly, of the two components, the trailing
  one has a flatter spectrum. Given the low S/N of the TMRT profile,
  it is not possible to identify systematic changes with frequency in
  the inner three components.

{\bf PSR J1644$-$4559 (B1641$-$45).} This is the second
brightest pulsar at 1.4~GHz and, at frequencies around 1~GHz, consists
of an asymmetric primary component with a weak leading component
\citep{mhm80,kj06}. At lower frequencies, the profile is broadened by
interstellar scattering.  Although the signal-to-noise ratio is low,
\citet{kjlb11} showed that the leading component is nearly as strong as the
main component at 17~GHz. Our 8.6~GHz profile (Fig.~\ref{fg:prf}) and
that of \citet{jkw06} clearly show the leading component and also a
similar trailing component. The leading and trailing components both
have relativly flat spectra and clearly are conal emission. Although
we do not know the impact parameter, the relatively flat PA variation
\citep{jkw06} as well as the strong core component suggests it must be
small so that the beam radius is probably about half of the separation of the
leading and trailing components, i.e., close to
$10\degr/\sin\alpha_B$, where $\alpha_B$ is the magnetic inclination
angle. 

{\bf PSR J1645$-$0317 (B1642$-$03).}  As Fig.~\ref{fg:4freq} shows, in
terms of the pulse profile and its frequency dependence, PSR
J1645$-$0317 is essentially a twin of PSR J1644$-$4559, except that
the outlier conal components are relatively stronger and are evident at a
lower frequency for PSR J1645$-$0317. At frequencies above 8.6~GHz, the
leading component dominates the profile whereas the central component
is weak, showing that it has a much steeper spectral index. The
apparent beam radius is a little smaller than for PSR J1644$-$4559,
about $6\degr/\sin\alpha_B$.

{\bf PSR J1705$-$1906 (B1702$-$19).} This pulsar has a relatively
short period ($\sim 0.299$~s) and, at frequencies around 1~GHz, a
strong interpulse separated from the main pulse by $180\degr$ of
longitude \citep{blh+88}. The observed polarisation variations can be
interpreted in terms of an orthogonal rotator with the main pulse and
interpulse from opposite poles \citep[e.g.,][]{kj04}. However, 
\citet{wws07} find a $10P$ modulation in both the main pulse and 
interpulse that remains phase-locked over years, a result that 
is difficult to explain in the two-pole model.

Our observations at 8.6~GHz (Fig.~\ref{fg:prf}, plotted with
256-bin/period resolution to improve the S/N) clearly shows the interpulse
separated from the main pulse by $179\degr \pm 1\degr$ and with a
peak flux density about 6\% of that of the main pulse. The
interpulse evidently has a spectrum relative to the main pulse which
peaks around 1 GHz, with flux density ratios of approximately 0.20,
0.24 and 0.45 at 408 MHz, 610 MHz and 1.4 GHz respectively
\citep{gl98}, and 0.30 at 4.85~GHz \citep{kkwj98}. At 8.6~GHz,
Fig.~\ref{fg:4freq} shows that the main pulse has at least two
components and that the pulse profile changes little between
frequencies of 900~MHz and 8.6~GHz.

{\bf PSR J1709$-$1640 (B1706$-$16).} The 8.6~GHz pulse profile shown
for this pulsar in Fig.~\ref{fg:prf} is dominated by a single strong
component with a weak component at its leading edge. These
components are also seen in the 4.75~GHz profile presented by
\citet{kxj+97}. Their 8.5~GHz profile is affected by systematic
baseline noise and these weaker components are not
visible. \citet{cw14} show that the width of the main component
decreases with increasing frequency, and our 8.6~GHz profile is
consistent with that.

{\bf PSR J1709$-$4429 (B1706$-$44).} PSR J1709$-$4429 is a young
Vela-like pulsar that shows both high-energy emission
\citep[e.g.,][]{aaa+10m} and intermittent strong micropulses from a
small phase range near the trailing edge of the pulse profile
\citep{jr02}. The radio pulse profile is simple with one dominant
component and is highly linearly polarised \citep{qmlg95,jkw06}. Our
8.6~GHz pulse profile is similar to that of \citet{jkw06}, showing a
single component of 50\% width about $16\degr$ of longitude. This is
significantly narrower than the 50\% width, about $21\degr$,
at 1.4 GHz \citep{hfs+04}.

{\bf PSR J1721$-$3532 (B1718$-$35).} This pulsar has a relatvely short
pulse period ($\sim 280$~ms) and a very high dispersion measure (496
cm$^{-3}$~pc) \citep{hlk+04}. At 8.6~GHz, Fig.~\ref{fg:prf} shows that
this pulsar has a simple but asymmetric profile with a slow rising
edge and a steeper falling edge \citep[cf.,][]{jkw06}. At 1.4 GHz, the
profile is highly scattered \citep{hlk+04} with a 1~GHz scattering
timescale of about 78~ms \citep{joh90}.

{\bf PSR J1740$-$3015 (B1737$-$30).} This pulsar is young
(characteristic age about 20,000 years) and has a large surface-dipole
magnetic field ($\sim 1.7\times 10^{13}$~G). Its main claim to fame is
the high rate of glitch occurrence, with more than 30 observed at an
average rate of about one per year \citep[][and the ATNF Pulsar
  Catalogue Glitch Table]{elsk11}. At frequencies around 1~GHz, the
profile for this pulsar is dominated by a single component
\citep[e.g.,][]{gl98} but as Fig.~\ref{fg:prf} shows, at 8.6 GHz the
peak of the profile splits into two components. Parkes observations at
a similar frequency \citep{kjlb11} show the same profile shape. Above
1~GHz, the profile 50\% width is approximately constant \citep{cw14},
but at lower frequencies it is affected by interstellar scattering
\citep{kmn+15}.

{\bf PSR J1745$-$3040 (B1742$-$30).} Fig.~\ref{fg:prf} shows that at
8.6~GHz this pulsar has a two-component profile, with a component
separation of slightly less than $10\degr$
(Table~\ref{tab:fluxpsr}). Fig.~\ref{fg:4freq}
shows that, toward lower frequencies, the leading component
becomes relatively weaker, a central component appears and the
separation between the leading and trailing components increases
greatly \citep[cf.,][]{gl98}.   

{\bf PSR J1752$-$2806 (B1749$-$28).} This strong pulsar, discovered by
\citet{tv68}, lies within $2\degr$ of the Galactic Centre. At
frequencies of a few GHz the pulse profile is dominated by a single
component. At frequencies below about 400~MHz, profile broadening due
to interstellar scattering is evident \citep{abs86}. Fig.~\ref{fg:prf}
shows that at 8.6~GHz there is a trailing component with a peak flux
density about 40\% that of the leading component. This trailing
component can be seen at 3.1~GHz and 4.75~GHz (Fig.~\ref{fg:4freq}),
becoming relatively weaker with decreasing frequency. At frequencies
around 1~GHz, a central, probably core, component is visible.

{\bf PSR J1803$-$2137 (B1800$-$21).} This young Vela-like pulsar has
an extraordinarily wide double profile with $W_{10}$ at 1.4 GHz of
43~ms or $115\degr$ of longitude \citep{hfs+04}. The spectrum has a
peak at around 1~GHz with a variable low-frequency spectrum probably
due to varying interstellar absorption \citep{brl+16}. 

At 8.6~GHz, Fig.~\ref{fg:prf} shows that the profile retains its
wide-double form. $W_{10}$ is $81\degr$ (Table~\ref{tab:fluxpsr})
showing that the component separation is a strong function of
frequency. Fig.~\ref{fg:4freq} illustrates this and shows that the
leading component has a relatively flatter spectrum, being barely
visible at 610~MHz \citep[cf.,][]{brl+16} and becoming prominent at
higher frequencies. At low frequencies, a central core component also
becomes evident.

{\bf PSR J1807$-$0847 (B1804$-$08).} At frequencies around 1~GHz, this
pulsar has a clear three-component pulse profile
\citep[e.g..][]{hfs+04}. As shown in Fig.~\ref{fg:prf}, at 8.6~GHz the
profile has a classic double form with a steep outer edges and a
connecting bridge of emission. Fig.~\ref{fg:4freq} shows that the
central component has a steeper spectrum and this along with its
central location marks it as a core component. There is no
  significant frequency dependence of component separation.

{\bf PSR J1829$-$1751 (B1826$-$17).} At 8.6~GHz Fig.~\ref{fg:prf}
shows that the mean pulse profile has two isolated components, with
the trailing one about twice as strong as the leading one. However, at
0.925 and 1.408~GHz \citep[Fig.~\ref{fg:4freq};][]{gl98}, a central
core component is clearly visible and the leading and trailing
components are of comparable strength, indicating a wide variation in
spectral index across the profile. Overall, the profile frequency
evolution is very similar to that of PSR J1807$-$0847, although there
is a more significant narrowing of the profile width with increasing
frequency for PSR J1829$-$1751.

{\bf PSR J1848$-$0123 (B1845$-$01).} Our 8.6~GHz profile has two
components with a bridge between them
(Fig.~\ref{fg:prf}). Fig.~\ref{fg:4freq} shows that the profile
frequency evolution in this pulsar is very similar to the preceding
two pulsars, with a central core component becoming dominant at
frequencies around 1.4~GHz. There is little change in pulse
width between 1~GHz and 10~GHz.

{\bf PSR J1932$+$1059 (B1929$+$10).} This is a nearby, bright and
isolated pulsar which has a relatively weak interpulse preceding the
main pulse by about $170\degr$ of longitude \citep[e.g.,][]{stc99}. At
low frequencies, emission can be seen over most of the pulse period
allowing detection of a double-notch feature trailing the main pulse
by about $100\degr$ of longitude \citep{mr04}. This feature similar to
those seen in PSR J0437$-$4715 and PSR J0953$+$0755 (B0950$+$08) which 
also have detectable emission over most of the pulse period. At 8.6~GHz 
the profile has broad wings and two identifiable components near the
profile peak (Fig.~\ref{fg:prf}). The multi-frequency profiles in
Fig.~\ref{fg:4freq} \citep[see also][]{hr86} show that the leading
component has a flatter spectrum than the central and trailing
components. At 1.4~GHz the main pulse has a 50\% width of about
$12\degr$ \citep{hfs+04}, whereas at 8.6~GHz the pulse is much
narrower with $W_{50}$ about $6\degr$ (Table~\ref{tab:fluxpsr}).

{\bf PSR J1935$+$1616 (B1933$+$16).} \citet{srw75} showed the
frequency evolution in this profile from a single dominant component
at 430~MHz to three components above 2~GHz. This is a classic
core-conal structure with the core region dominating at low
frequencies and conal outriders appearing at higher
frequencies. Our 8.6 GHz profile (Fig.~\ref{fg:prf}) shows
  that this evolution continues to higher frequencies, with the
  central and trailing components of comparable strength while the
  leading component is stronger than at lower frequencies
  (Fig.~\ref{fg:4freq}) but only about 30\% as strong as the other
  two components. The higher S/N profiles at 925~MHz \citep{gl98} and
1418~MHz \citep{wcl+99} given in Fig.~\ref{fg:4freq} show that the
core emission consists of two overlapping components. This is
consistent with the gradual evolution from core emission to conal
emission across the profile discussed by \citet{lm88}.

{\bf PSR J1948$+$3540 (B1946$+$35).} Even more than PSR J1935$+$1616
discussed above, PSR J1948$+$3540 shows the evolution from
core-dominated at low frequencies to cone-dominated at high
frequencies (Fig.~\ref{fg:4freq}). The 8.6~GHz integrated pulse
profile given in Fig.~\ref{fg:prf} has a basically double pulse
profile although there is evidence for a weak central or core
component. At least above 1~GHz, the  pulse phases of three components
and the overall pulse width are stable (Fig.~\ref{fg:4freq}).

{\bf PSR J2022$+$5154 (B2021$+$51).} PSR J2022$+$5154 is one of the
strongest pulsars at high frequencies with detections up to 43~GHz
\citep[e.g.,][]{kjdw97}. At low frequencies, e.g., around 400~MHz, the
profile is double-peaked with the trailing component about twice as
strong as the leading one and a component separation of about
$8\degr$ \citep{gl98}. Around 1.4~GHz, the leading component evidently
disappears and the trailing component bifurcates to two overlapping
components. Fig.~\ref{fg:prf} shows that, at 8.6~GHz, only a single
component with a 50\% width of $6.7\degr$ (Table~\ref{tab:fluxpsr}) is evident. 

{\bf PSR J2048$-$1616 (B2045$-$16).} This well known pulsar,
discovered by \citet{tv68}, has a triple component profile at
frequencies of a few GHz and below and a position-angle swing
indicating a traverse of the polar cap with low impact angle, i.e.,
the line of sight passing close to the magnetic axis
\citep[e.g.,][]{man71b}. Fig.~\ref{fg:prf} shows that at 8.6~GHz the
central component has almost disappeared. It therefore has a steeper
spectrum and is consistent with core emission despite being offset from
the profile centre \citep[cf.,][]{lm88}. Fig.~\ref{fg:4freq} shows
that the trailing component becomes relatively stronger at high
frequencies and that the component separation is a decreasing function
of frequency \citep[cf.,][]{cw14}.

\section{Discussion}\label{sec:discn}

Pulsar integrated profiles come in many different forms with different
numbers of pulse components, different separations, sometimes
interpulses and different spectral behaviour for the different
components. In most cases, profiles and components are narrower and
weaker at higher frequencies. These different behaviours give
important information about the structure of pulsar emission regions
and the radiation mechanisms.

Even though there are very many observations of integrated pulse
profiles, the sample at high frequencies is relatively limited,
largely because of the typically steep spectrum of pulsar
emission. The 8.6~GHz observations of 26 pulsars reported here nearly
double the number of published high-quality high-frequency pulse
profiles.

\subsection{Frequency dependence of profile widths}\label{sec:w-nu}
It has long been known that for most pulsars the profile width, or
component separation for multiple-component pulsars, decreases with
increasing radio frequency \citep[e.g.,][]{mt77,cor78}, at least at
frequencies below about 1~GHz. This was often modelled as a power law
$\Delta\phi \sim \nu^\beta$ with $\beta$ typically about $-0.25$. At
higher frequencies, pulse widths become more frequency-independent,
leading to two-component power-law models with a break at some
frequency, typically about 1~GHz \citep{sba87}. As discussed above in
\S\ref{sec:results}, an alternative model (Eq. 6) with a single power law
combined with a minimum pulse width was shown by \citet{tho91a} to
accurately describe the frequency dependence for many pulsars.

Our 8.6~GHz observations add significantly to the available data on
high-frequency pulse widths. Of the 26 pulsars, 19 have two or more
clearly resolved components, allowing a measurement of $\Delta\phi$ as
listed in Table~\ref{tab:fluxpsr}.\footnote{PSR J1740$-$3015 is
  omitted since the components are not clearly resolved.} These
component separations are compared with other measurements made over a
range of frequencies \citep{sgg+95,hx97,gl98,wcl+99,dhm+15} in
Fig.~\ref{fg:comp-sep}. The observed frequency dependencies are
closely power-law and appear to divide into two groups, those with a
signficant frequency dependence and those which are essentially
frequency-independent. The frequency range fitted, the power-law
indices ($\beta$) and the rms residuals from the fit ($\sigma$) are
given in Table~\ref{tb:pl1} for the first group and Table~\ref{tb:pl2}
for the second group. Table~\ref{tb:pl2} also gives the mean
separation of the outermost components, $\langle\Delta\phi\rangle$,
for the pulsars with frequency-independent component
separations. These power-law frequency dependencies are shown in
Fig.~\ref{fg:4freq} for each pulsar, with the phase of the dashed
lines for each profile representing the predicted phase for the
frequency of that profile.

\begin{table}[h]
\caption{Power-law ($\nu^{\beta}$) indices for component separation
  for the nine pulsars with decreasing separation at higher
  frequencies.}\label{tb:pl1}
\begin{tabular}{clcc}
\hline
PSR  & Frequency& $\beta$ & $\sigma$ \\
 Name  & ~~(MHz)  &   & ($\degr$) \\
 \hline\\
 J0738$-$4042 & 1375$-$8600      & $-0.15\pm0.04$  & 0.4 \\
 J0742$-$2822 & 1375$-$10550     & $-0.10\pm0.03$  & 0.2 \\
 J1136$+$1551 & 2250$-$10450     & $-0.09\pm0.01$  & 0.1 \\
 J1239$+$2453 & $\,\ $610$-$8600 & $-0.07\pm0.01$  & 0.2 \\
 J1745$-$3040 & $\,\ $610$-$8600 & $-0.17\pm0.04$  & 0.9 \\
 J1803$-$2131 & 1642$-$8600      & $-0.30\pm0.05$  & 2.0 \\
 J1829$-$1751 & $\,\ $925$-$8600 & $-0.07\pm0.02$  & 0.3 \\
 J1932$+$1059 & 1414$-$10450     & $-0.18\pm0.02$  & 0.1 \\
 J2048$-$1616 & $\,\ $408$-$8600 & $-0.09\pm0.02$  & 0.5 \\
\hline
\end{tabular}
\end{table}

\begin{table}[h]
\caption{Power-law indices ($\nu^{\beta}$) for component separation
  the ten pulsars with essentially constant component separation at
  higher frequencies.}\label{tb:pl2}
\begin{tabular}{clrcr}
\hline
PSR    & Frequency& $\beta$~~~  & $\sigma$ & $\langle\Delta\phi\rangle$ \\
 Name  & ~~(MHz)  &  &($\degr$)~  & ($\degr$)~ \\
 \hline\\
 J0437$-$4715 & $\,\ $728$-$8600 & $0.01\pm0.01$		& 1.2  & 141.5\\
 J0837$-$4135 & 1375$-$8600		& $0.04\pm0.04$ 		& 0.6  & 11.3 \\
 J1644$-$4559 & 3100$-$8600		& $-0.03\pm0.03$ 	& 0.3  & 21.4 \\
 J1645$-$0317 & 1410$-$10550	 	& $0.003\pm0.009$ 	& 0.1  & 12.1 \\
 J1705$-$1906 & $\,\ $925$-$8600 & $0.00\pm0.04$		& 0.3  & 4.9  \\
 J1752$-$2806 & 3100$-$8600		& $-0.02\pm0.04$ 	& 0.1  & 5.1  \\
 J1807$-$0847 & $\,\ $925$-$8600 & $0.02\pm0.01$		& 0.2  & 15.6 \\
 J1848$-$0123 & 1418$-$10550	 	& $-0.02\pm0.01$ 	& 0.2  & 13.1 \\
 J1935$+$1616 & 1418$-$8600		& $-0.03\pm0.01$ 	& 0.1  & 15.2 \\
 J1948$+$3540 & $\,\ $925$-$8600 & $-0.01\pm0.01$	& 0.2  & 12.8 \\
 \hline
\end{tabular}
\end{table}

Component separation indices for the first group range between $-0.07$ and
$-0.3$, whereas for the second group they are generally $0.00\pm
0.03$. For the pulsars in the first group there is no sign of a
flattening at the highest frequencies plotted and for those in the
second group, there is no sign of a steepening for the lowest
frequencies plotted. This is not necessarily inconsistent with earlier
results that indicate a change of power-law index as this is only seen
in a subset of all pulsars \citep[e.g.,][]{cw14} and, when present,
generally occurs around 1~GHz. Most of the plotted points are at
frequencies higher than this.

Most magnetic-pole emission models interpret the decreasing profile
width with increasing frequency in terms of radius-to-frequency
apping \citep[e.g.,][]{cor78}. Predicted indices range between $-0.14$
\citep{bgi88} and $-0.45$ \citep{vj73} with the well-known \citet{rs75}
model giving $-0.33$. Within the context of radius-to-frequency mapping,
the flattening out at high frequencies is interpreted in terms of a lower
limit to the altitude of the emission region, possibly at or close to
the neutron-star surface \citep[e.g.,][]{kxj+97}. 

 Interestingly, \citet{mr02a} found from an analysis of
  multiple-component profiles that, while outer conal pairs followed
  the \citet{tho91a} frequency dependence of component separation,
  inner conal pairs did not and have an essentially
  frequency-independent component separation. A similar dependence of
  frequency dependence for inner and outer cones was found by
  \citet{wgr+98} for PSR B1451$-$68. These observations provide an
  alternate explanation for the two types of frequency dependence that
  we observed, namely that for the frequency-independent cases, the
  emission is from an inner cone. However, as Fig.~\ref{fg:comp-sep}
  shows, the frequency-independent component separations are typically
  about the same as the frequency-dependent ones, which would not be
  expected if the former were inner cones. This simple comparison
  ignores the effects of line-of-sight impact parameter and magnetic
  inclination. Most of these pulsars show core emission, and so impact
  parameters should be small and have little effect. The effects of
  magnetic inclination are difficult to reliably quantify since most
  estimates are based on observed pulse widths and the assumption of a
  well-defined beam opening angle \citep{lm88,ran90}. Another relevant
  point is that at least two of the frequency-independent group (PSRs
  J1807$-$0847 and J1848$-$0123) have multiple components and the
  frequency independence appears to extend across all
  components. Millisecond pulsars such as PSR J0437$-$4715 are also
  exceptional and are discussed below.

  Most emission models assume that the radiation is emitted
  tangentially to the local magnetic field, giving a simple relation
  between the beam opening angle, the emission height and the radial
  position (from the magnetic axis) of the field line on the polar
  cap \citep[e.g.,][]{ran93,mr02a}. However, in the inverse Compton
  scattering (ICS) model \citep{ql98}, the different conal
  components result from different beam angles relative to the field
  direction on a given field line. The ICS model can account for the
  different observed frequency dependence of component separations
  for inner and outer cones \citep{qlz+01}.

For most millisecond pulsars \citep[e.g.,][]{dhm+15}, including PSR
J0437$-$4715 described in \S\ref{sec:results}, and some other pulsars with wide profiles, e.g., PSR J0953+0755 (B0950+08) also described in 
\S\ref{sec:results}, the observed component separation is frequency
independent. For these pulsars the emission region may be close to the 
light cylinder and caustic effects may be important in defining the observed
profile shape \citep{rmh10}, thereby negating the effects of
radius-to-frequency mapping and providing an alternative explanation
for frequency-independent component separations. 

\begin{table}[h]
\caption{Spectral indices for leading, central and trailing   components}\label{tb:comp-spec}
\begin{tabular}{cccc}
\hline
 PSR  & $\alpha_{\rm l}$ & $\alpha_{\rm c}$ & $\alpha_{\rm t}$ \\
 Name  &    &    &    \\
 \hline\\
  J0837$-$4135 & $-1.3\pm0.2$ & $\,\ -2.2\pm0.01$ & $\,\ -1.3\pm0.01$\\
 J1645$-$0317 & $-2.0\pm0.2$ & $-5.0\pm0.2$ & $-2.5\pm0.3$\\
 J1807$-$0847 & $-1.0\pm0.1$ & $-1.5\pm0.1$ & $-1.2\pm0.1$\\
 J1829$-$1751 & $-1.7\pm0.2$ & $-2.1\pm0.2$ & $-1.2\pm0.1$\\
 J1848$-$0123 & $-1.4\pm0.2$ & $-1.9\pm0.4$ & $-1.5\pm0.2$\\
 J1948$+$3540 & $-1.5\pm0.1$ & $-2.7\pm0.1$ & $-1.1\pm0.1$\\
 J2048$-$1616 & $\,\ -1.9\pm0.05$ & $\,\ -2.3\pm0.06$ & $\,\ -1.6\pm0.09$\\
\hline
\end{tabular}
\end{table}

\subsection{Spectral properties of components in integrated pulse
  profiles}\label{sec:comp-spec}
There is ample evidence that central and outer regions of observed
pulse profiles have different spectral properties, with the central
regions generally having steeper spectra
\citep[e.g.,][]{ran83,ran93,lm88}. Within the context of the
magnetic-pole model, this is generally interpreted as differing
properties for emission from the ``core'' and ``conal'' or outer 
regions of the emission cone. There is debate about whether the mechanism
for the core and cone emission is different \citep{ran83} or basically the
same with a gradation of properties across the polar cap
\citep{lm88}.

In Table~\ref{tb:comp-spec} we present spectral indices for the
leading, centre and trailing components for the \textbf{seven} of our 26
pulsars where these components are clearly visible over a range of
frequencies. These results combine data from our 8.6~GHz observations
with the observations at other frequencies as used in the analysis of
component separation (\S\ref{sec:w-nu}). Component flux densities were
estimated by fitting Gaussian profiles to the relevant components at
each frequency and taking the product of the component amplitude and
width. This procedure gives a better estimate of the total component 
flux density since component widths tend to be greater at lower
frequencies \citep[cf.,][]{wgr+98}.

In all seven cases, the spectral indices of the central components are
close to or more negative than those of the leading and trailing
components, further emphasizing this property of the pulsar emission
mechanism. In some cases the spectral index difference is large. For
example, for PSR J1645$-$0317 (B1642$-$03), the spectral index of
the central component is $-5.0$ compared to a mean spectral index for the 
outer components of about $-2.3$.

\subsection{Period dependence of core components}\label{sec:core}

  On the assumption that pulsar emission beams are bounded by
  the open field lines emanating from a polar cap, the observed pulse
  widths are determined by the altitude of the emission region
  relative to the light-cylinder radius, the line-of-sight impact
  parameter relative to the beam radius and the magnetic inclination
  angle $\alpha_B$ \citep[e.g.,][]{lm88}. To obtain beam radii,
  emission altitudes and magnetic inclination angles from these
  relations, the (linear) polarisation properties must be
  known. \citet{ran90} proposed a simpler relation based on the
  observed width of core components. If a gaussian beam profile is
  assumed for the components, then the observed half-power component
  width is independent of the impact parameter. Consequently,
  total-power measurements are sufficient. \citet{ran90} used the
  observed core-component half-power width $W_{50}^c$ at 1.0~GHz of 59
  triple or multi-component pulsars to establish a relation between
  the pulsar period and the lower bound of the observed widths:
  \begin{equation}\label{eq:core}
    W_{50}^{c,i} = 2\fdg45 P^{-1/2}.
  \end{equation}
  Larger observed widths are interpreted as magnetic inclination
  angles $\alpha_B \neq 90\degr$, that is non-orthogonal rotators, for
  which the observed pulse widths are greater than the intrinsic
  widths by a factor $1/\sin\alpha_B$. On the assumption that
  Equation~\ref{eq:core} accurately describes the intrinsic core
  beamwidth in pulsars, the angles $\alpha_B$ can be estimated from
  $\sin\alpha_B = W_{50}^{c,i}/W_{50}^c$.
  
  In Table \ref{tb:core-width} we give measured core-component
  half-power widths at 8.6 GHz, $W_{50}^{c,8.6}$, for pulsars with
  multiple components at this frequency. We also give measured core
  widths from \citet{ran90} for these pulsars (excepting the Crab
  pulsar, where \citet{ran90} gives the width of the precursor
  component, whereas we adopt the interpulse width at 8.4~GHz from
  \citet{mh99}). The average ratio of 8.6~GHz width to 1.0~GHz width
  for these pulsars is about 0.85, so we adopt a scale factor for the
  8.6~GHz intrinsic core widths of $2\fdg1$, viz.,
  \begin{equation}\label{eq:core8}
    W_{50}^{c,i,8.6} = 2\fdg1 P^{-1/2}.
  \end{equation}
  Figure~\ref{fg:wp} shows measured 8.6~GHz core widths
  $W_{50}^{c,8.6}$ and the two width-period relations. We then use
  Equation~\ref{eq:core8} to compute the magnetic inclination angles
  $\alpha_B$ given in the second-last column of
  Table~\ref{tb:core-width}. In some cases these angles are similar to
  those reported by \citet{ran90}, given in the final column of
  Table~\ref{tb:core-width}, but in other cases, e.g., PSRs
  J0742$-$2822 (B0740$-$28) and J1752$-$2806 (B1749$-$28), there are
  wide discrepancies. These discrepancies illustrate the considerable
  uncertainties in magnetic inclination angles derived using the the
  core-width method. Interestingly, the inclination angle of $57\degr$
  derived from the 8.6~GHz interpulse width for the Crab pulsar is the
  same (within the uncertainties) as that derived by \citet{mh99} from
  a fit of the rotating-vector model to the observed 1.4~GHz polarisation
  position angle variations.

\begin{table}[h]
\caption{Core widths and magnetic inclination angles for 12 pulsars with multiple 
components}\label{tb:core-width}

\begin{tabular}{l r r r r r r r r r r r}
\hline

PSR             &       $P$   &$W_{50}^{c,8.6}$ & Ref. &
$W_{50}^{c,1.0}$ & $\alpha_B^{8.6}$ & $\alpha_B^{1.0}$\\
Name            &       (s) &	(deg)          &      & (deg) & (deg)
& (deg)\\
\hline\\                                                              
J0534+2200i     & 0.0331    &   $13.7\pm0.8$   & 1    &  -- 			& $57^{+6}_{-5}$ & 86 \\
J0742$-$2822    & 0.1668    &   $5.2\pm0.2$    & 3    &  10          &$80^{+10}_{-9}$ & 37\\
J0835$-$4510    & 0.0892    &   $6.3\pm0.1$    & 2    &  $8.2\pm0.3$ & 90 & 90 \\
J0837$-$4135    & 0.7516    &   $2.7\pm0.1$    & 3    &  3.7         & $63\pm3$ & 50\\
J1644$-$4559    & 0.4551    &   $4.81\pm0.03$  & 3    &   6.7        &$40.2\pm0.3$ & 33\\
J1645$-$0317    & 0.3876    &   $3.2\pm0.2$    & 3    &  4.2         &$90_{-10}$ & 70\\
J1705$-$1906i   & 0.2990    &   $4.5\pm1.1$    & 3    &  $4.5\pm0.3$ &$58^{+23}_{-15}$ & 85\\
J1752$-$2806    & 0.5292    &   $2.9\pm0.1$    & 3    &  5.0         &$82^{+8}_{-9}$ & 41\\
J1807$-$0847    & 0.1637    &   $7.6\pm1.1$    & 3    &  $\sim 7$    &$43^{+10}_{-6}$ & 60\\
J1848$-$0123    & 0.6594    &   $7.8\pm0.8$    & 3    &  --          &$19^{+3}_{-2}$ & --\\
J1935$+$1616    & 0.3587    &   $4.7\pm0.3$    & 3    &  5.25        &$48\pm4$ & 51\\
J2048$-$1616    & 0.1961    &   $5.9\pm1.2$    & 3    &  4.0         &$53^{+29}_{-11}$ & 26\\

\hline
\hline
\end{tabular}\\
 1: \citet{mh99}; 2: \citet{jkw06}; 3: This paper.\\
\\
\end{table}

\section{Conclusions}\label{sec:concl}

Pulsars are usually very weak at high frequency because of their steep
power-law spectra, so high frequency observations are relatively
difficult. At frequencies around 8.6~GHz, less than 50 pulse profiles
have been published up to now. In this paper, we have presented
integrated pulse profiles at 8.6~GHz for 26 pulsars observed with the
Shanghai TianMa Radio Telescope, 11 of which have not been previously
published.  Comparison of 8.6~GHz profiles with those at lower
frequencies for 19 pulsars shows two distinct behaviours in the
profile width or, more specifically, the separation of the outermost
components. In nine cases, the component separation decreases with
increasing frequency, whereas in ten other pulsars, there is no
significant change in separation between about 1 GHz and 10 GHz. 
 For \textbf{seven} pulsars over the same frequency range we showed 
 that the spectral index of  the central component is steeper than 
 for the outer components. We give the observed
 core half-power widths of 12 pulsars around 8.6~GHz and obtain a
 modified width-period relation for 8.6~GHz observations. Magnetic
 inclination angles derived using this relation are in some cases very
 different from those derived from lower-frequency data. Evidence for mode 
 changing in the high-frequency profile of PSR J0742$-$2822 was found 
 by comparing our 8.6~GHz profile with the 8.4~GHz profile of \citet{jkw06}.

\section*{Acknowledgements}
This work was supported in part by the National Natural Science
Foundation of China (grants 11173046, 11403073, U1631122, 11633007,
11373011 and 11673002), the Natural Science Foundation of
Shanghai No. 13ZR1464500, the Strategic Priority Research Program “The
Emergence of Cosmological Structures” of the Chinese Academy of
Sciences (grant No. XDB09000000 and XDB23010200), the Knowledge 
Innovation Program of the Chinese Academy of Sciences (grant 
No. KJCX1-YW-18), and the Scientific Program of Shanghai Municipality
(08DZ1160100). Pulsar profile data and parameters were obtained from the
European Pulsar Network profile database and the Australia Telescope
National Facility Pulsar Catalogue. We thank Ao-Bo GONG for valuable
suggestions during the data analysis.

\label{lastpage}

%\bibliographystyle{apj}
%\bibliography{journals,modrefs,psrrefs,crossrefs}

\clearpage

\begin{figure}[h]
\centering
\begin{center}
\begin{tabular}{cc}
\resizebox{0.53\hsize}{!}{\includegraphics[angle=0]{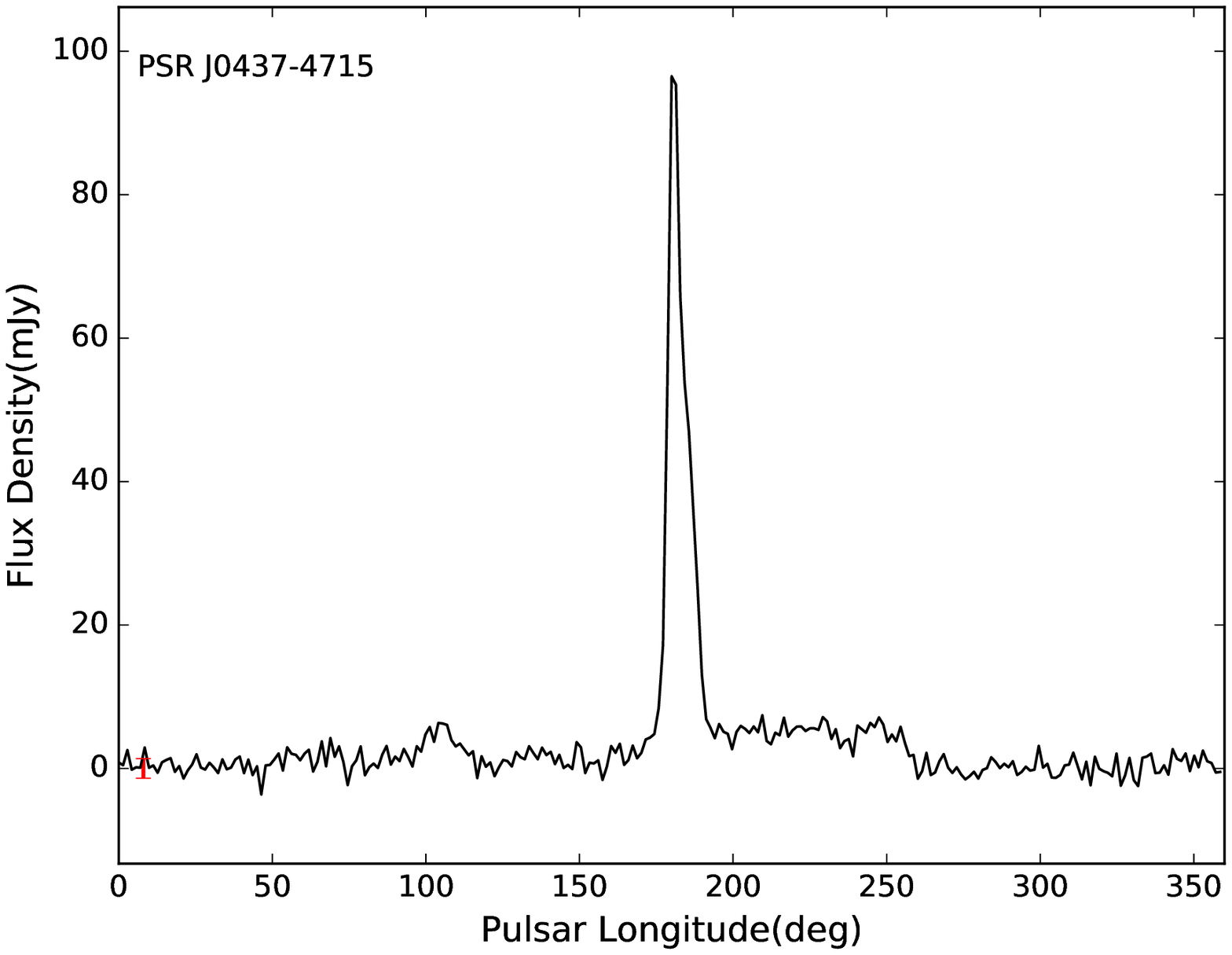}}&
\resizebox{0.53\hsize}{!}{\includegraphics[angle=0]{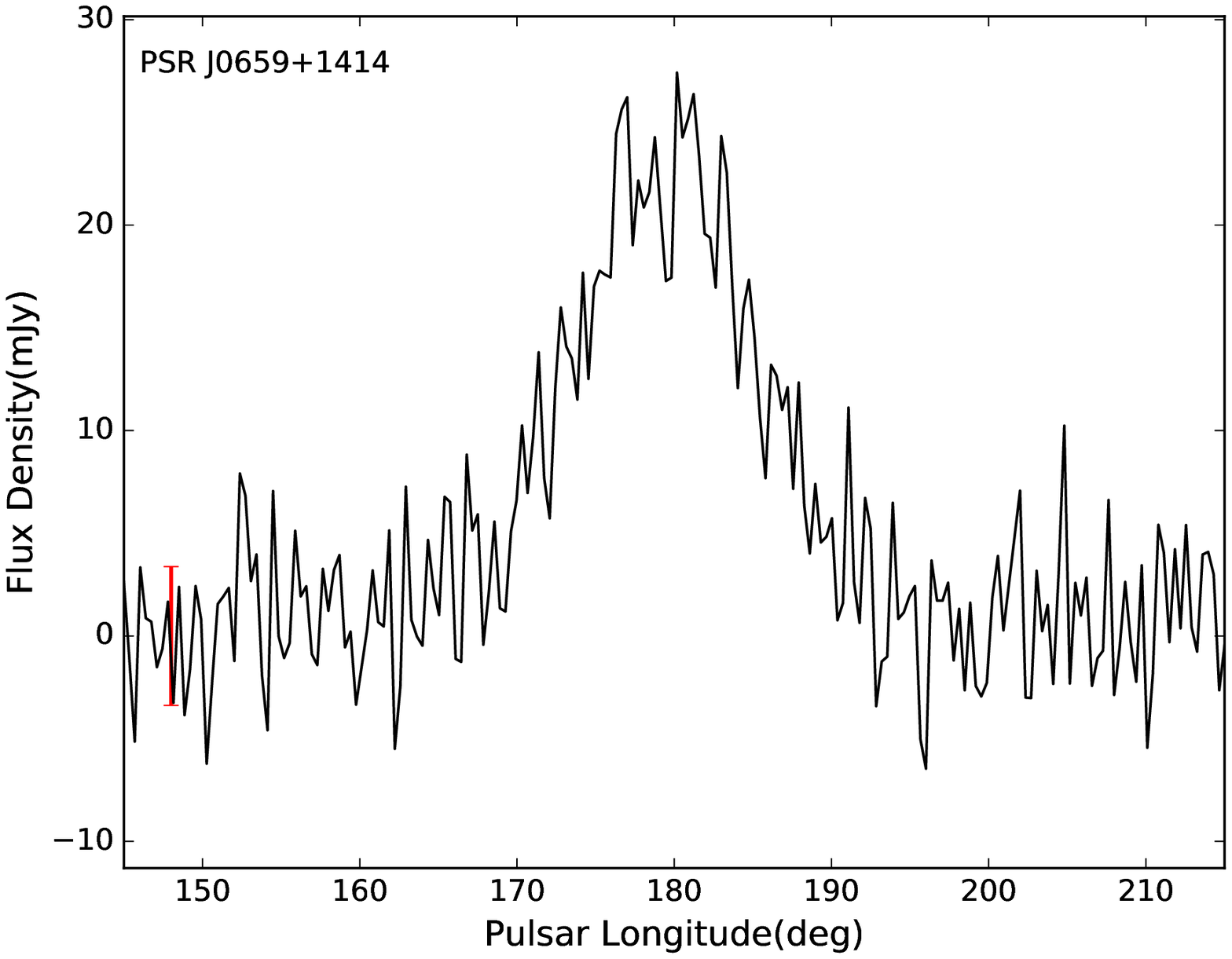}}\\
\resizebox{0.53\hsize}{!}{\includegraphics[angle=0]{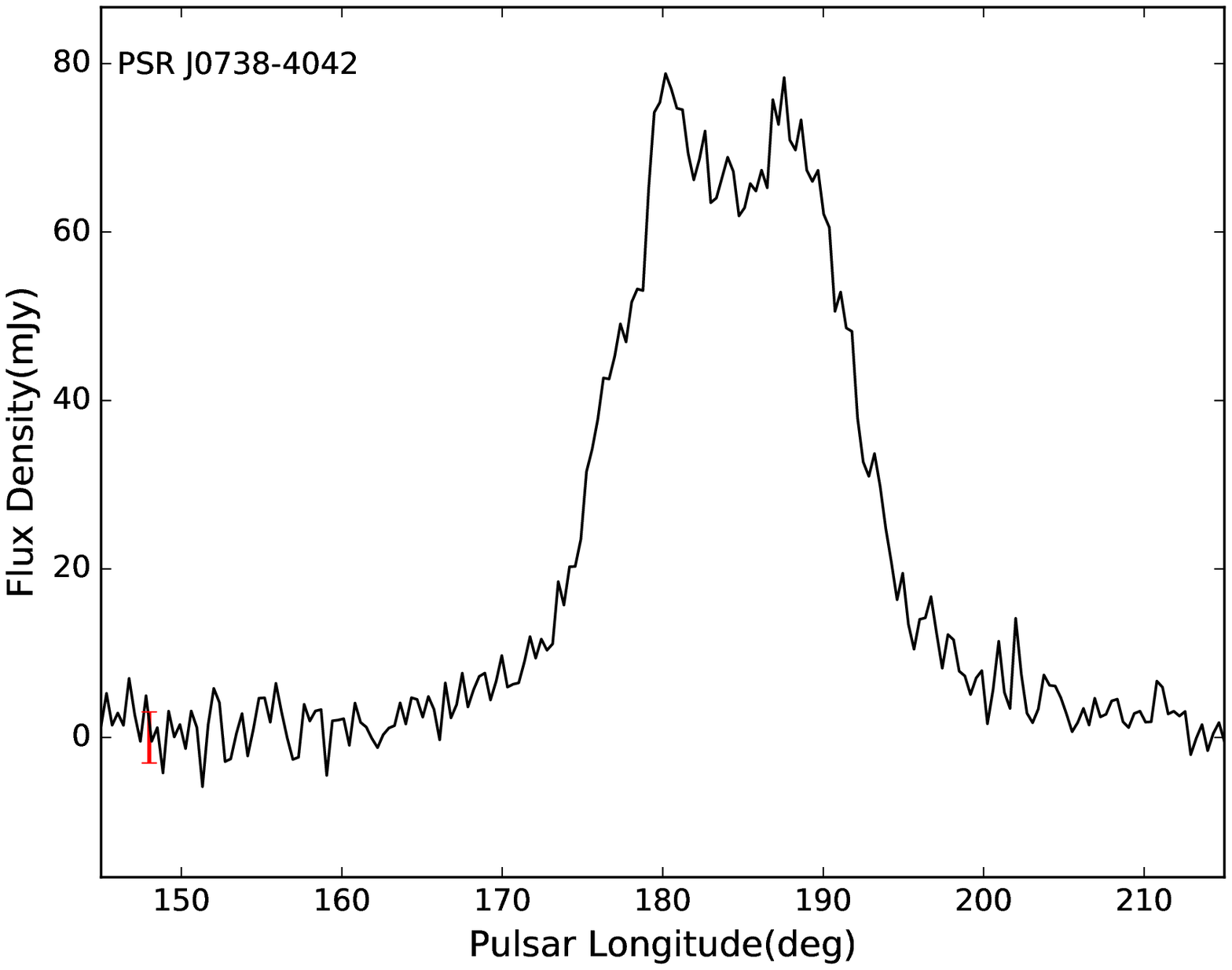}}&
\resizebox{0.53\hsize}{!}{\includegraphics[angle=0]{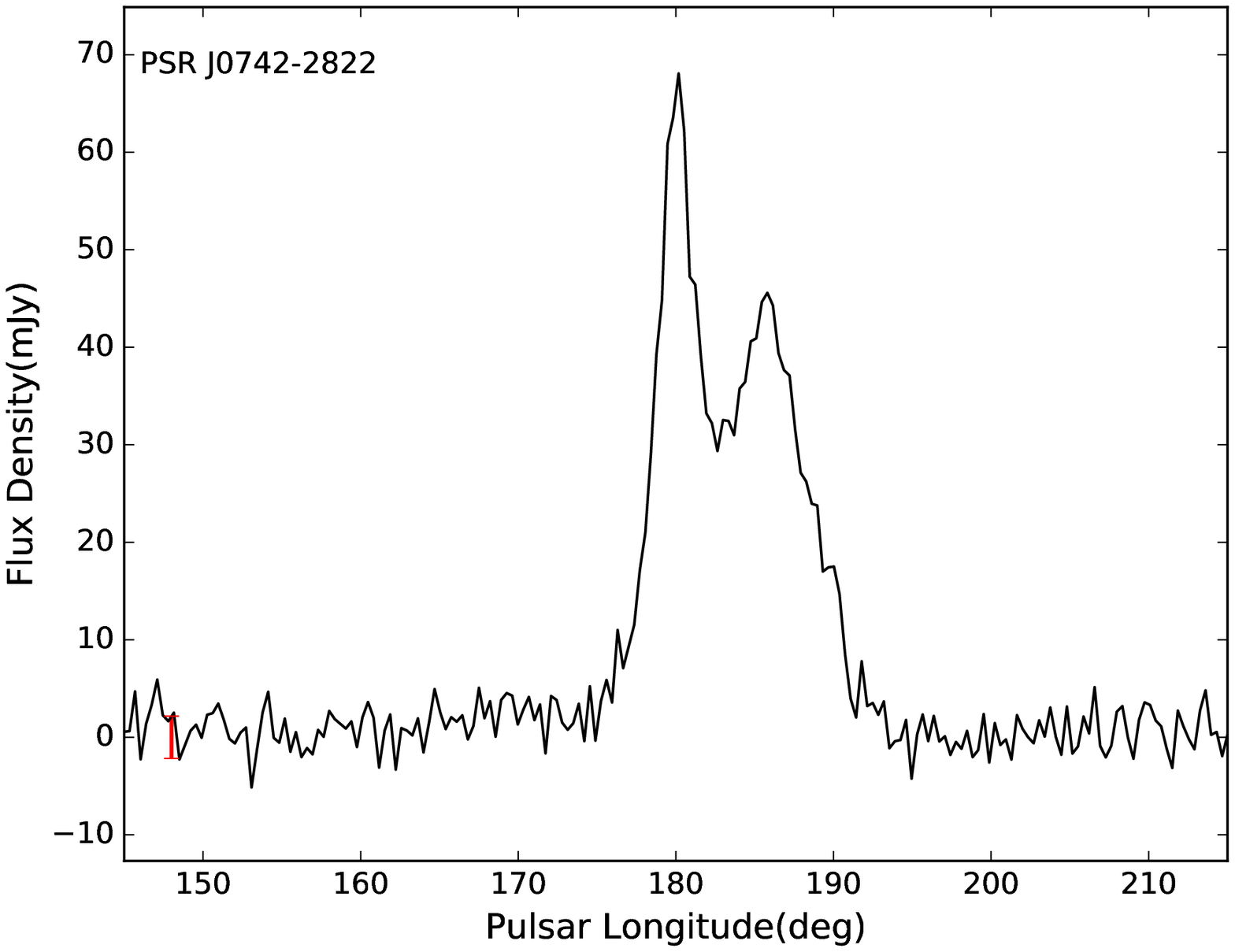}}\\
\resizebox{0.53\hsize}{!}{\includegraphics[angle=0]{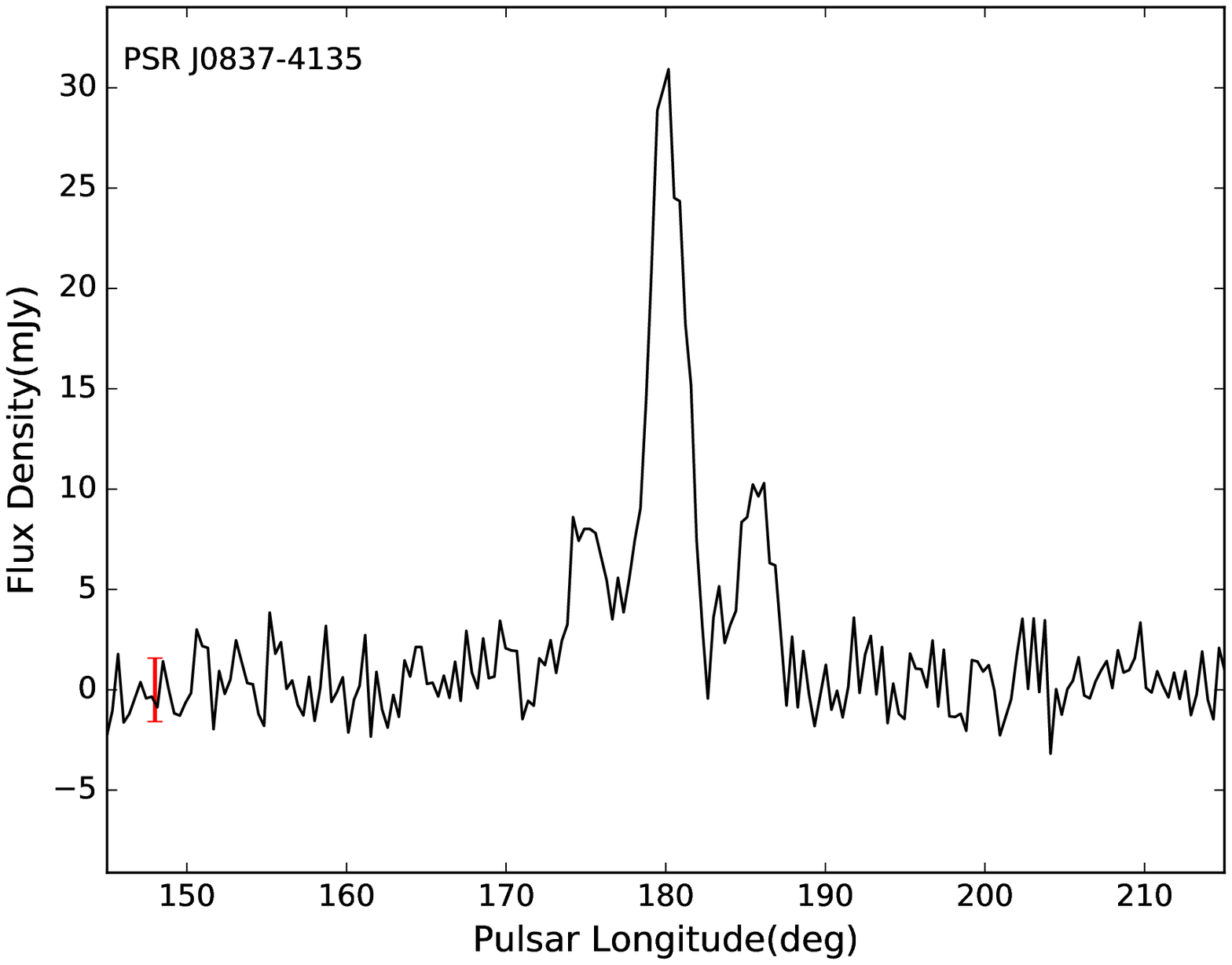}}&
\resizebox{0.53\hsize}{!}{\includegraphics[angle=0]{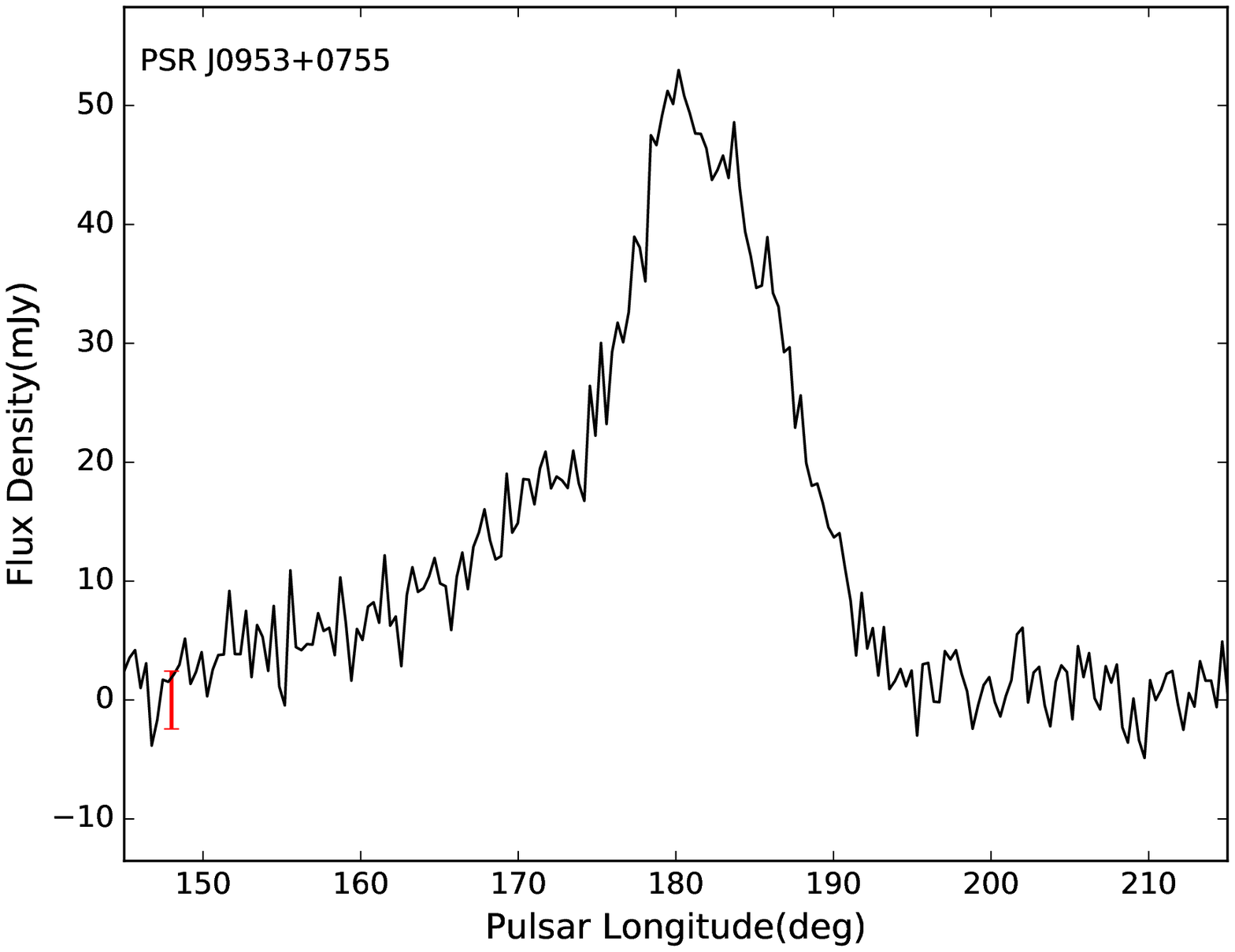}}\\

\end{tabular}
\end{center}
\caption{Integrated pulse profiles at 8.6~GHz obtained using the TMRT for 26
  pulsars. The red bar represents $\pm\sigma_b$, the rms
  baseline noise. \textbf{Except for PSRs J0437$-$4715 and J1705$-$1906, where there are
    256 bins across the pulse period, all profiles have 1024-bins/period resolution.}}\label{fg:prf}
\end{figure}
\addtocounter{figure}{-1}
\begin{figure}
\begin{center}
\begin{tabular}{cc}
\resizebox{0.53\hsize}{!}{\includegraphics[angle=0]{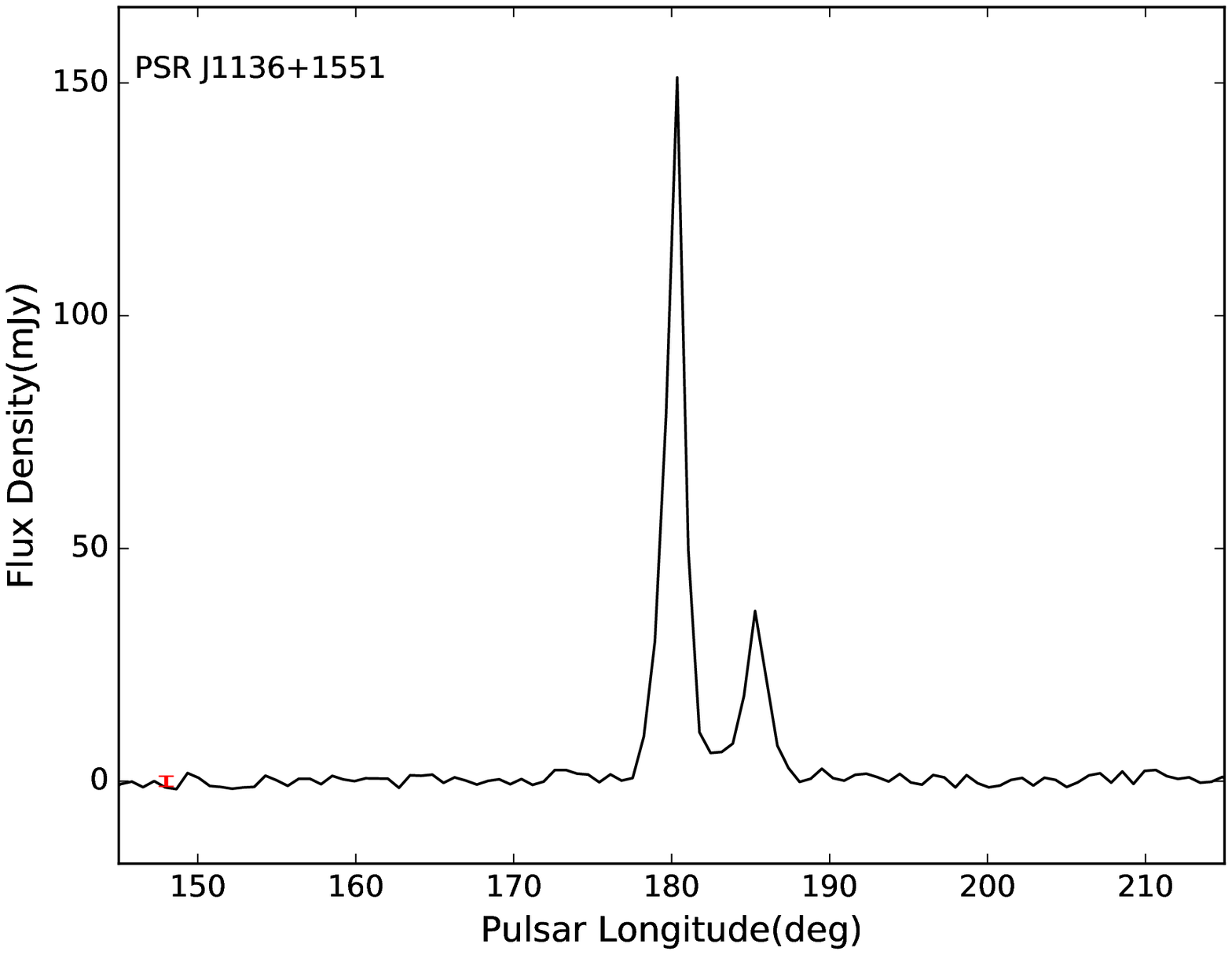}}&
\resizebox{0.53\hsize}{!}{\includegraphics[angle=0]{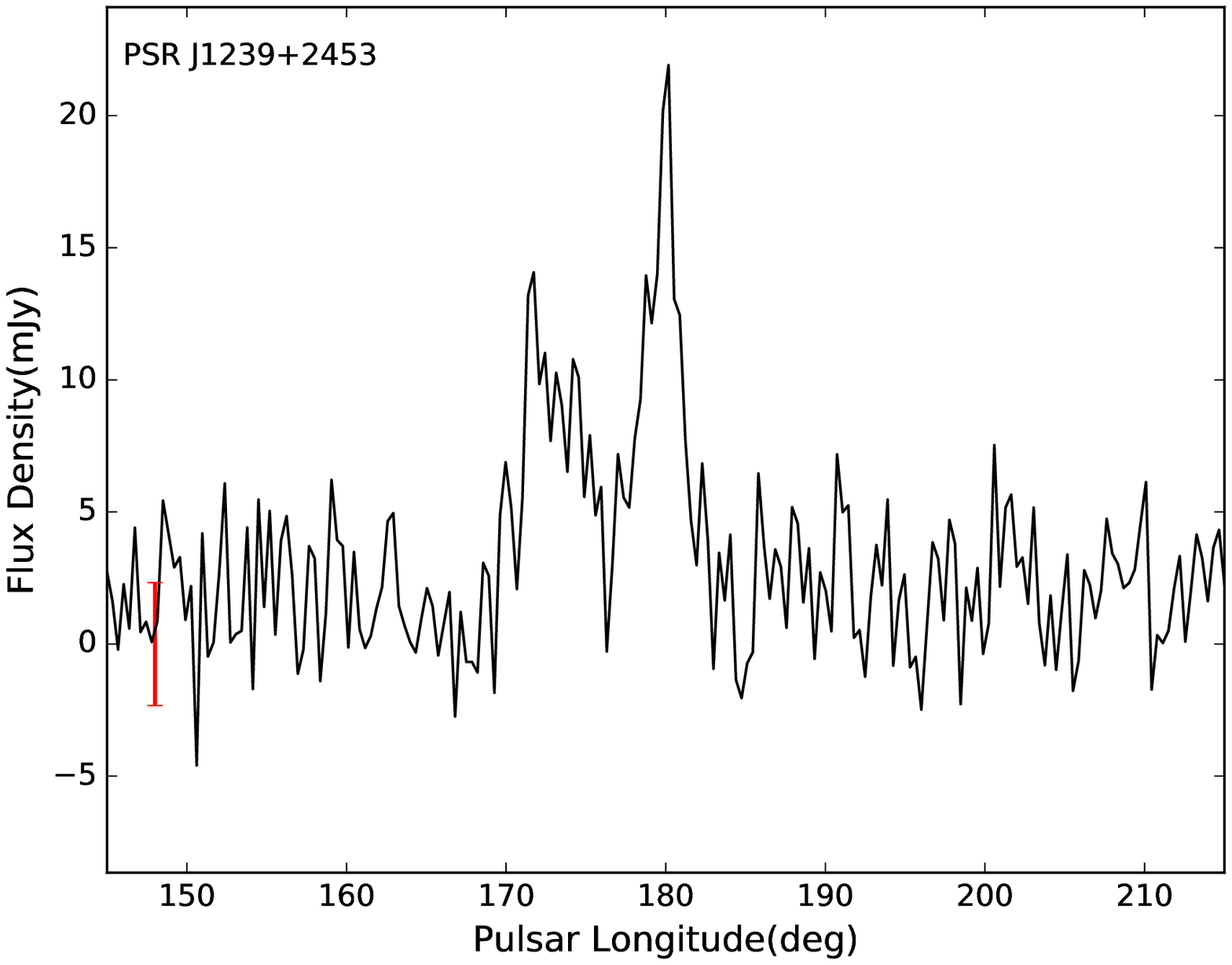}}\\
\resizebox{0.53\hsize}{!}{\includegraphics[angle=0]{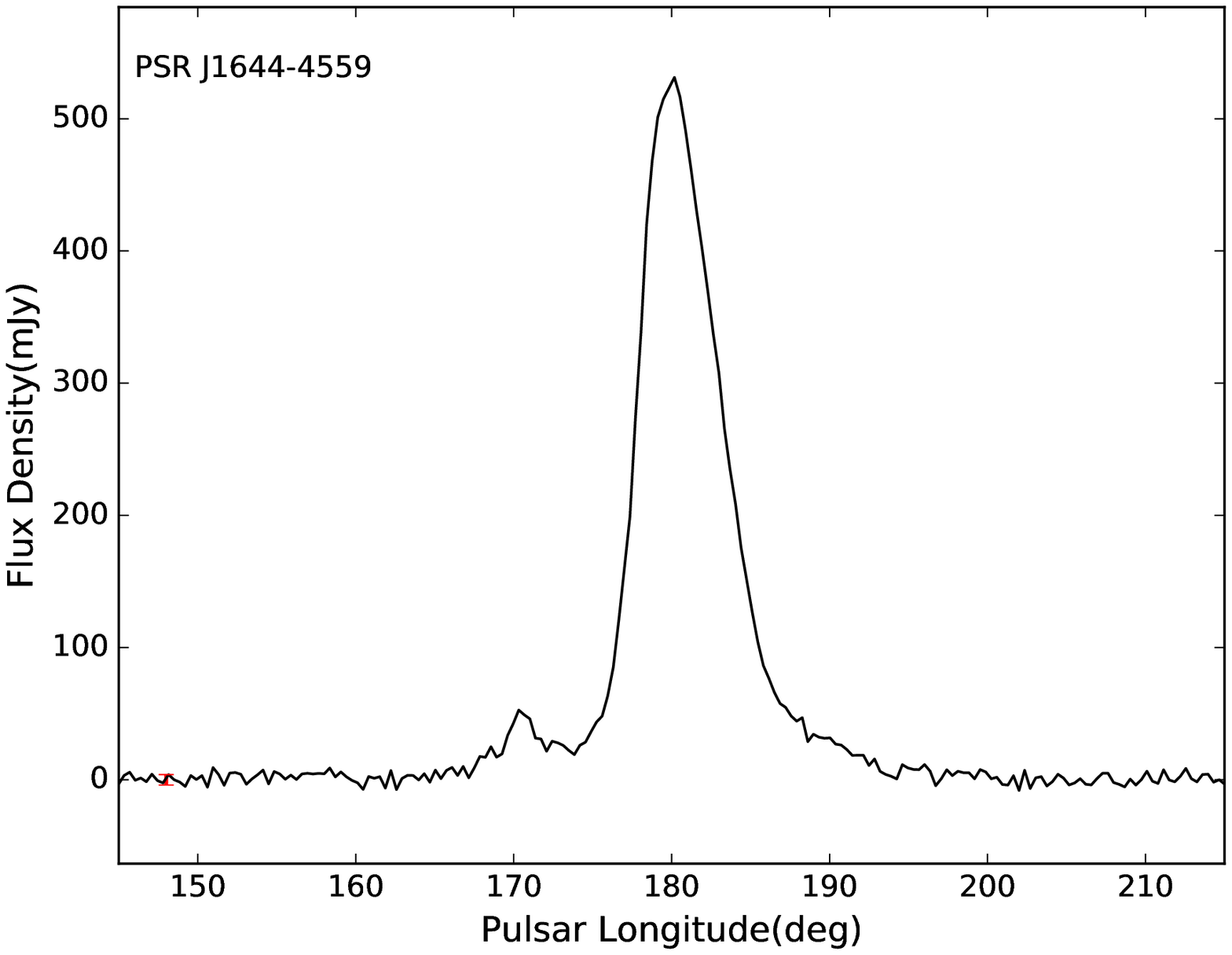}}&
\resizebox{0.53\hsize}{!}{\includegraphics[angle=0]{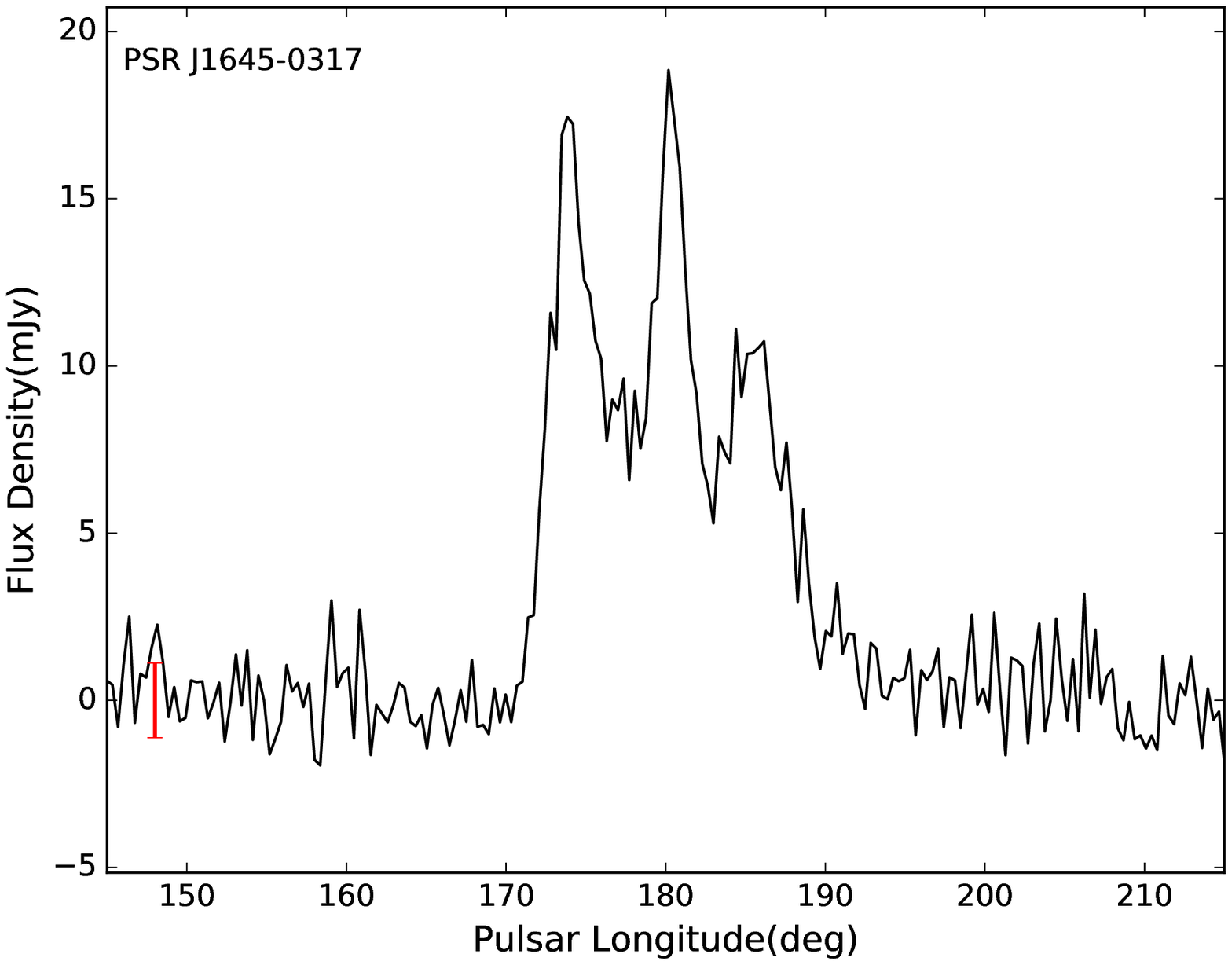}}\\
\resizebox{0.53\hsize}{!}{\includegraphics[angle=0]{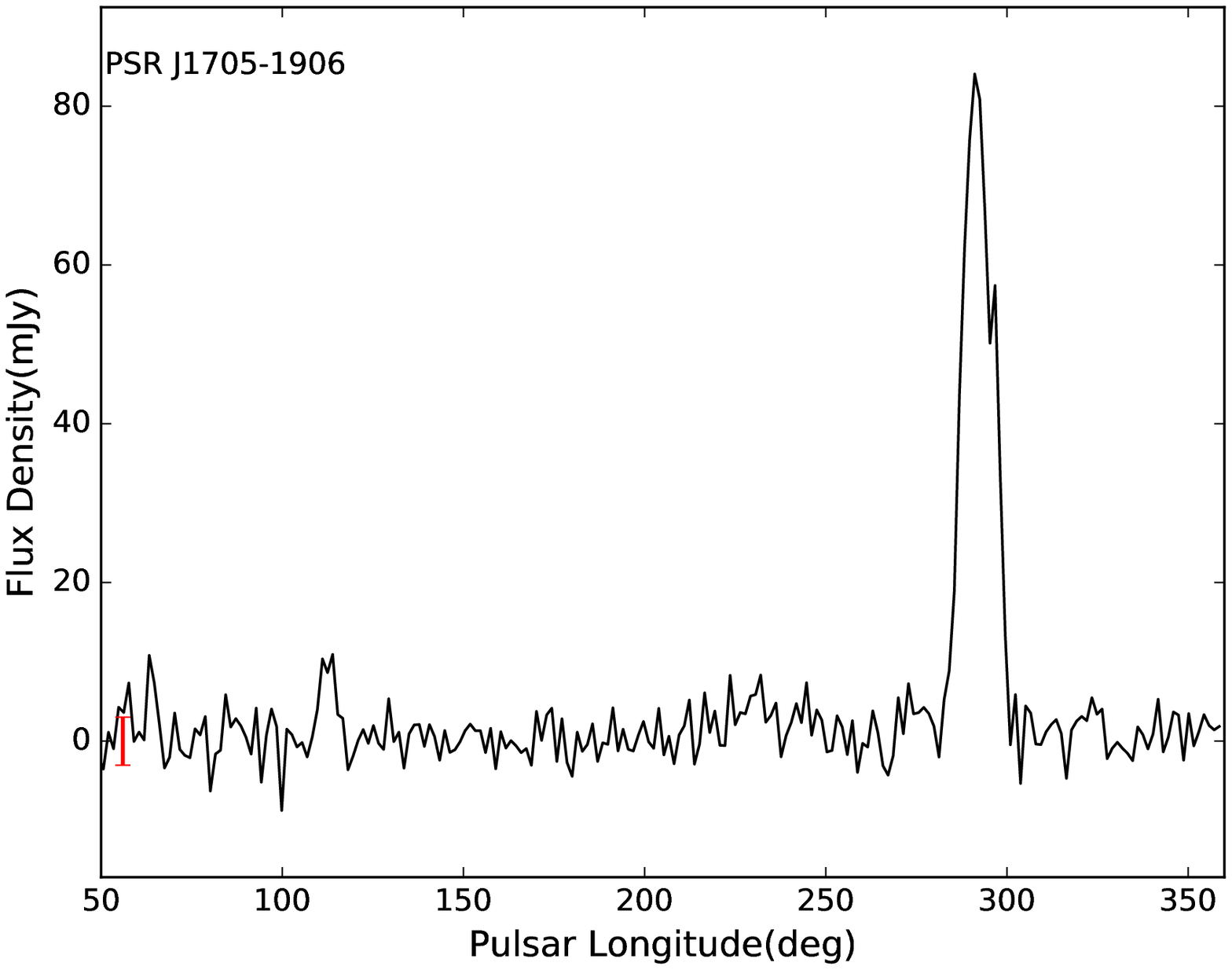}}&
\resizebox{0.53\hsize}{!}{\includegraphics[angle=0]{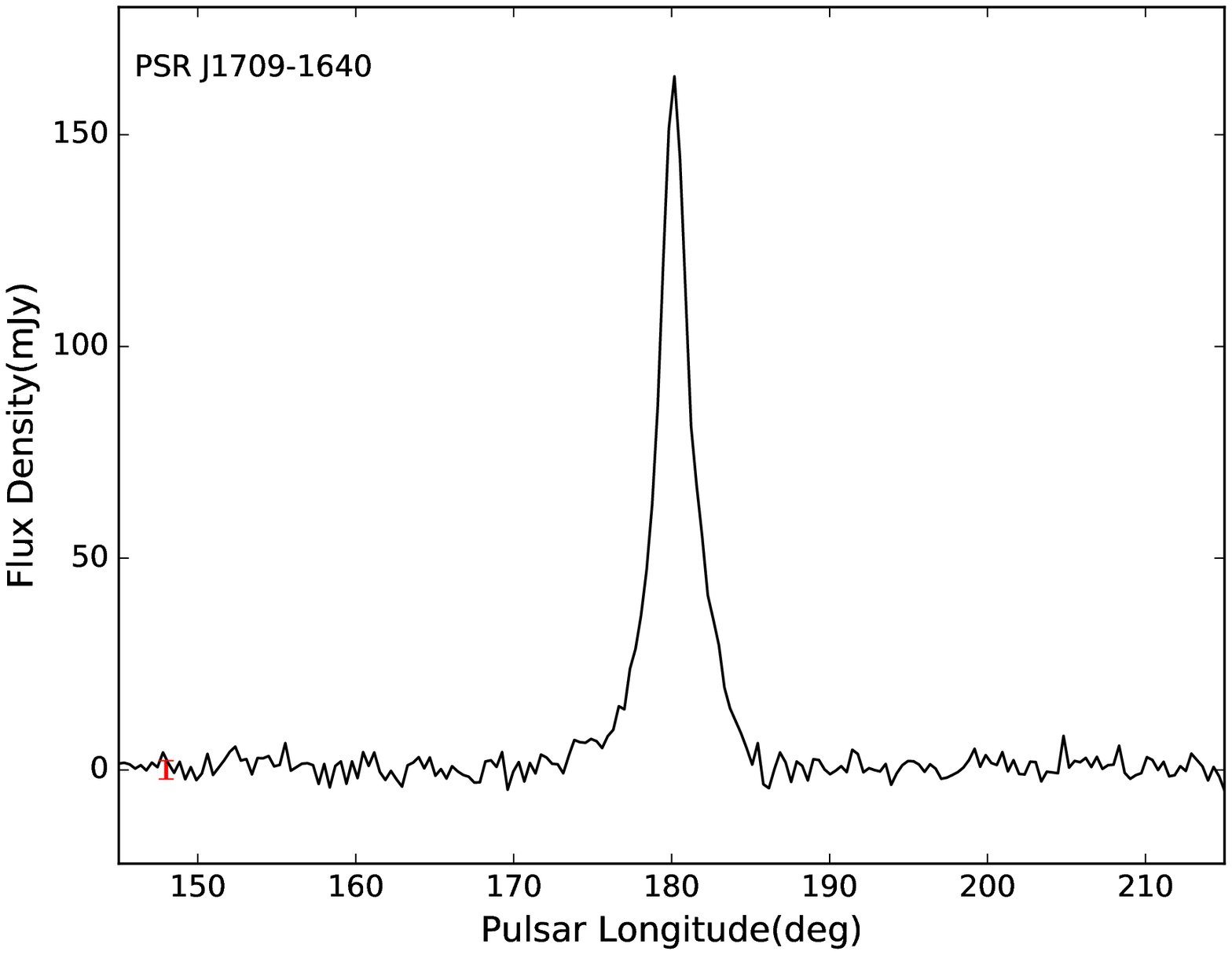}}\\

\end{tabular}
\end{center}
\caption{- continued}
\end{figure}
\addtocounter{figure}{-1}
\begin{figure}
\begin{center}
\begin{tabular}{cc}
\resizebox{0.53\hsize}{!}{\includegraphics[angle=0]{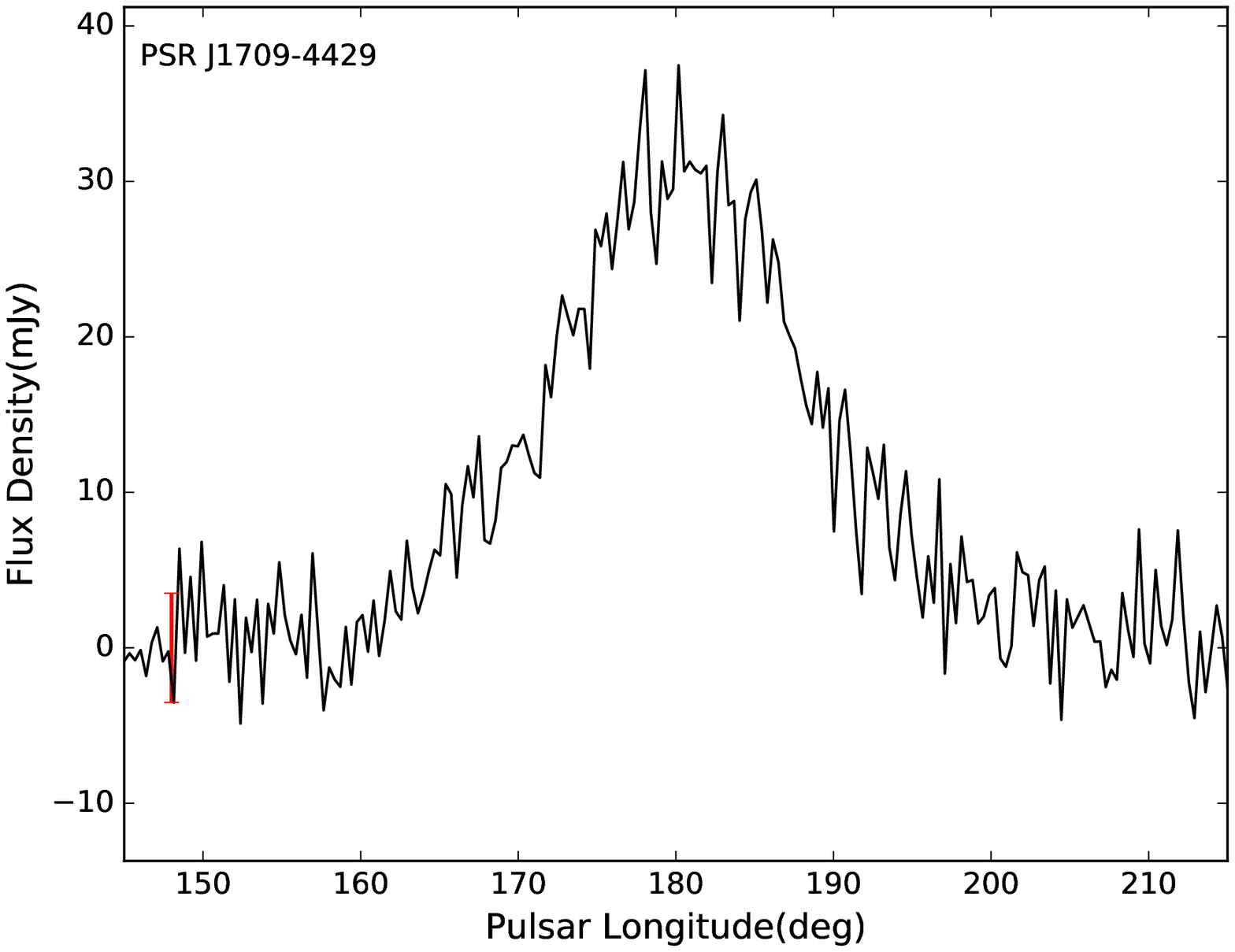}}&
\resizebox{0.53\hsize}{!}{\includegraphics[angle=0]{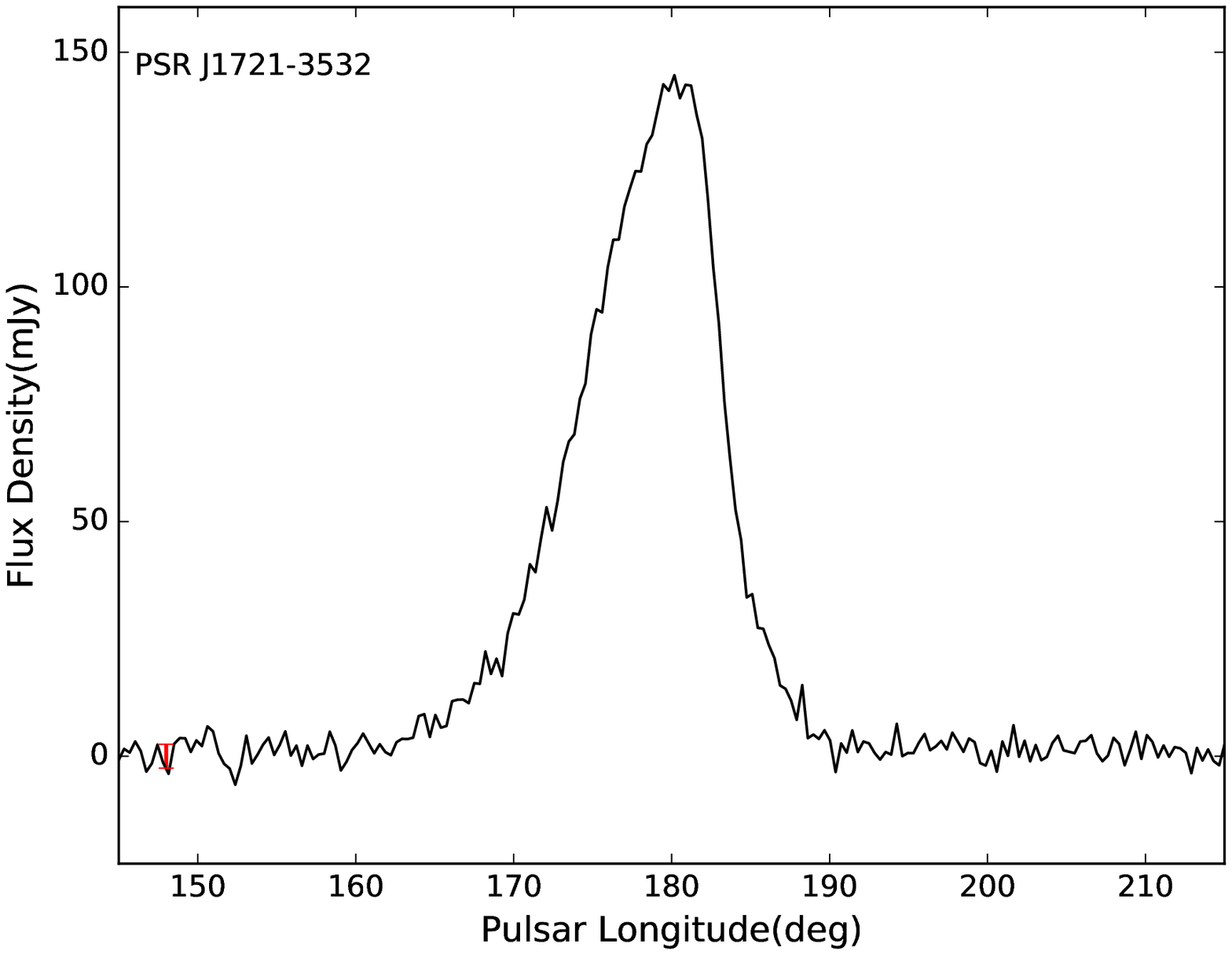}}\\
\resizebox{0.53\hsize}{!}{\includegraphics[angle=0]{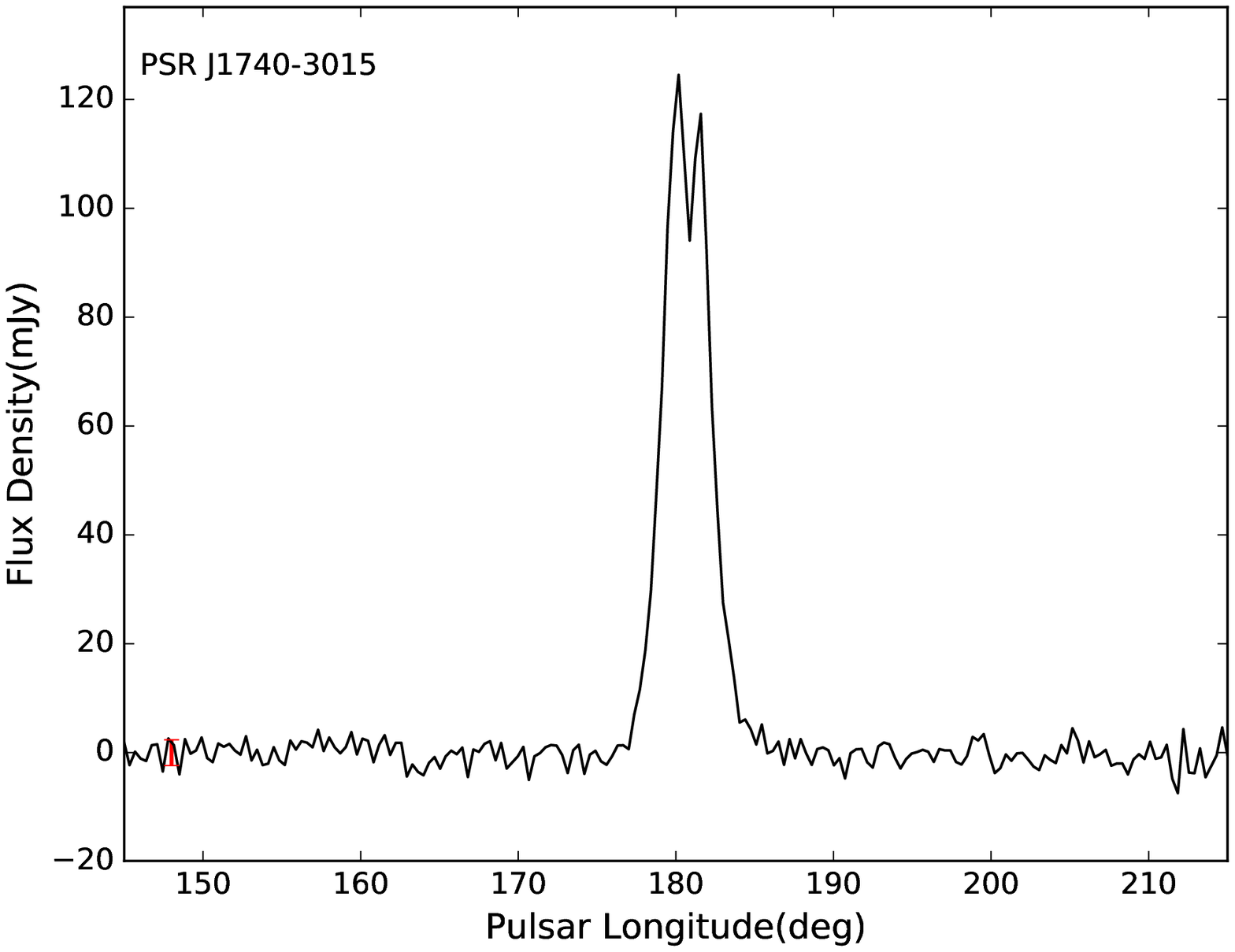}}&
\resizebox{0.53\hsize}{!}{\includegraphics[angle=0]{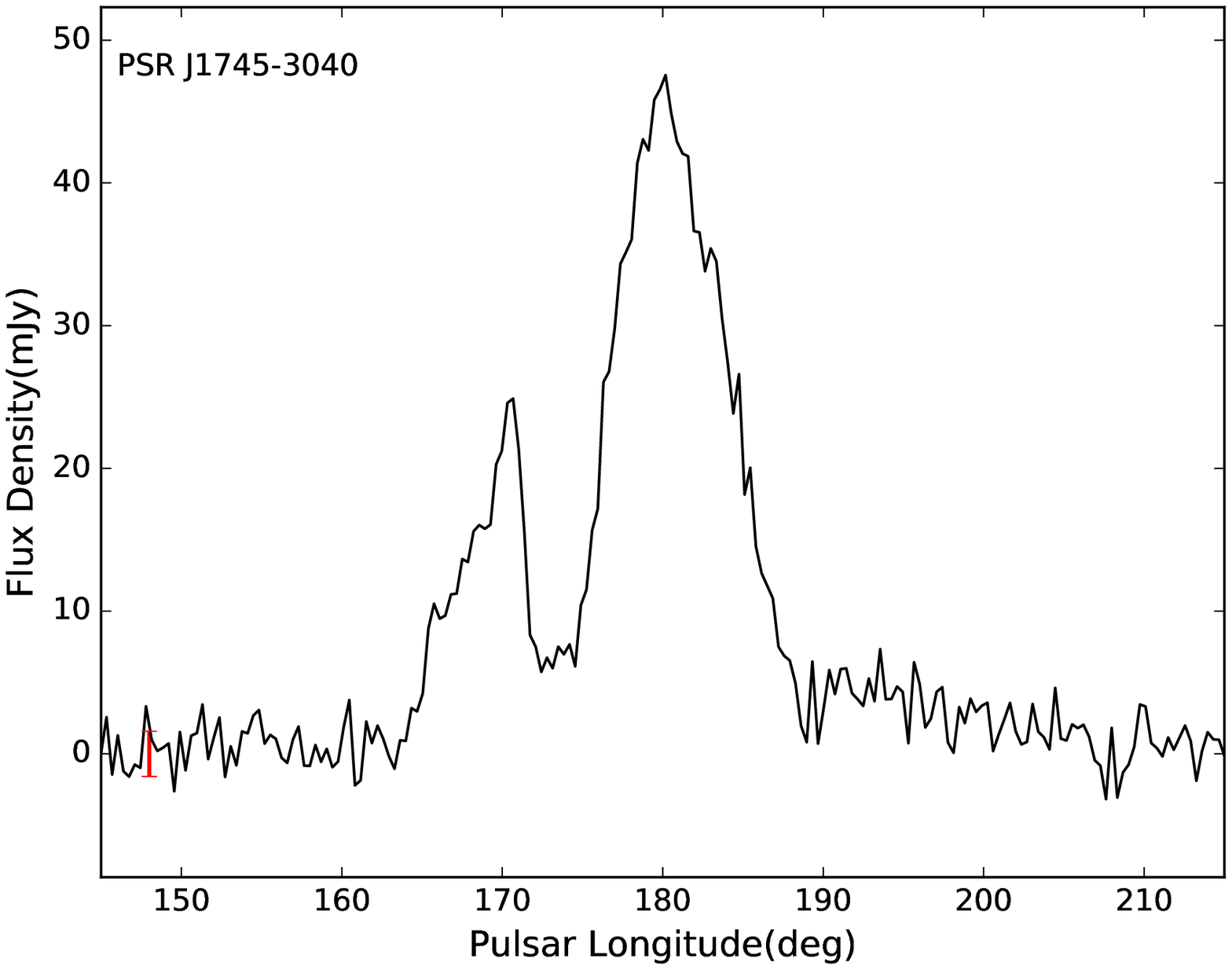}}\\
\resizebox{0.53\hsize}{!}{\includegraphics[angle=0]{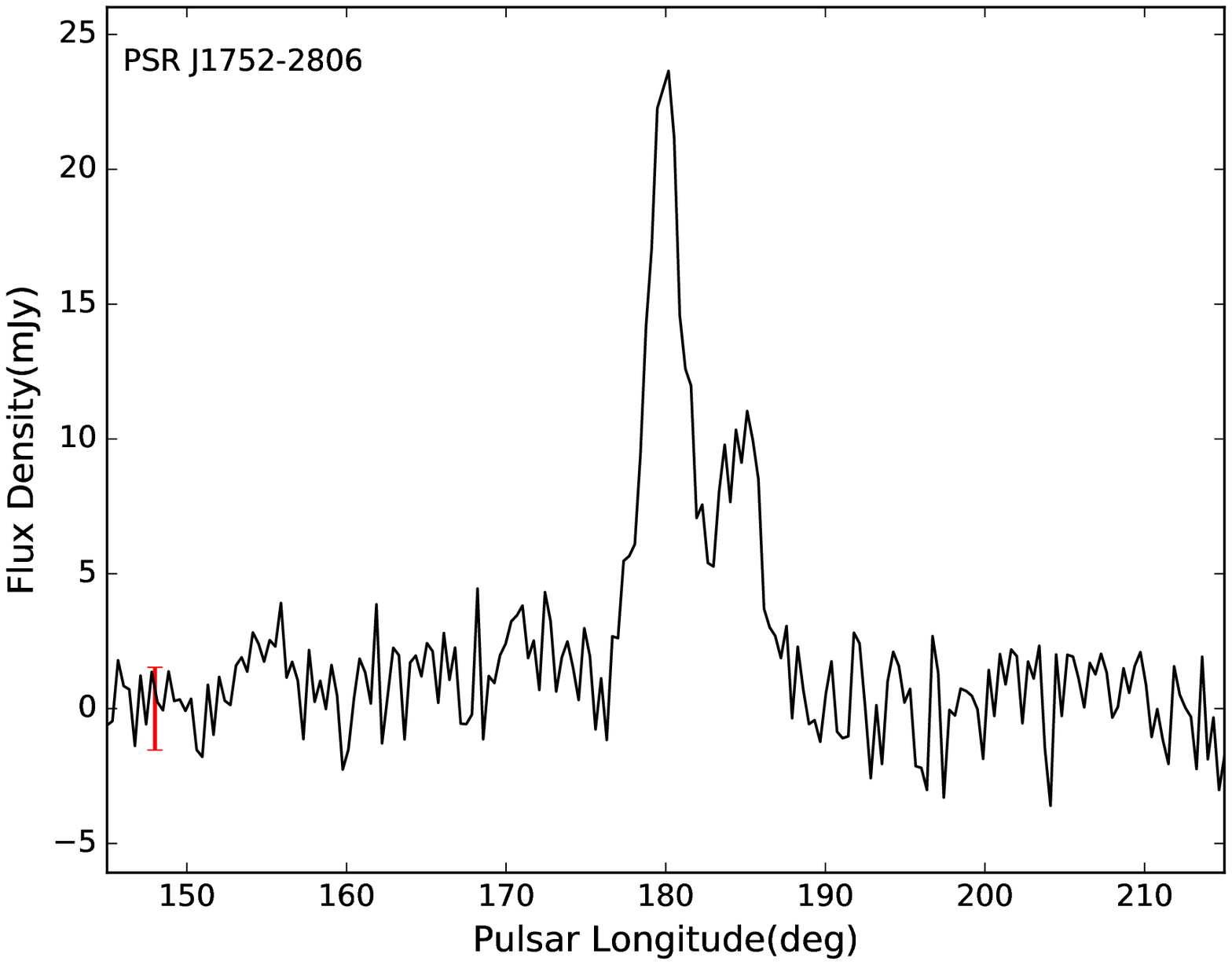}}&
\resizebox{0.53\hsize}{!}{\includegraphics[angle=0]{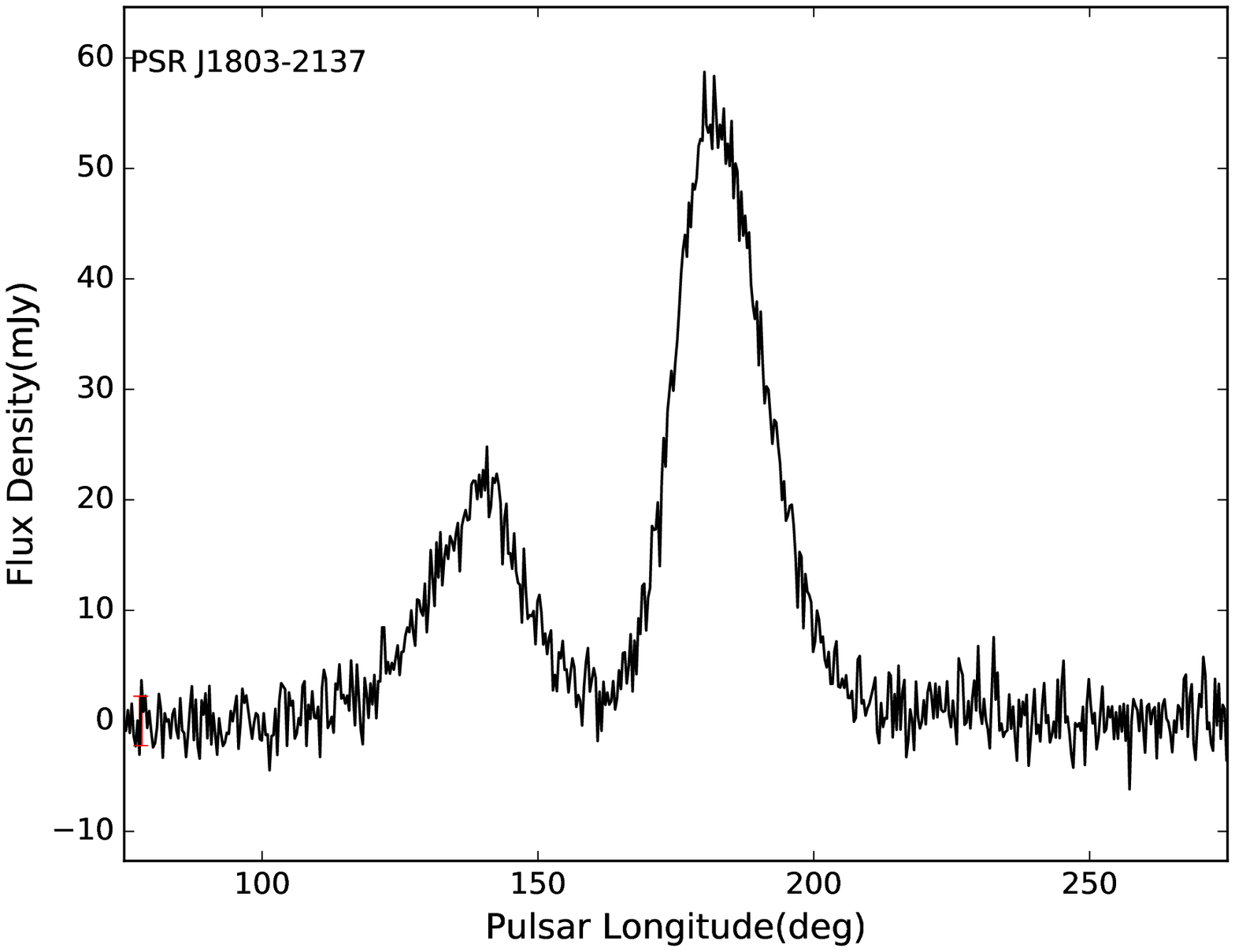}}\\
\end{tabular}
\end{center}
\caption{continued}
\end{figure}
\addtocounter{figure}{-1}
\begin{figure}
\begin{center}
\begin{tabular}{cc}

\resizebox{0.53\hsize}{!}{\includegraphics[angle=0]{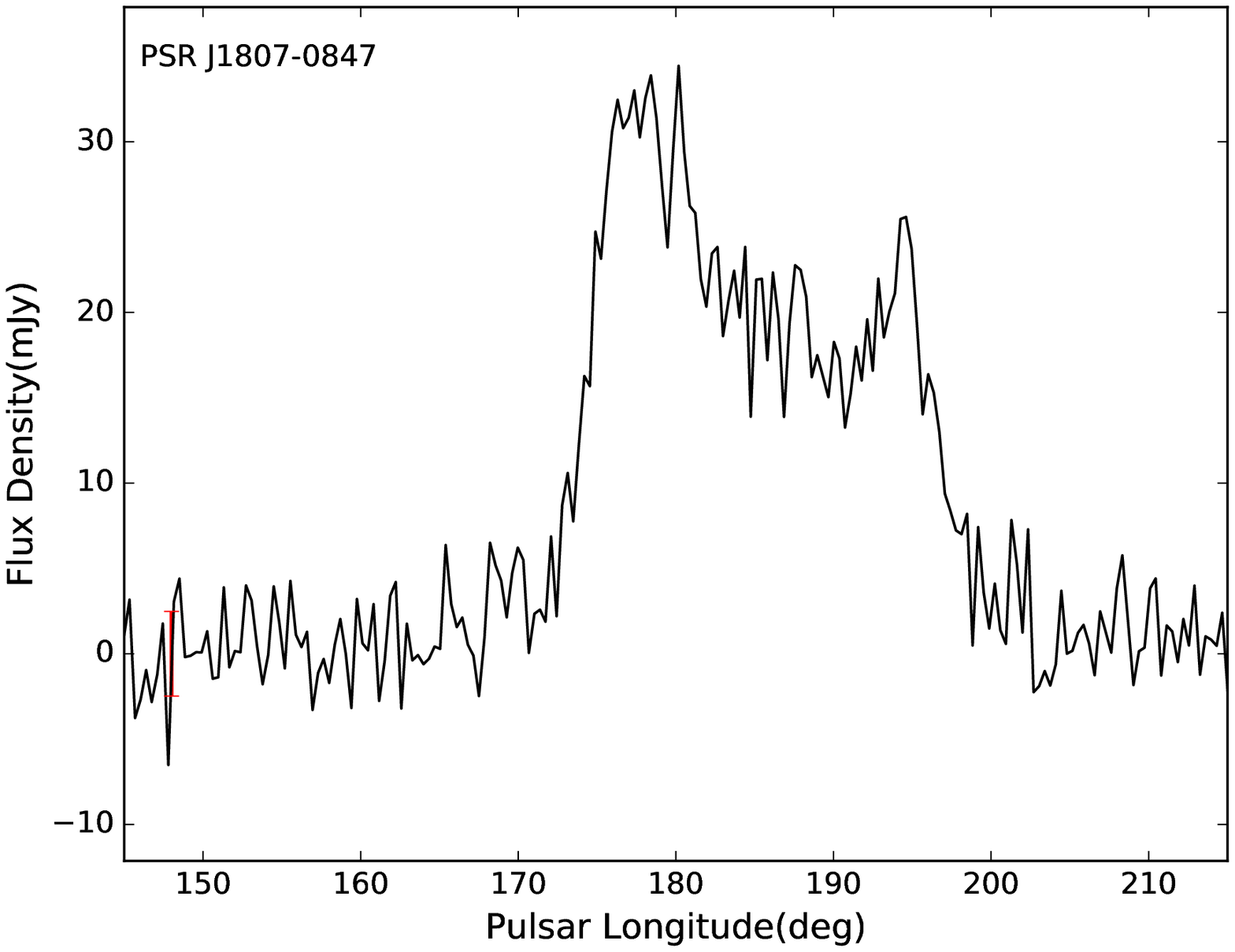}}&
\resizebox{0.53\hsize}{!}{\includegraphics[angle=0]{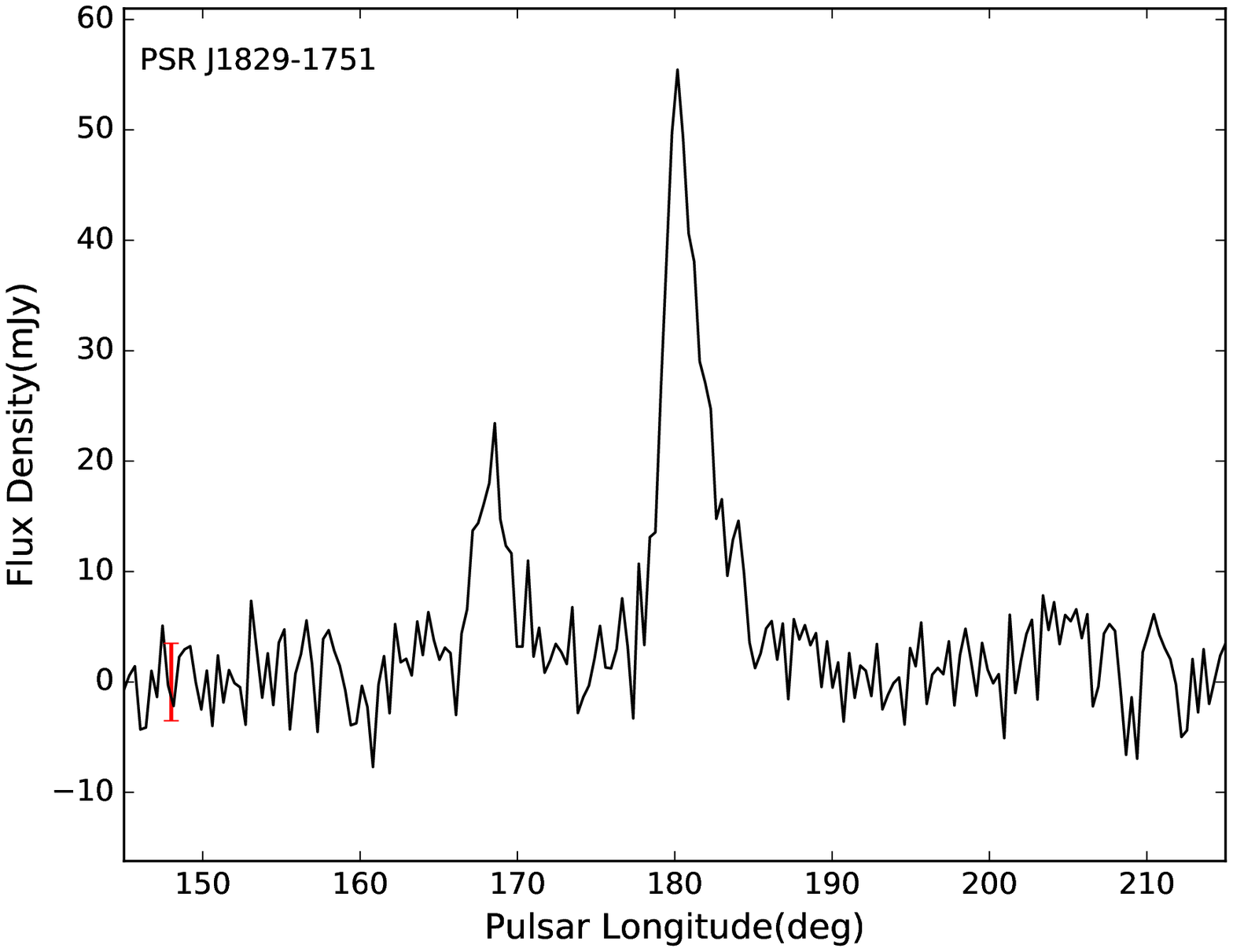}}\\
\resizebox{0.53\hsize}{!}{\includegraphics[angle=0]{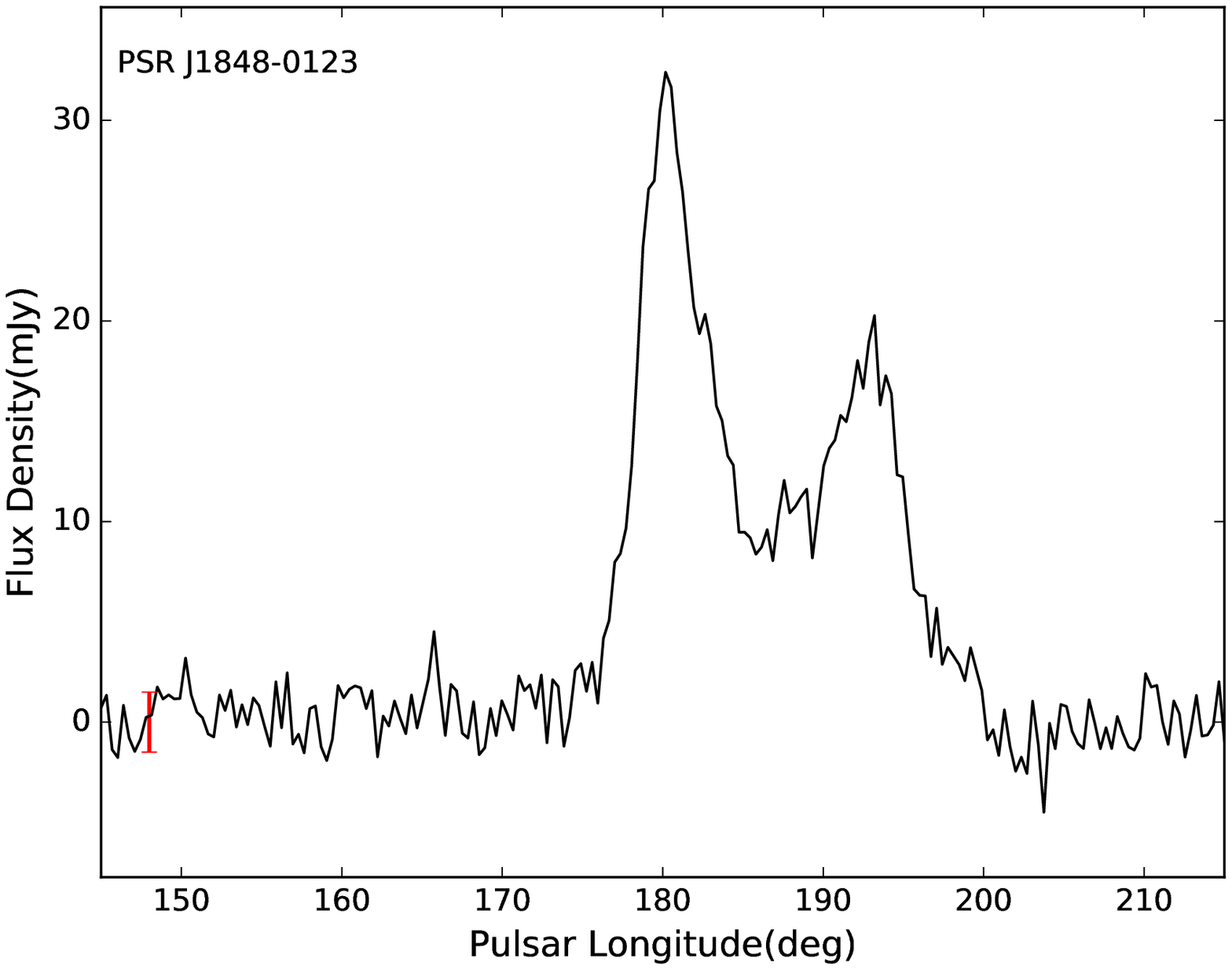}}&
\resizebox{0.53\hsize}{!}{\includegraphics[angle=0]{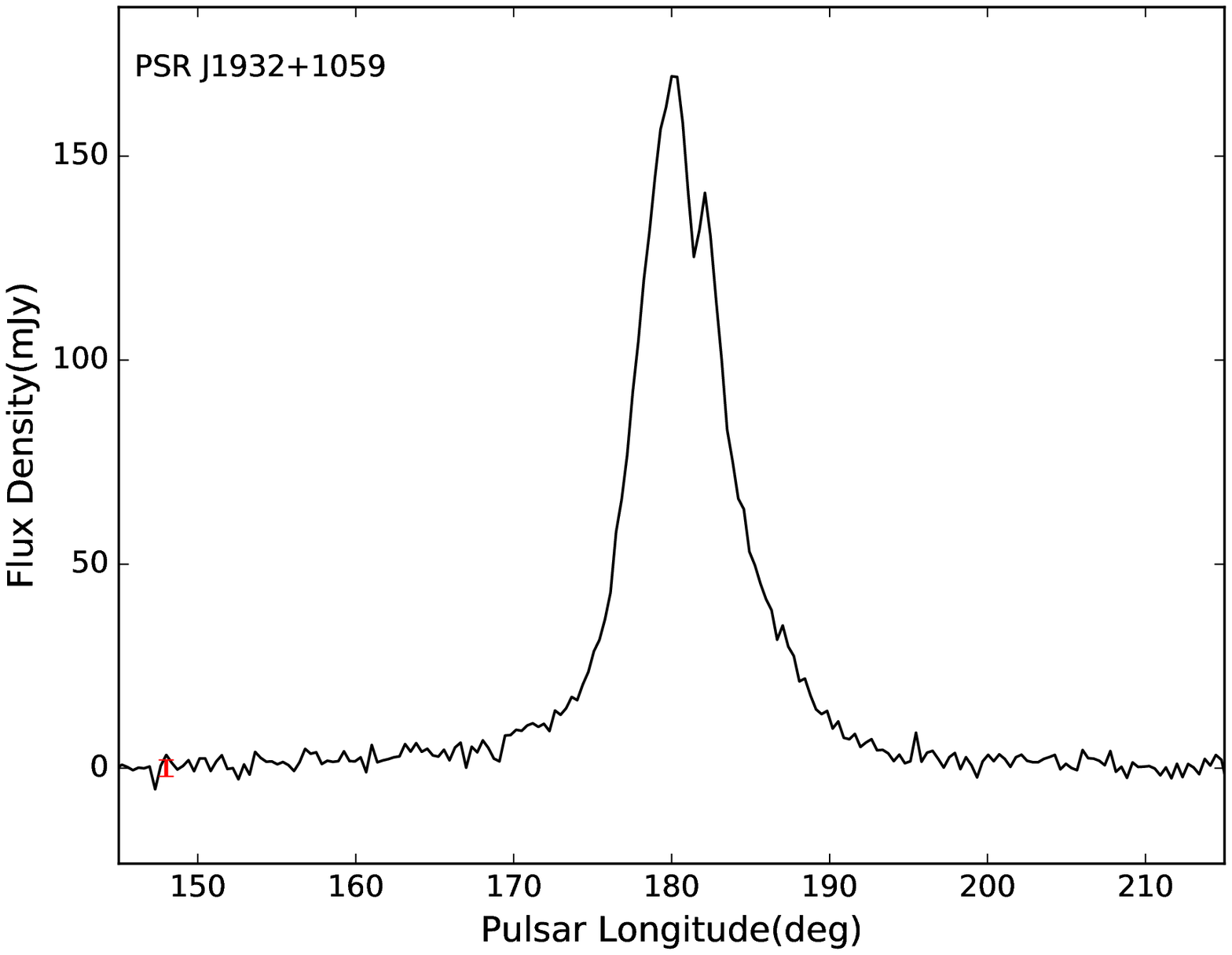}}\\
\resizebox{0.53\hsize}{!}{\includegraphics[angle=0]{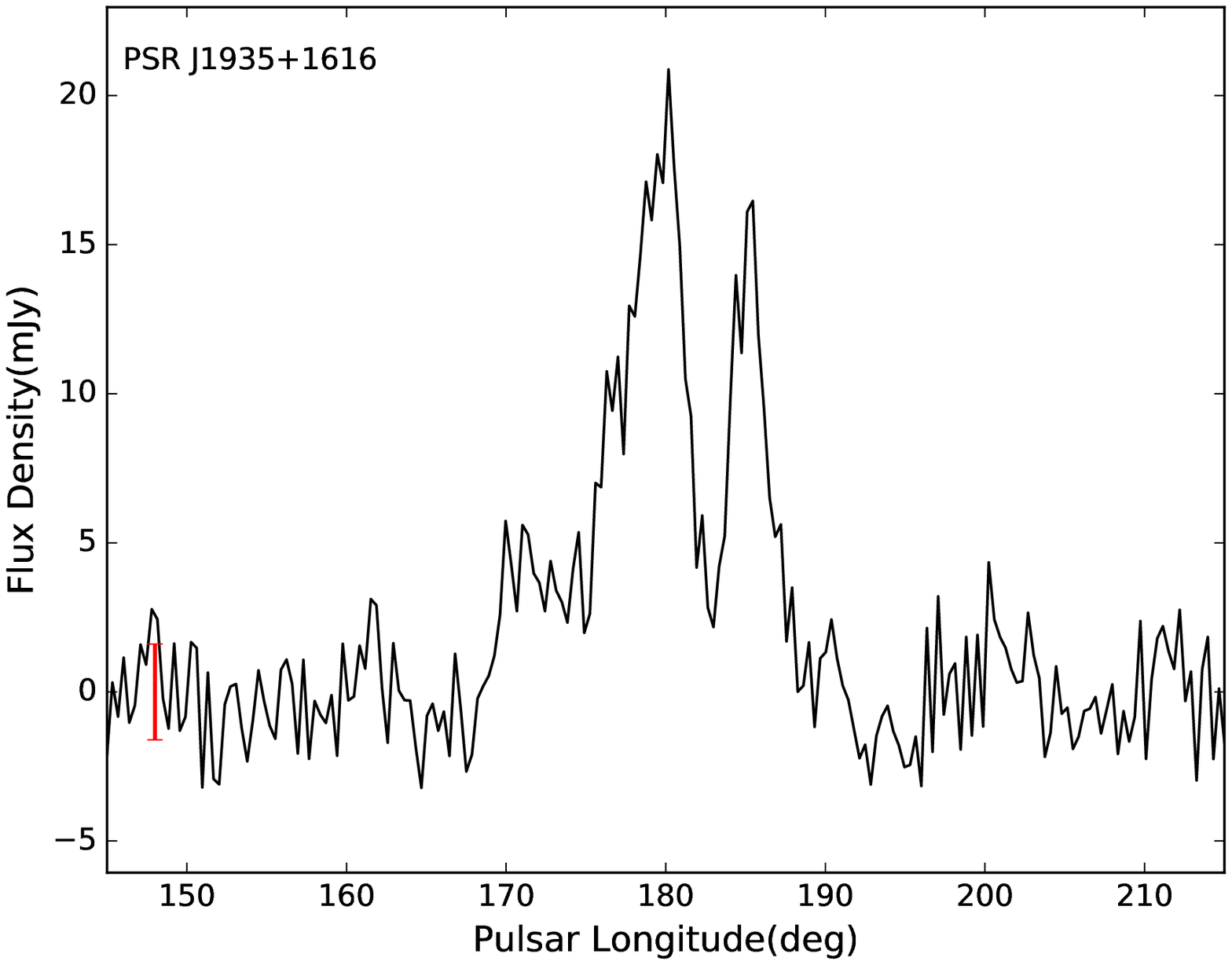}}&
\resizebox{0.53\hsize}{!}{\includegraphics[angle=0]{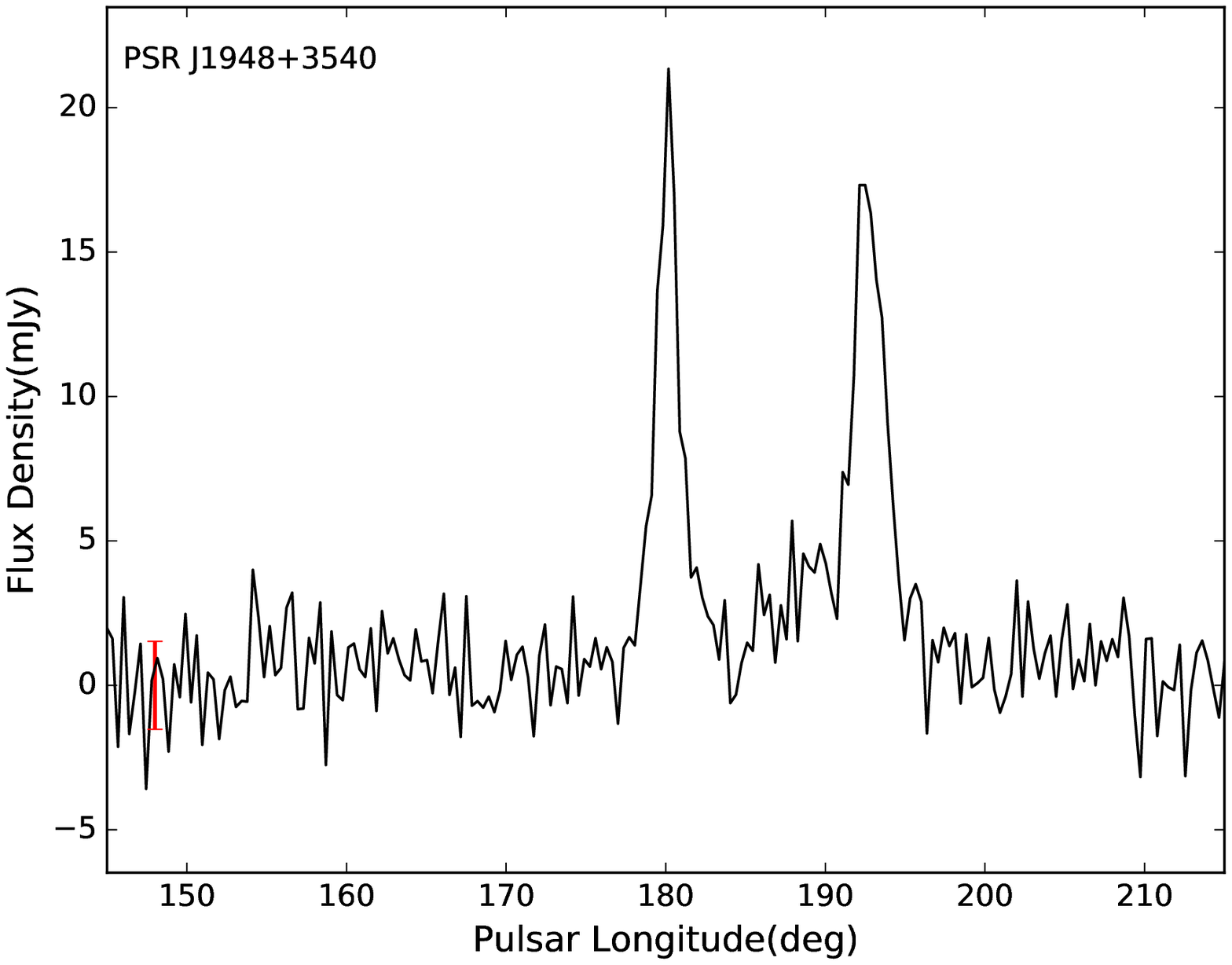}}\\
\end{tabular}
\end{center}
\caption{- continued}
\end{figure}
\addtocounter{figure}{-1}
\begin{figure}
\begin{center}
\begin{tabular}{cc}

\resizebox{0.53\hsize}{!}{\includegraphics[angle=0]{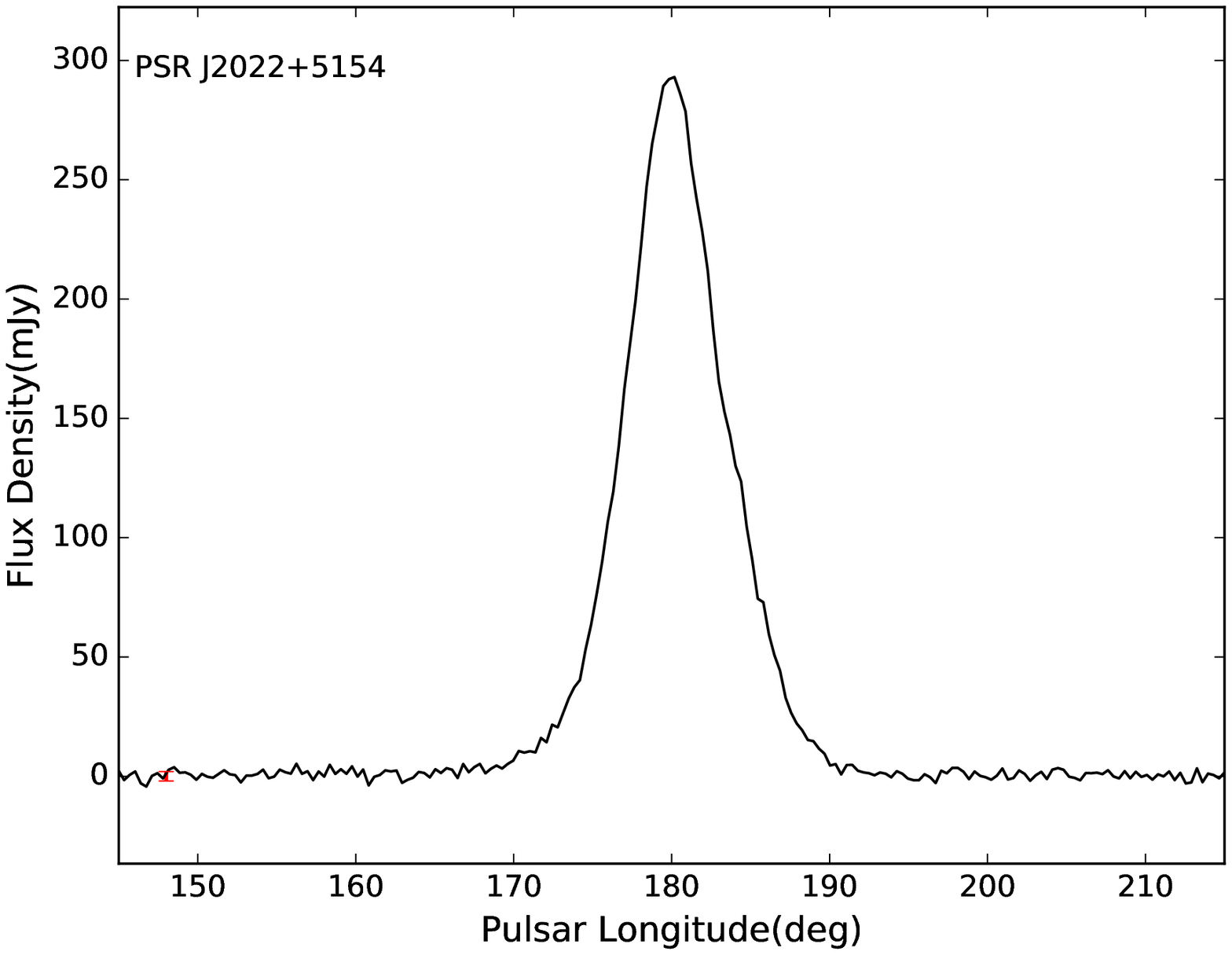}}&
\resizebox{0.53\hsize}{!}{\includegraphics[angle=0]{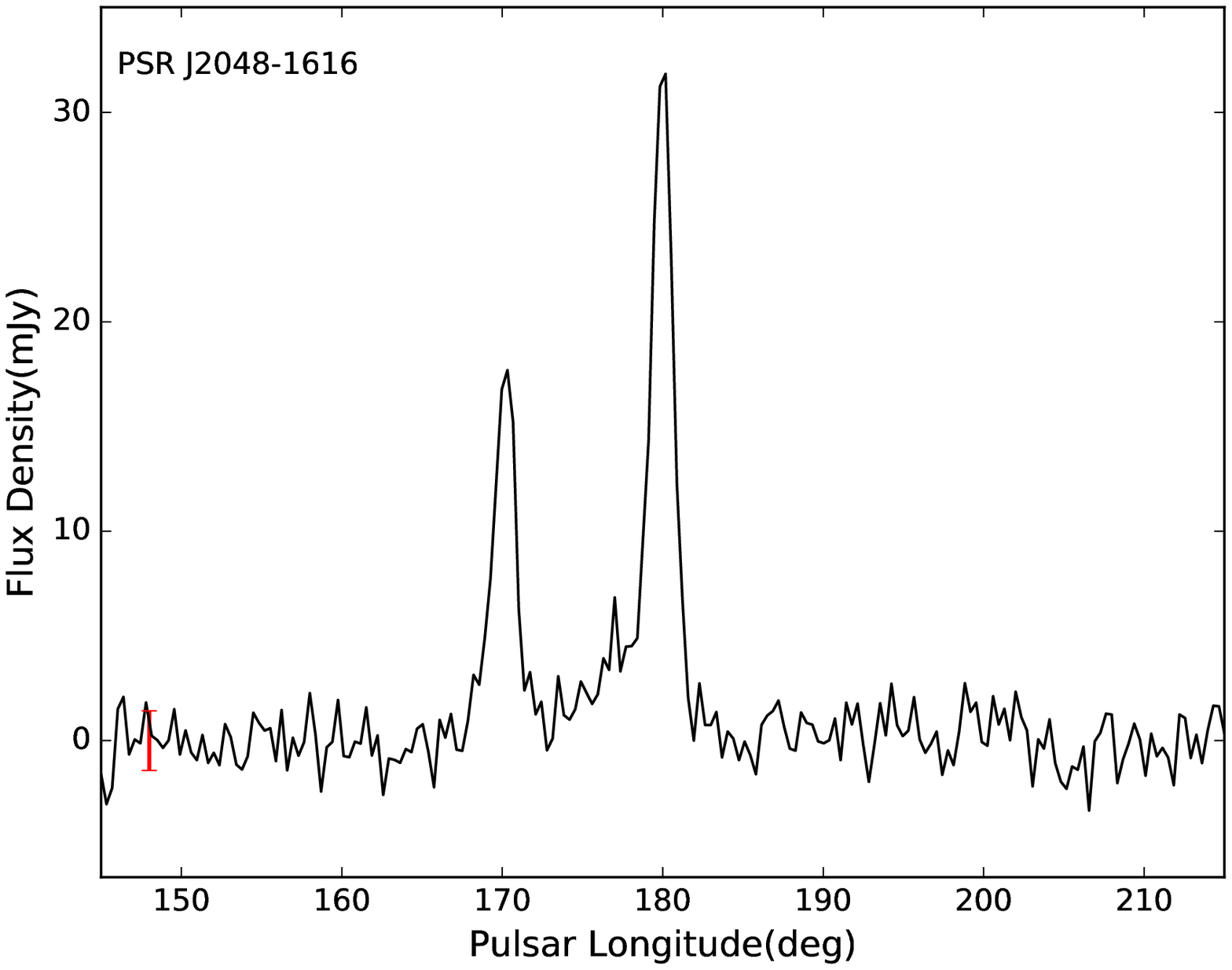}}\\
\end{tabular}
\end{center}
\caption{- continued}
\end{figure}

\clearpage

\begin{figure}[h]
\begin{center}
\begin{tabular}{cc}

\resizebox{0.53\hsize}{!}{\includegraphics[angle=0]{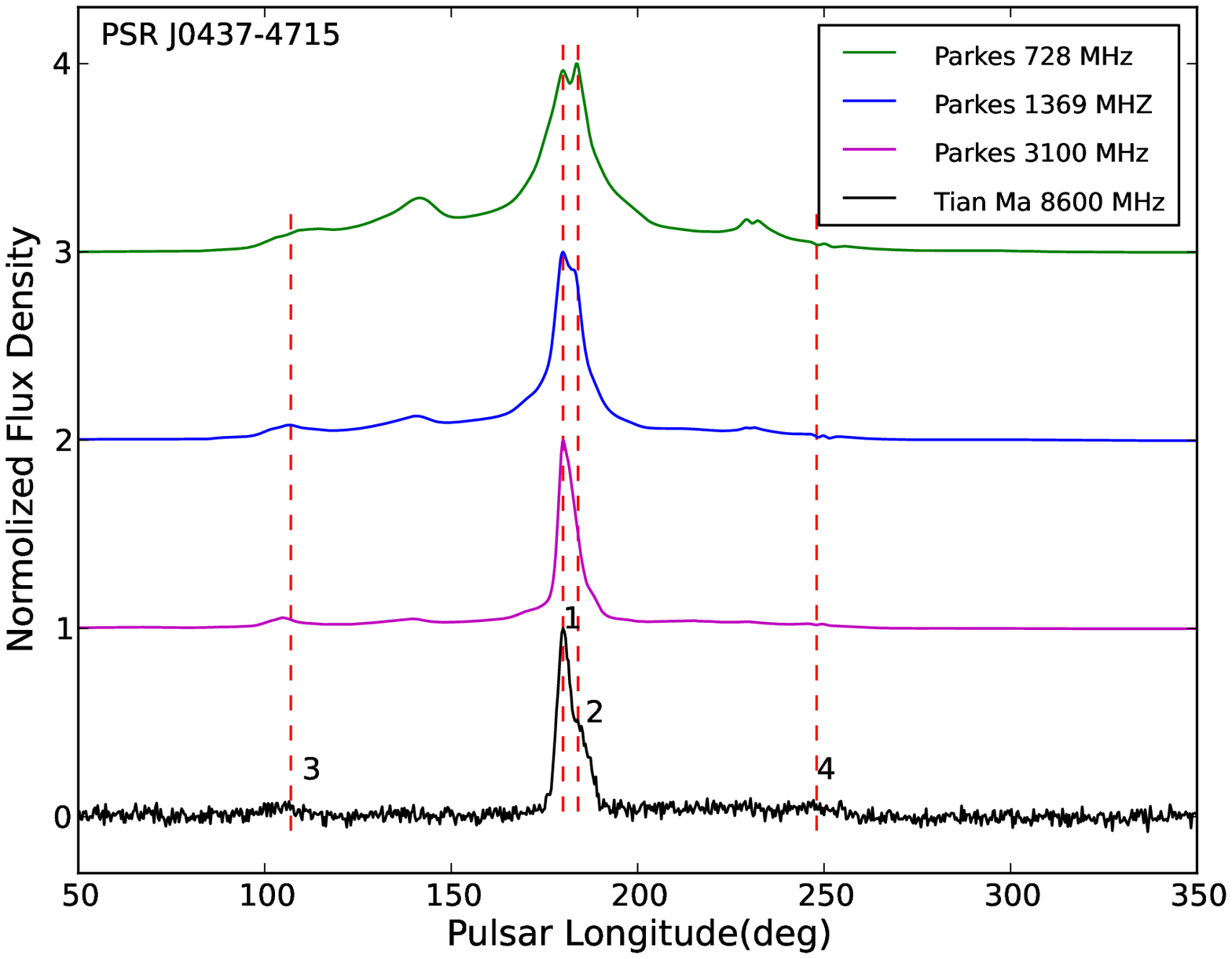}}&
\resizebox{0.53\hsize}{!}{\includegraphics[angle=0]{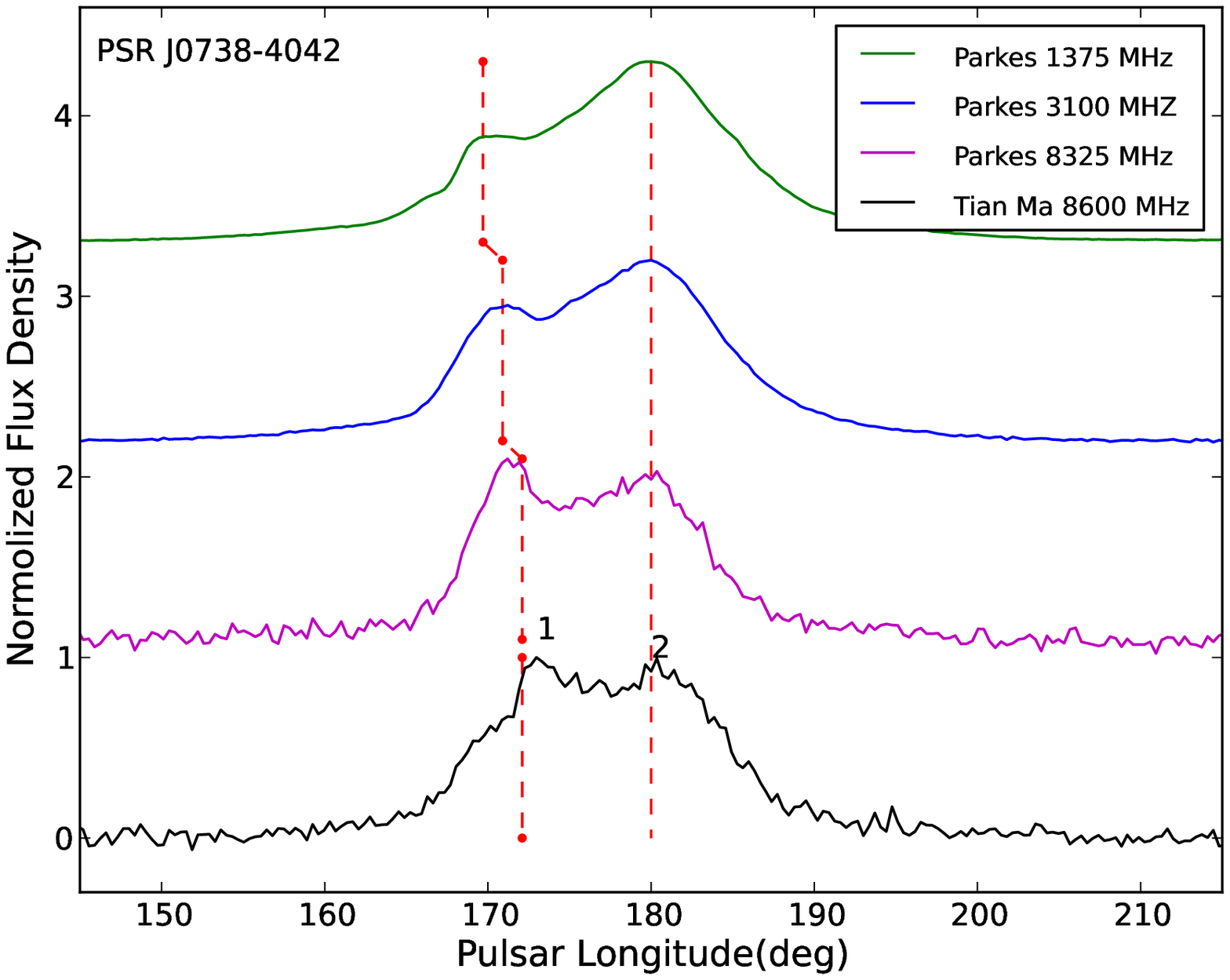}}\\
\resizebox{0.53\hsize}{!}{\includegraphics[angle=0]{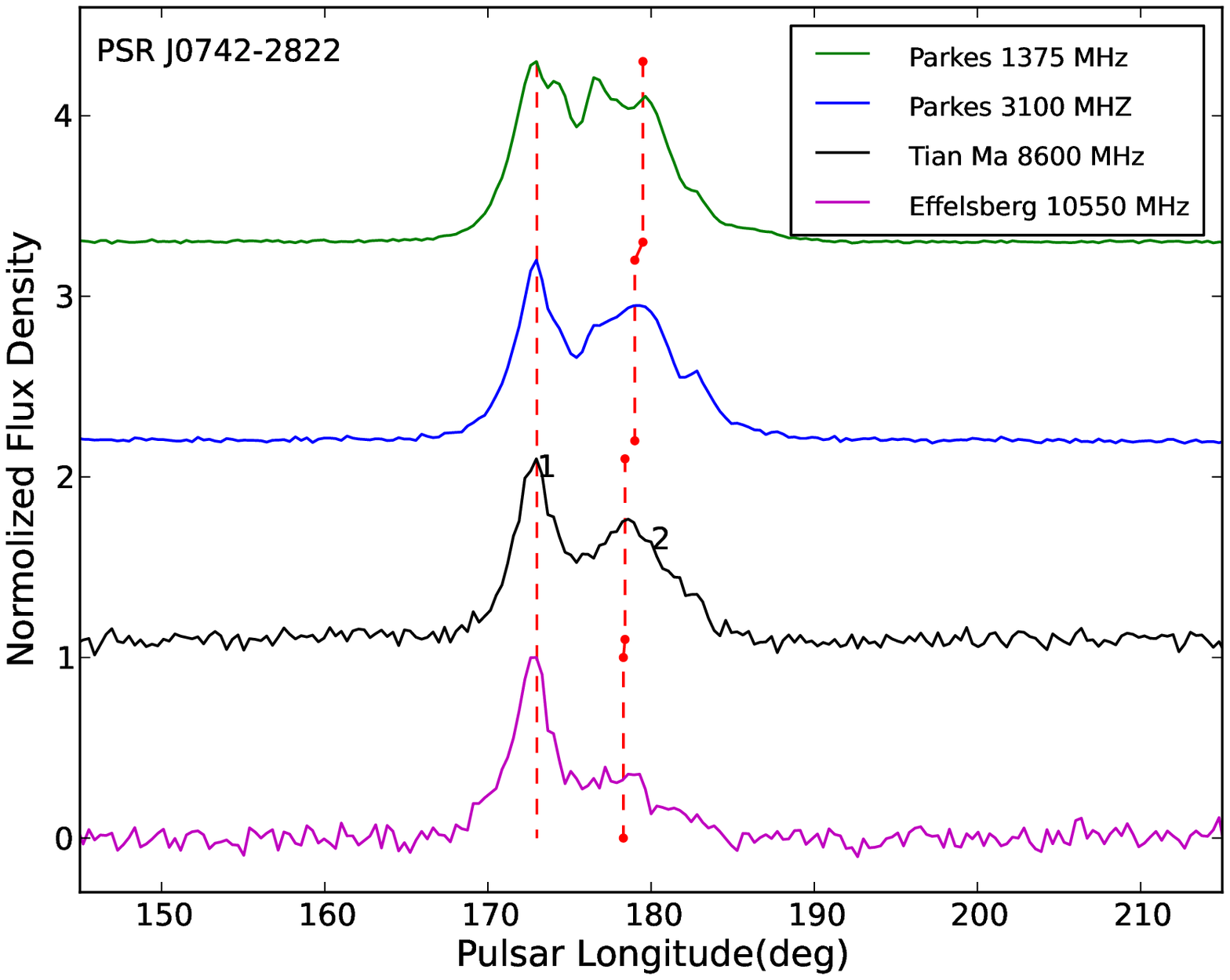}}&
\resizebox{0.53\hsize}{!}{\includegraphics[angle=0]{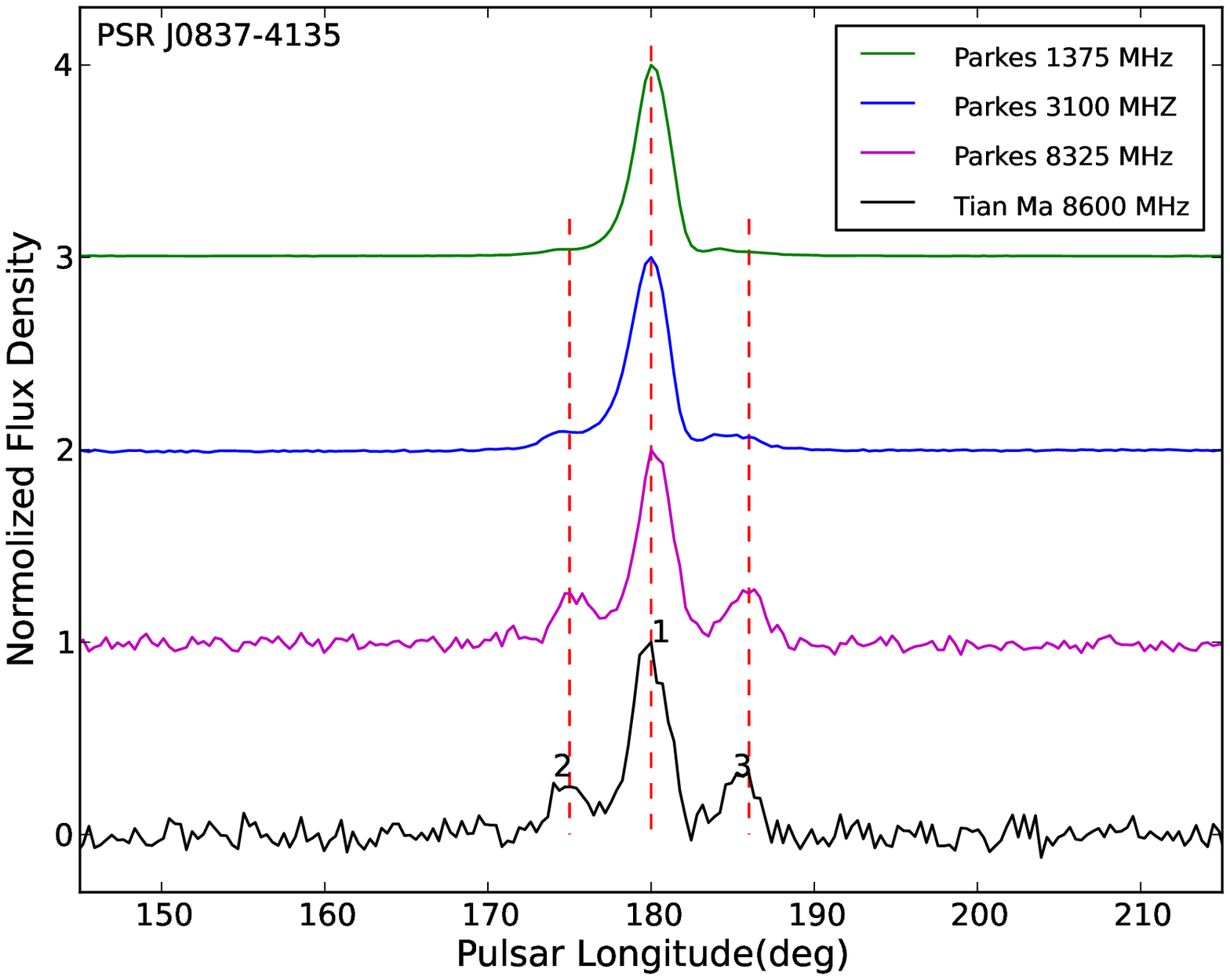}}\\
\resizebox{0.53\hsize}{!}{\includegraphics[angle=0]{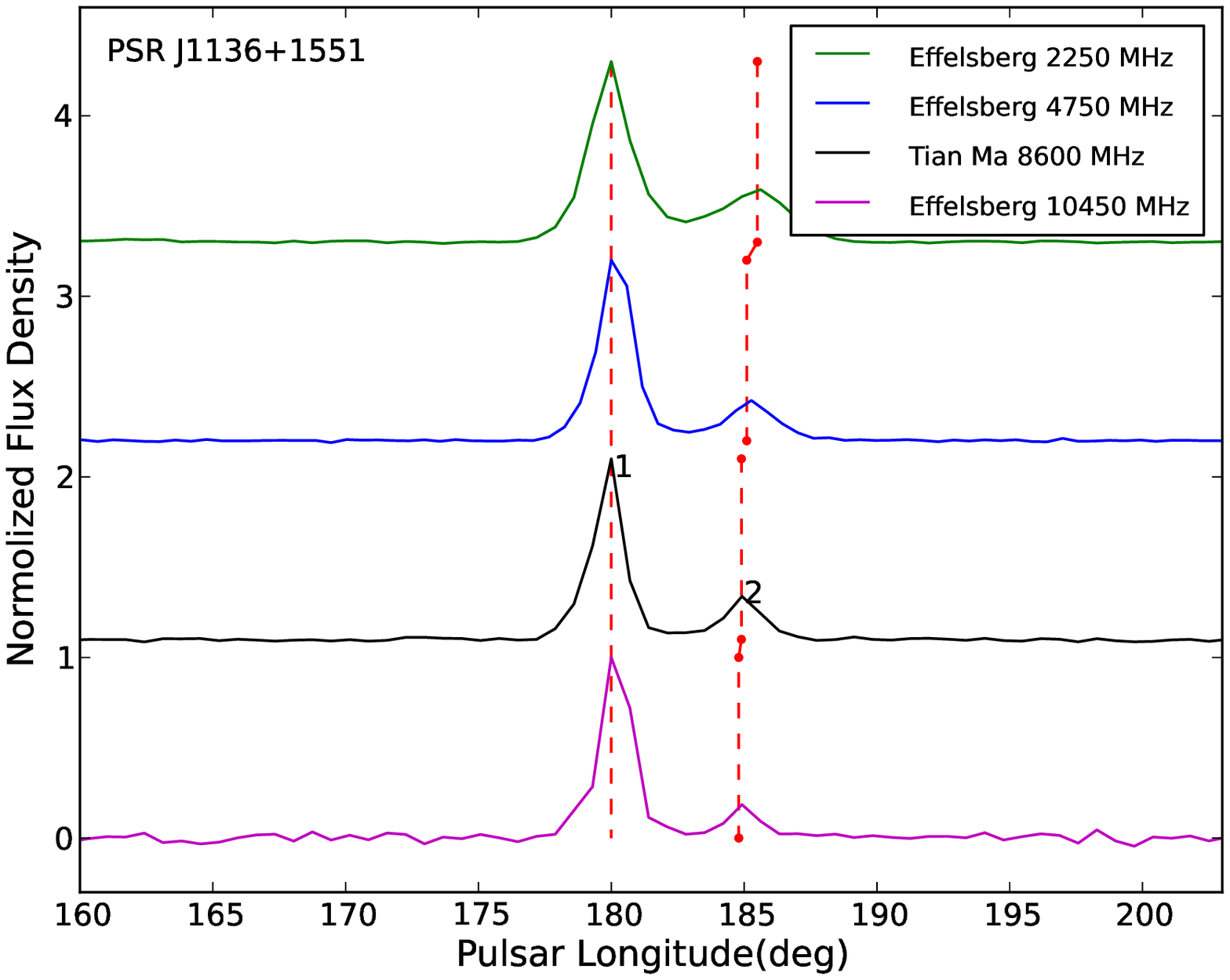}}&
\resizebox{0.53\hsize}{!}{\includegraphics[angle=0]{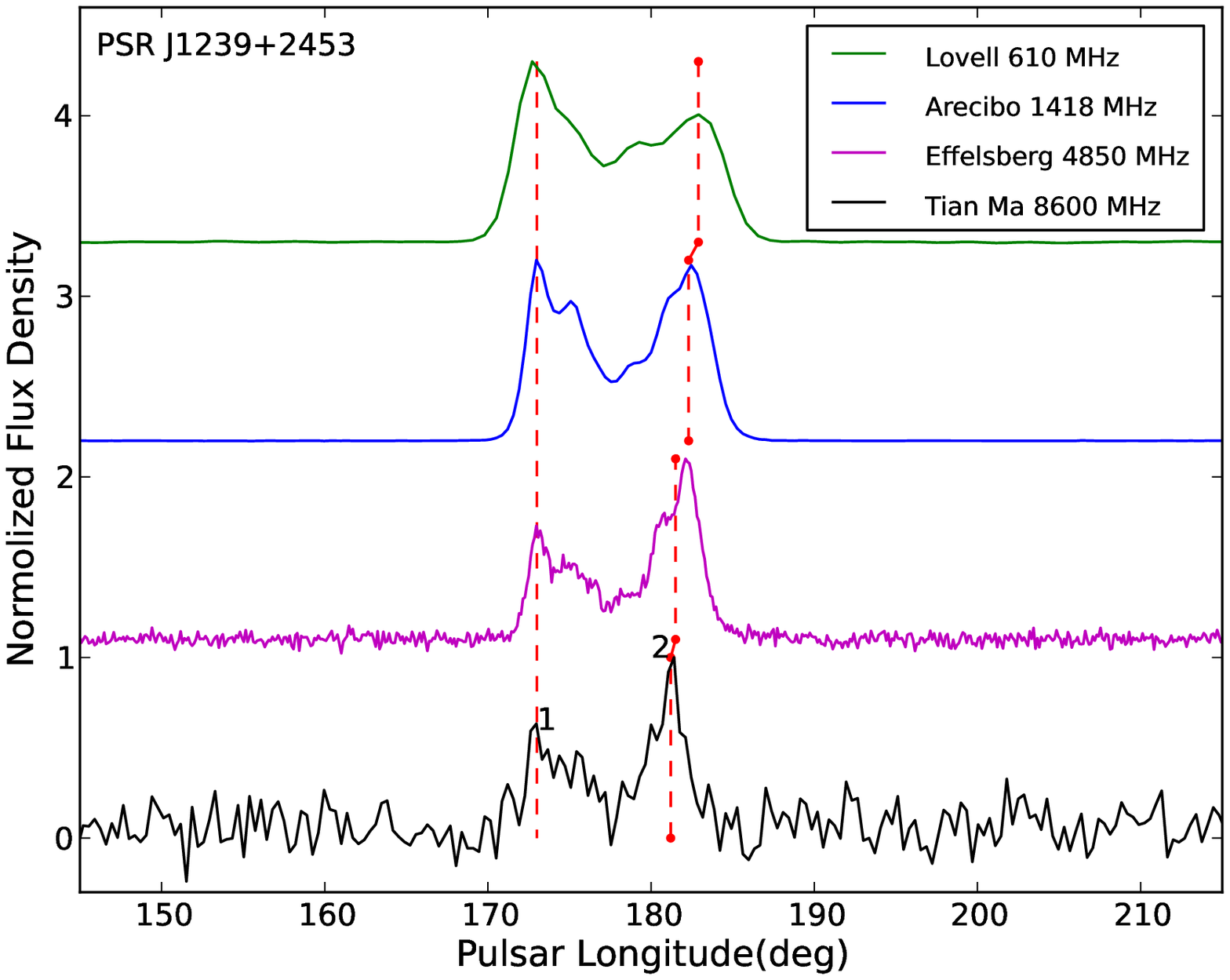}}\\
\end{tabular}
\end{center}
\caption{Observed integrated pulse profiles for 19 pulsars at four
  frequencies: 8.6~Hz from the TMRT and three other frequencies from
  the EPN profile database. Components are labelled on the TMRT
  profiles and corresponding components are connected by dashed lines.}\label{fg:4freq}
\end{figure}
\addtocounter{figure}{-1}
\begin{figure}
\begin{center}
\begin{tabular}{cc}
\resizebox{0.53\hsize}{!}{\includegraphics[angle=0]{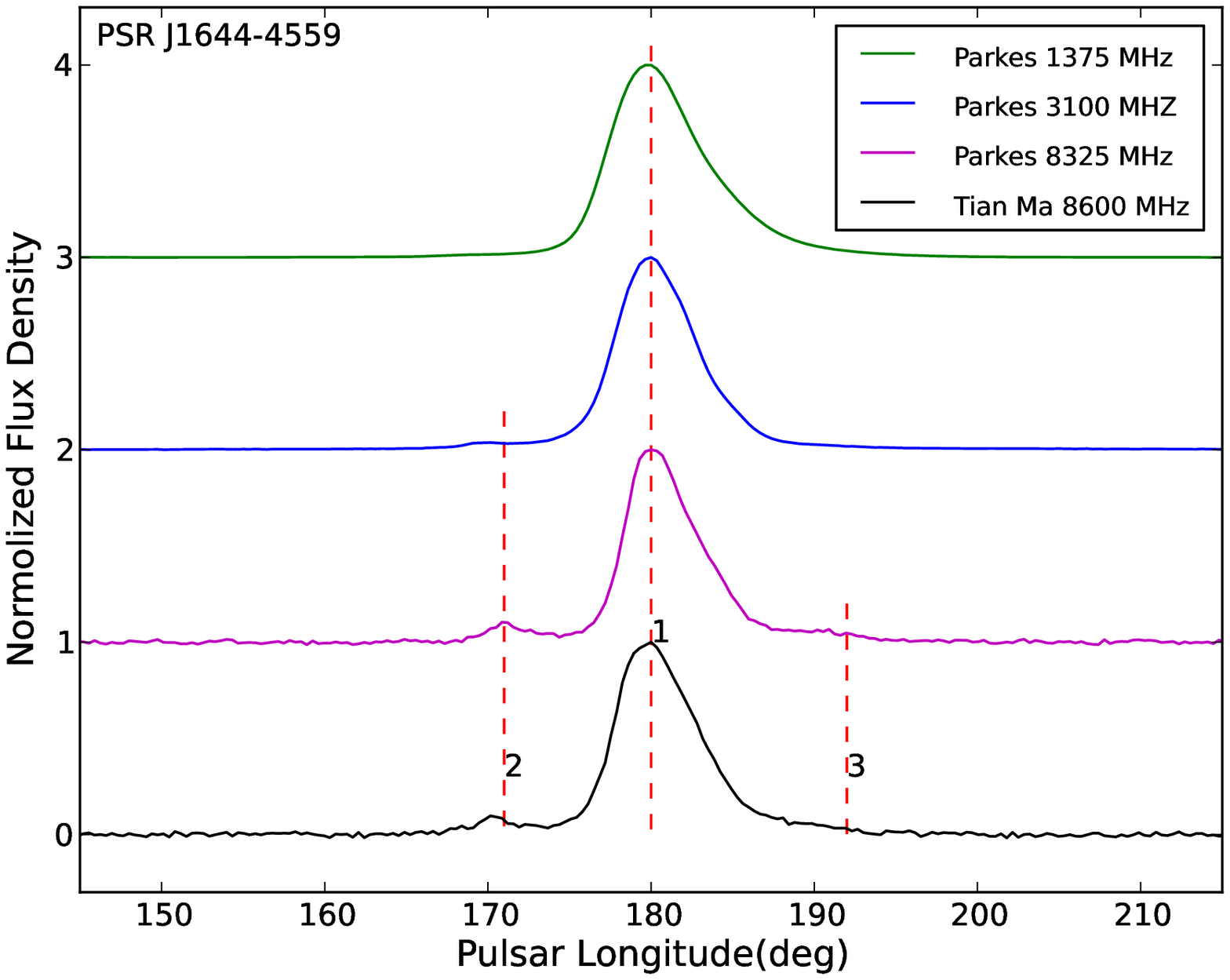}}&
\resizebox{0.53\hsize}{!}{\includegraphics[angle=0]{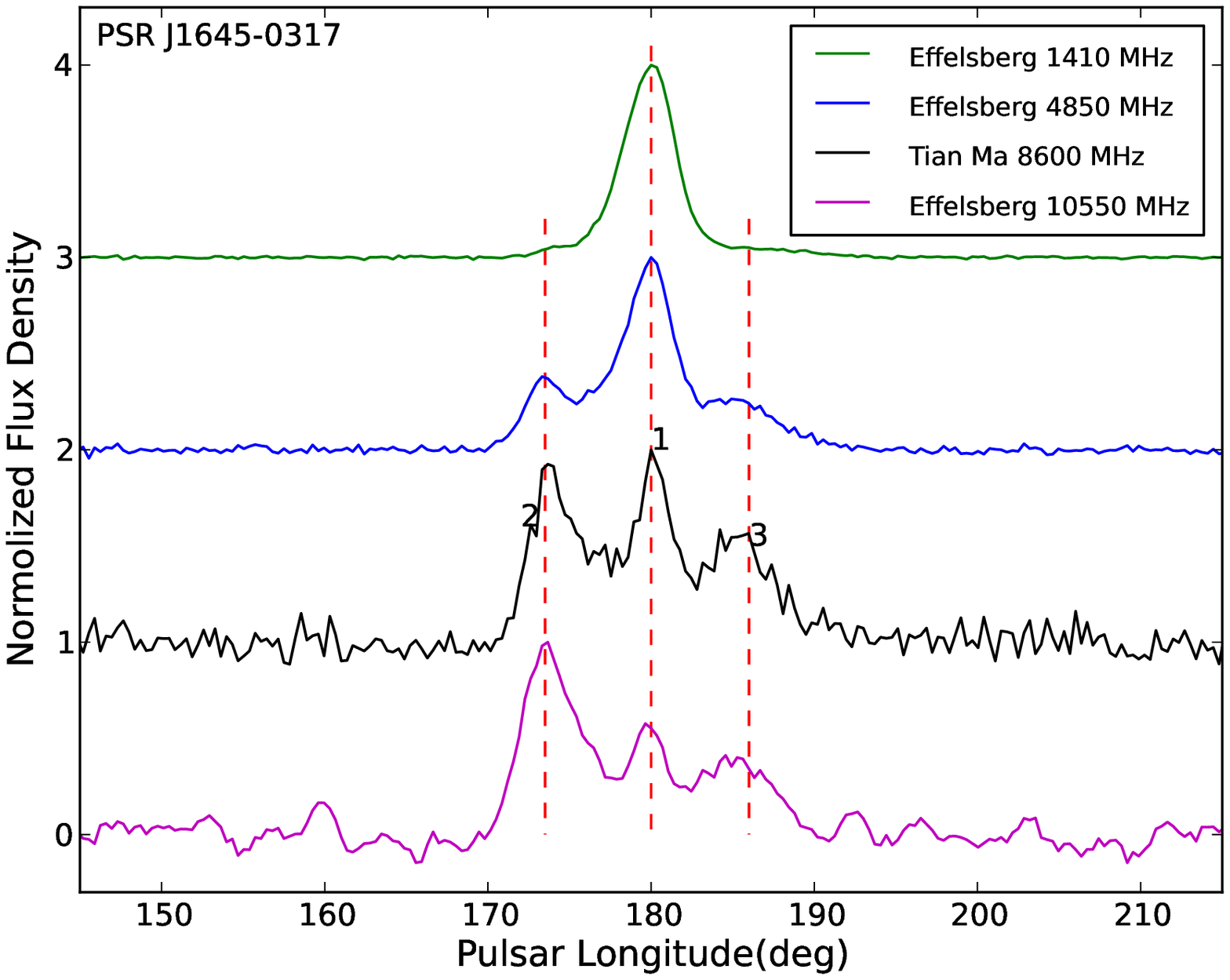}}\\
\resizebox{0.53\hsize}{!}{\includegraphics[angle=0]{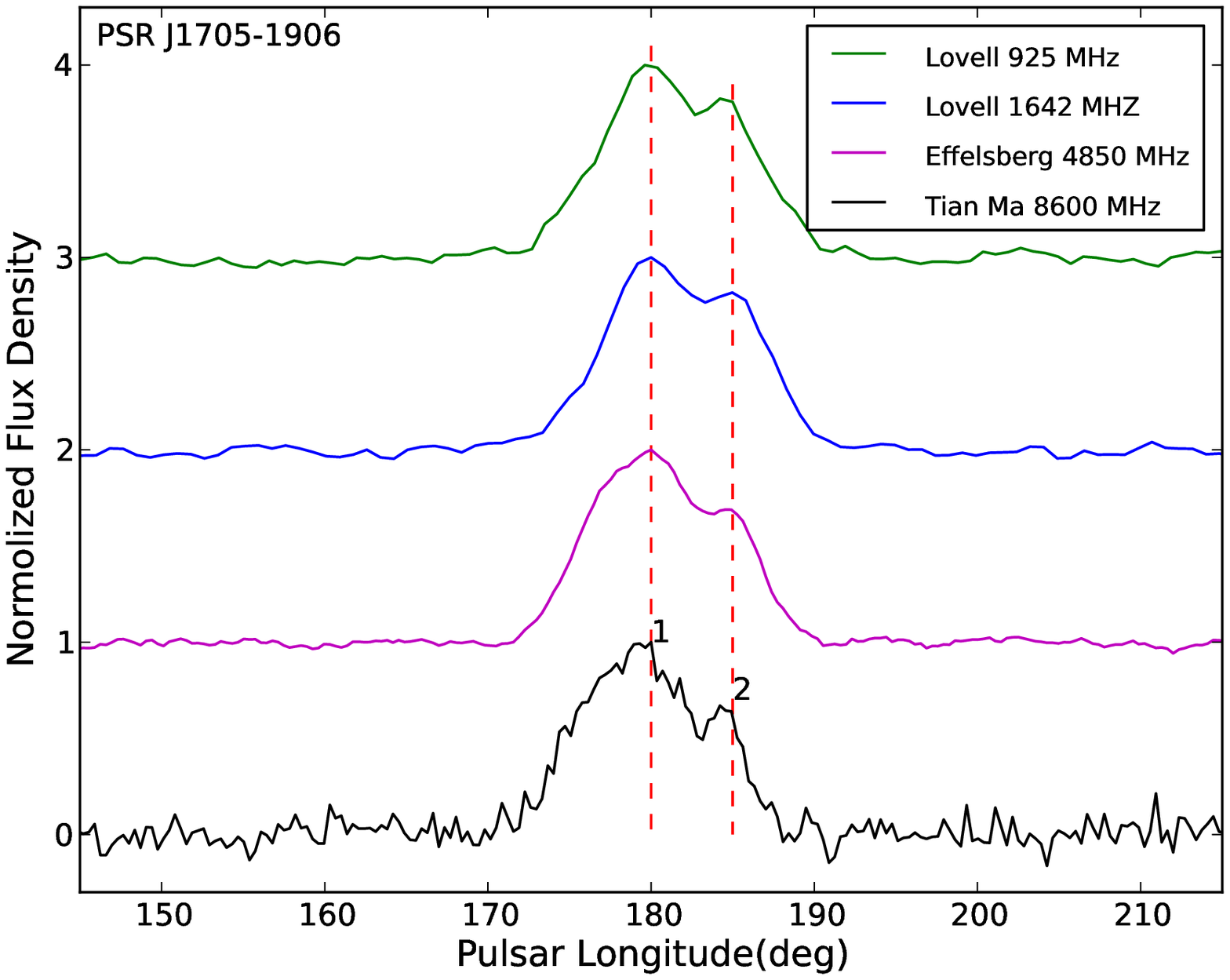}}&
\resizebox{0.53\hsize}{!}{\includegraphics[angle=0]{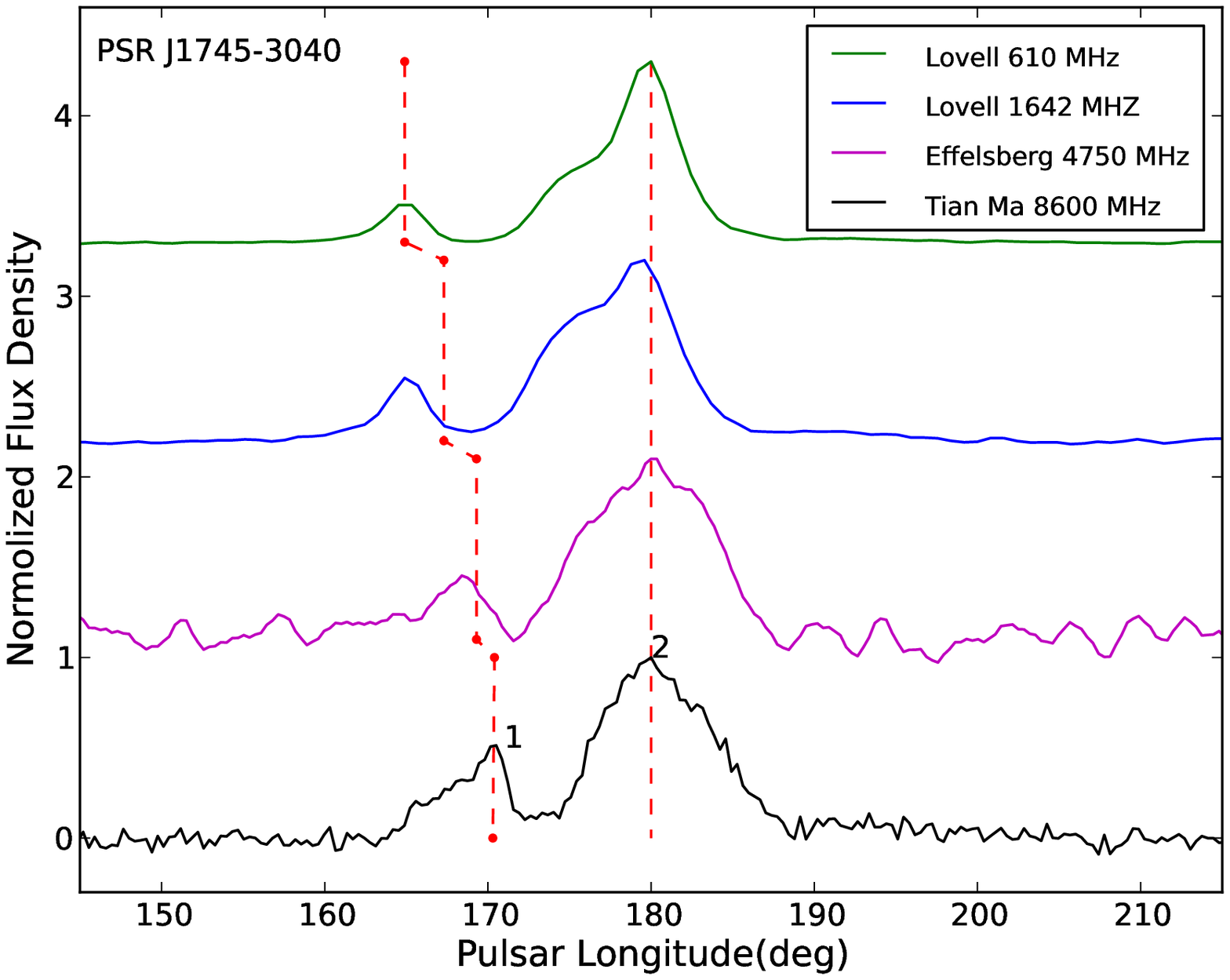}}\\
\resizebox{0.53\hsize}{!}{\includegraphics[angle=0]{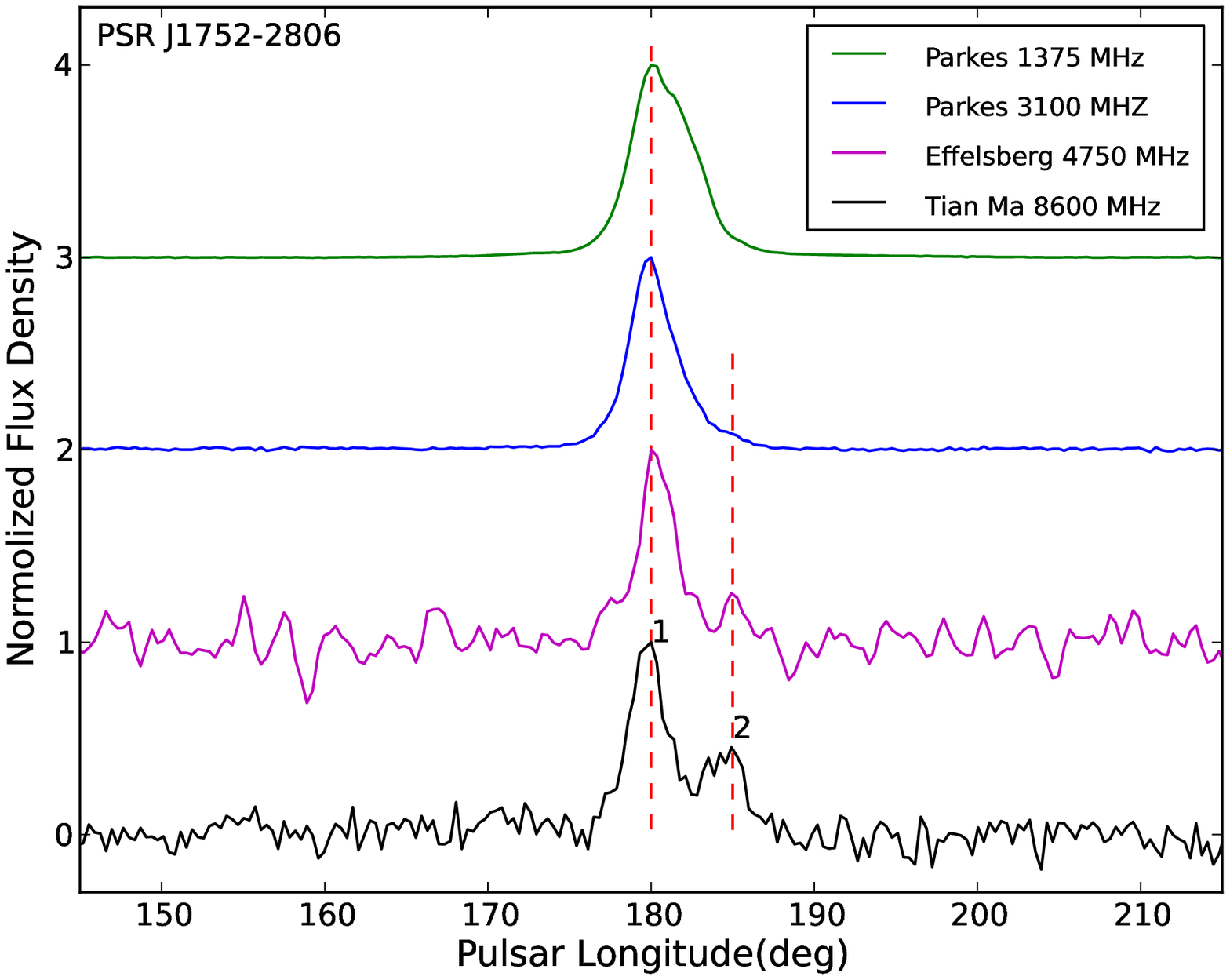}}&
\resizebox{0.53\hsize}{!}{\includegraphics[angle=0]{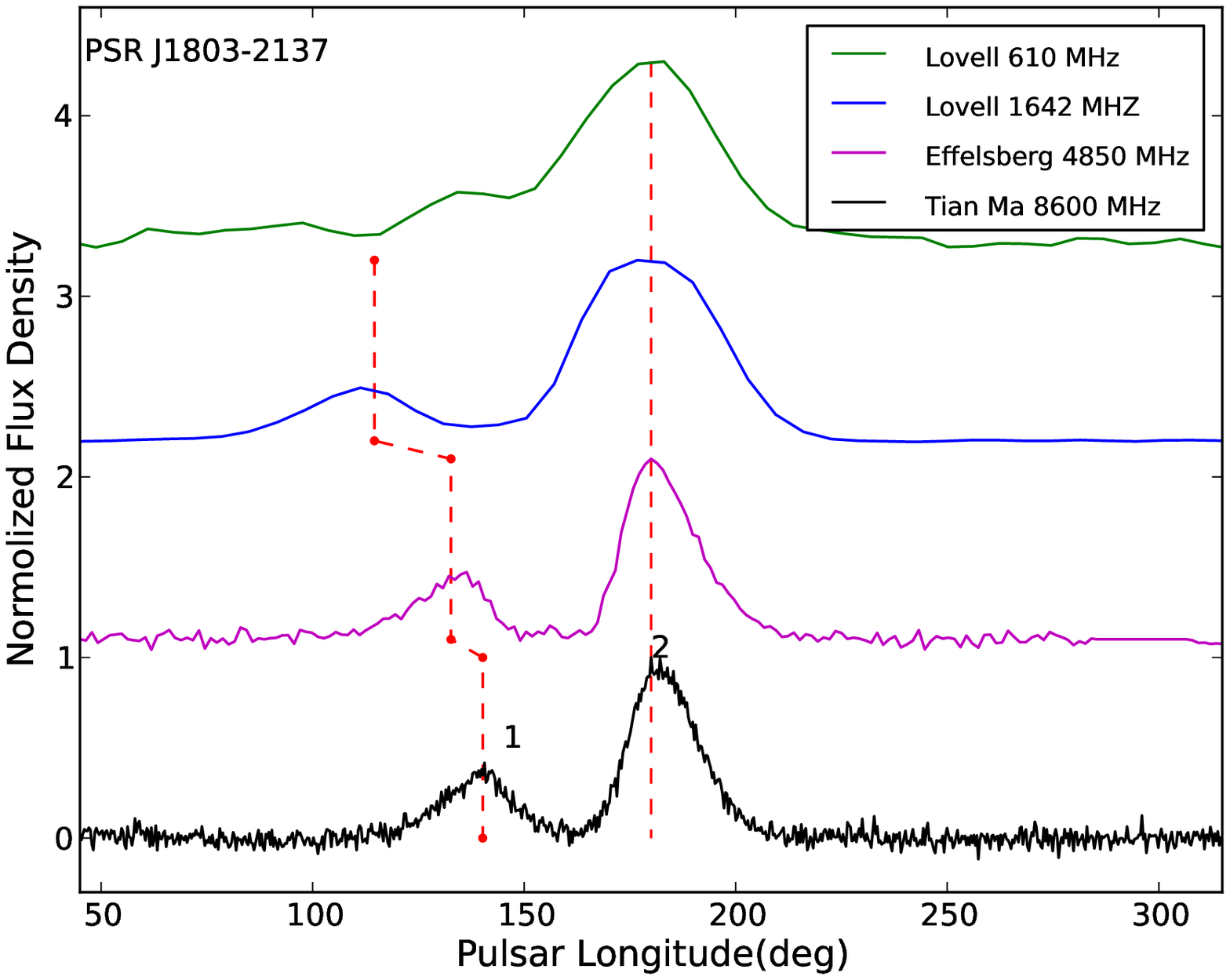}}\\
\end{tabular}
\end{center}
\caption{- continued}
\end{figure}
\addtocounter{figure}{-1}
\begin{figure}
\begin{center}
\begin{tabular}{cc}
\resizebox{0.53\hsize}{!}{\includegraphics[angle=0]{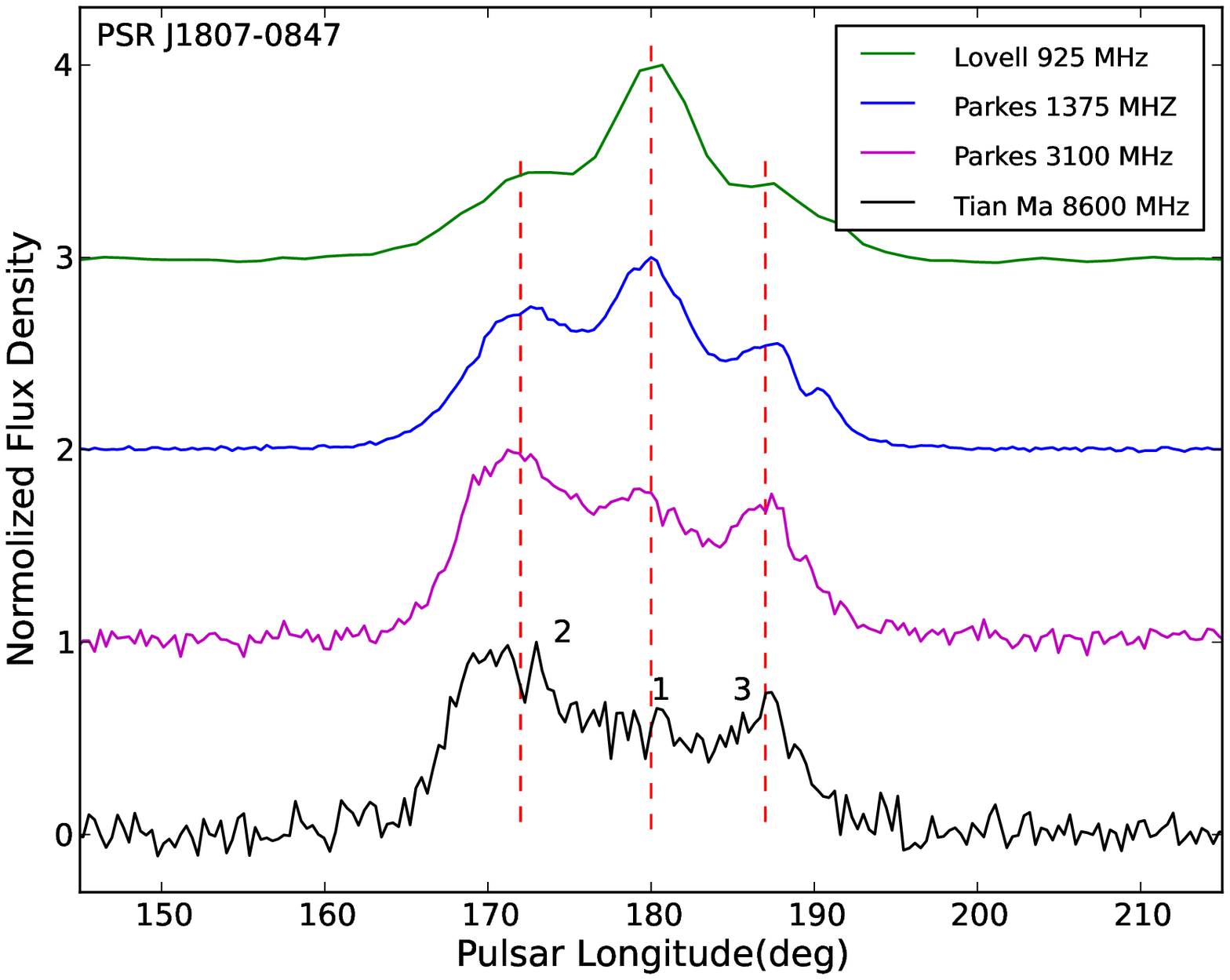}}&
\resizebox{0.53\hsize}{!}{\includegraphics[angle=0]{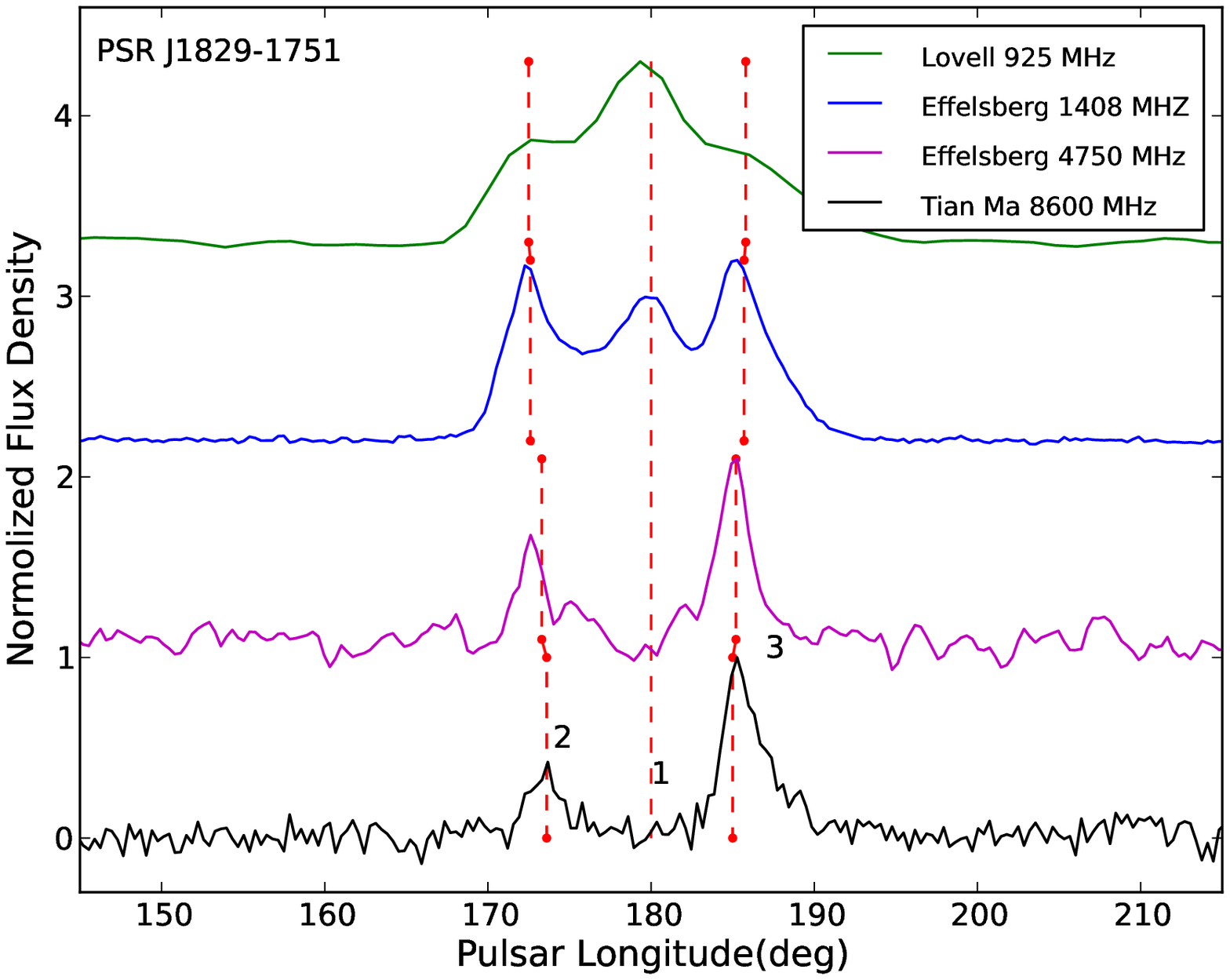}}\\
\resizebox{0.53\hsize}{!}{\includegraphics[angle=0]{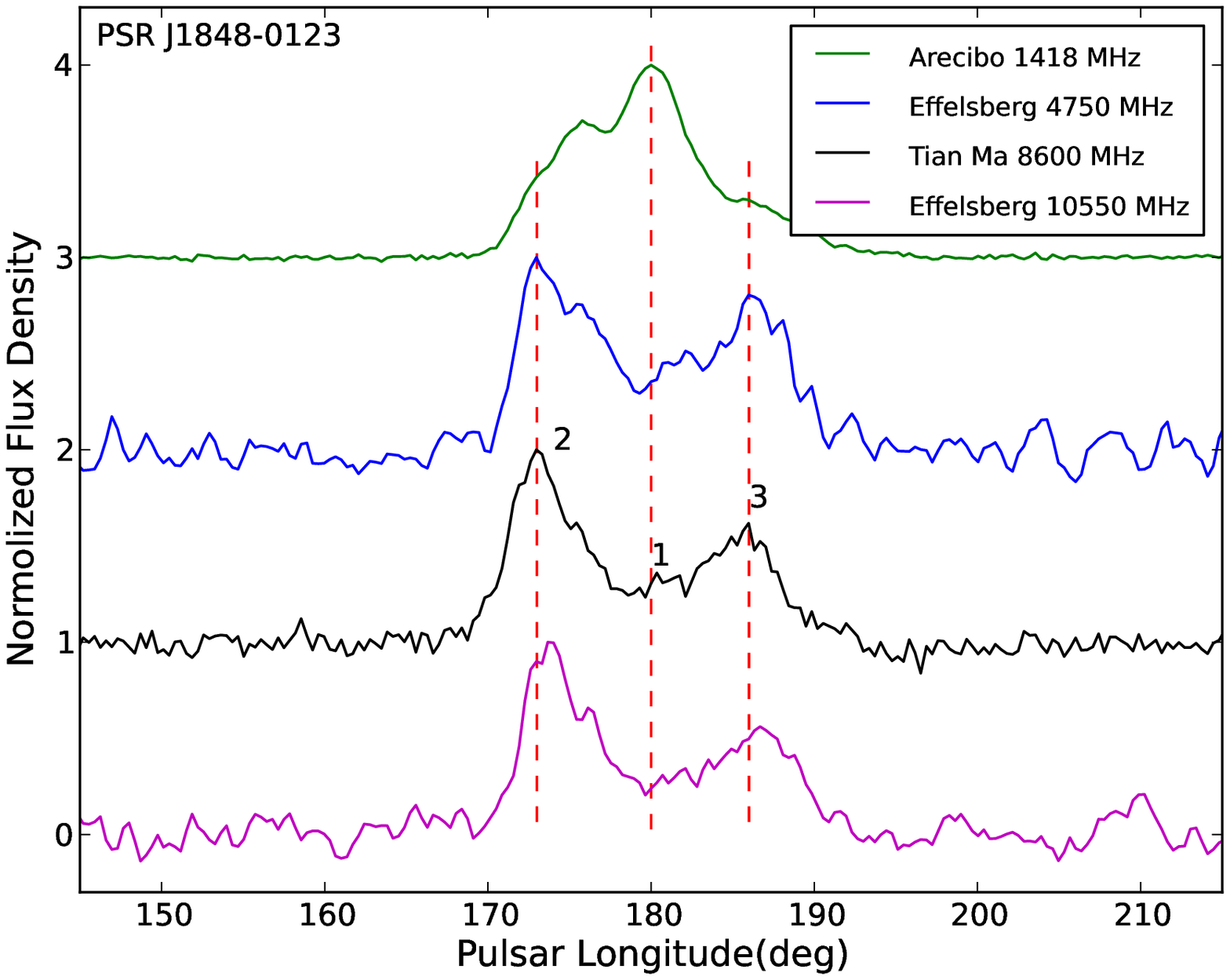}}&
\resizebox{0.53\hsize}{!}{\includegraphics[angle=0]{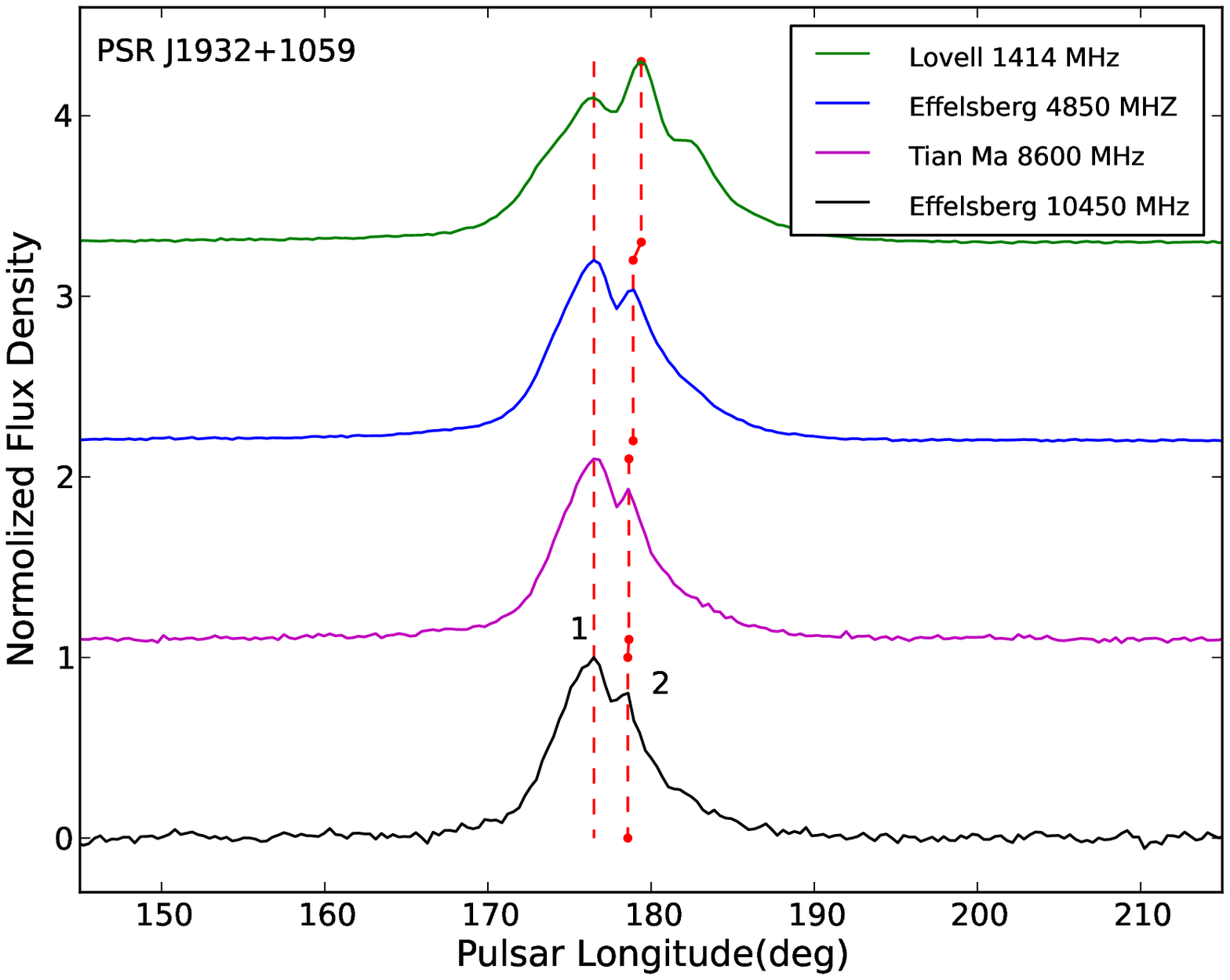}}\\
\resizebox{0.53\hsize}{!}{\includegraphics[angle=0]{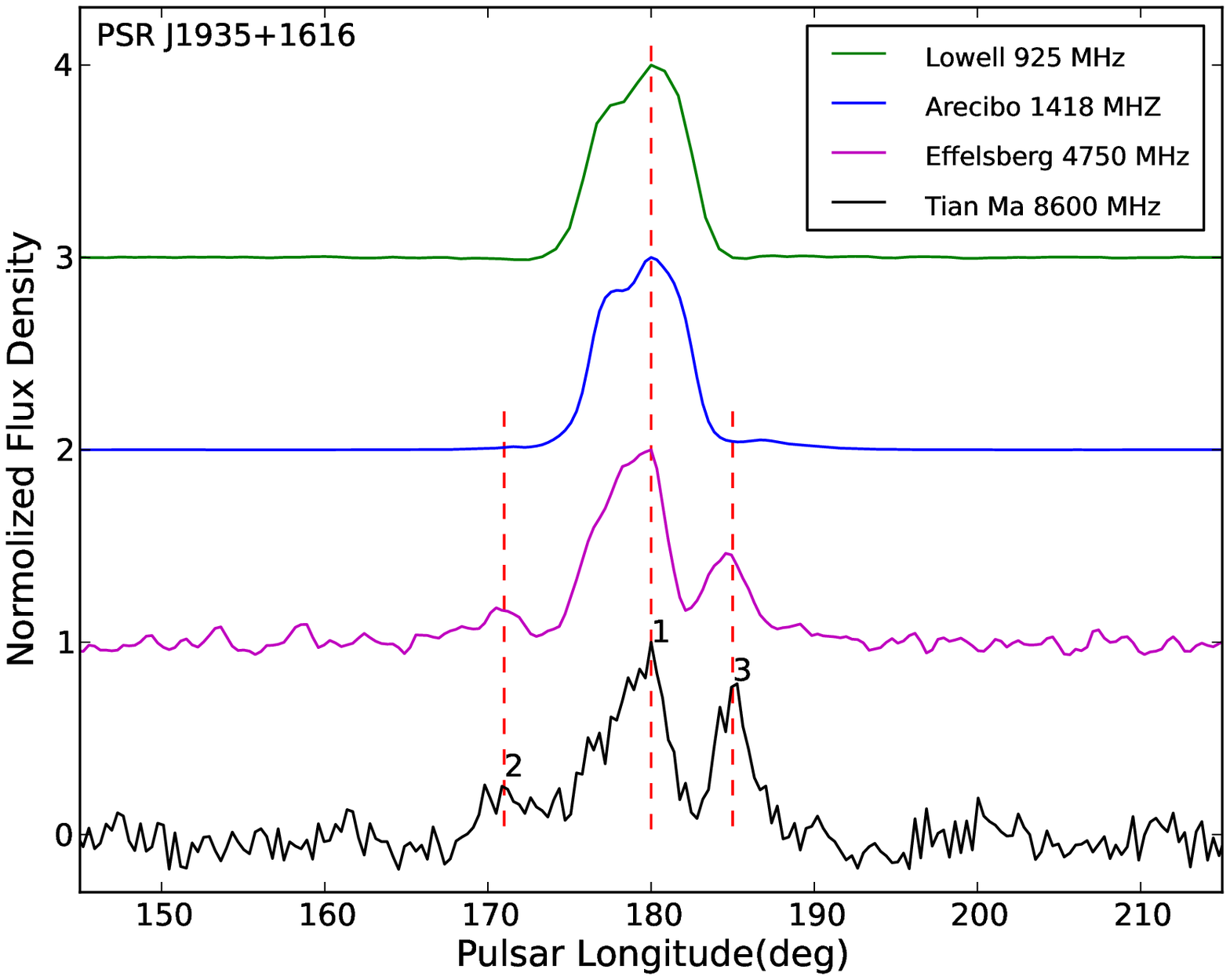}}&
\resizebox{0.53\hsize}{!}{\includegraphics[angle=0]{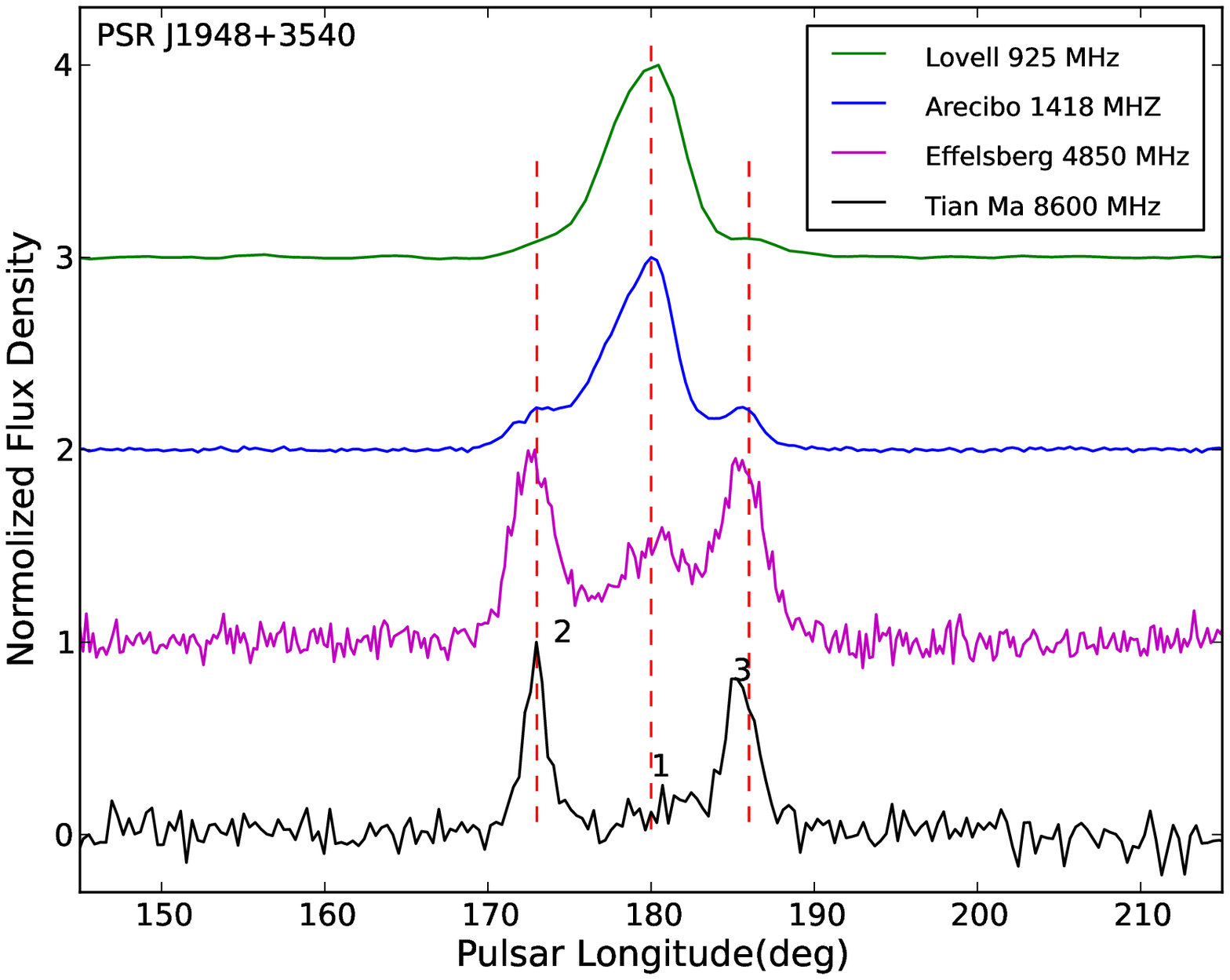}}\\
\end{tabular}
\end{center}
\caption{- continued}
\end{figure}
\addtocounter{figure}{-1}
\begin{figure}
\begin{center}
\begin{tabular}{cc}
\resizebox{0.53\hsize}{!}{\includegraphics[angle=0]{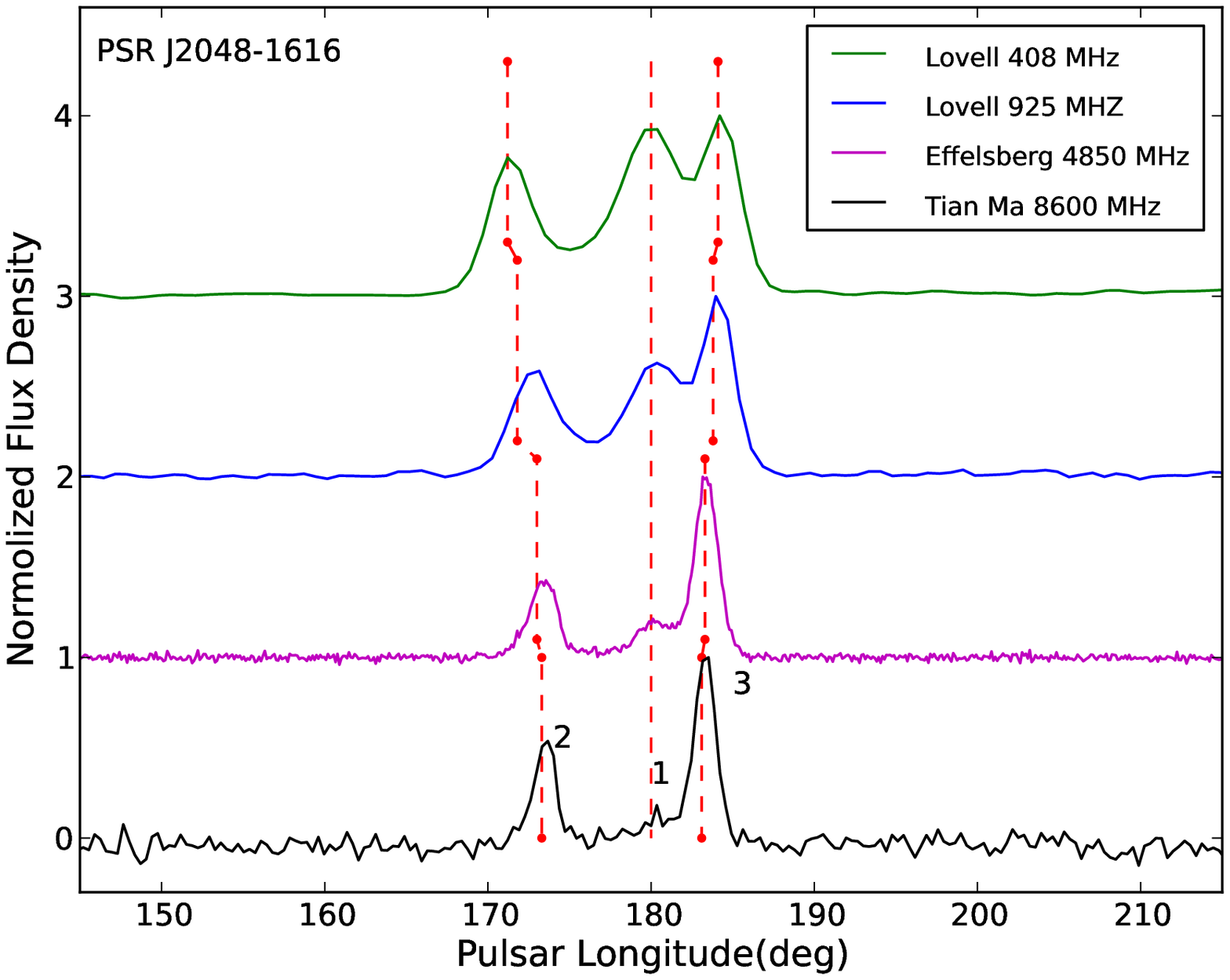}}\\
\end{tabular}
\end{center}
\caption{- continued}
\end{figure}

\clearpage

%\begin{figure}[h]
%\begin{center}
%\begin{tabular}{cc}
%\resizebox{0.53\hsize}{!}{\includegraphics[angle=0]{J0437-4715_256b.eps}}
%\end{tabular}
%\end{center}
%\caption{Integrated pulse profile of PSR J0437$-$4715 with 256 bins (compared to the 1024 bins in Fig.~\ref{fg:prf}), showing the leading component at $104 \degr$.}\label{fg:J0437_256b}
%\end{figure}

\clearpage

%\begin{figure}[h]
%\begin{center}
%\begin{tabular}{cc}
%\resizebox{0.53\hsize}{!}{\includegraphics[angle=0]{J0437-4715_spc.eps}}\\
%\end{tabular}
%\end{center}
%\caption{Spectra of the components of PSR J0437-4715.}
%\label{fg:0437}
%\end{figure}

\begin{figure}[h]
\caption{Comparison of Parkes and TMRT 8.6~GHz integrated profiles for
  PSR J0742-2822}
\label{fg:mode}
\begin{center}
\begin{tabular}{cc}
\resizebox{0.53\hsize}{!}{\includegraphics[angle=0]{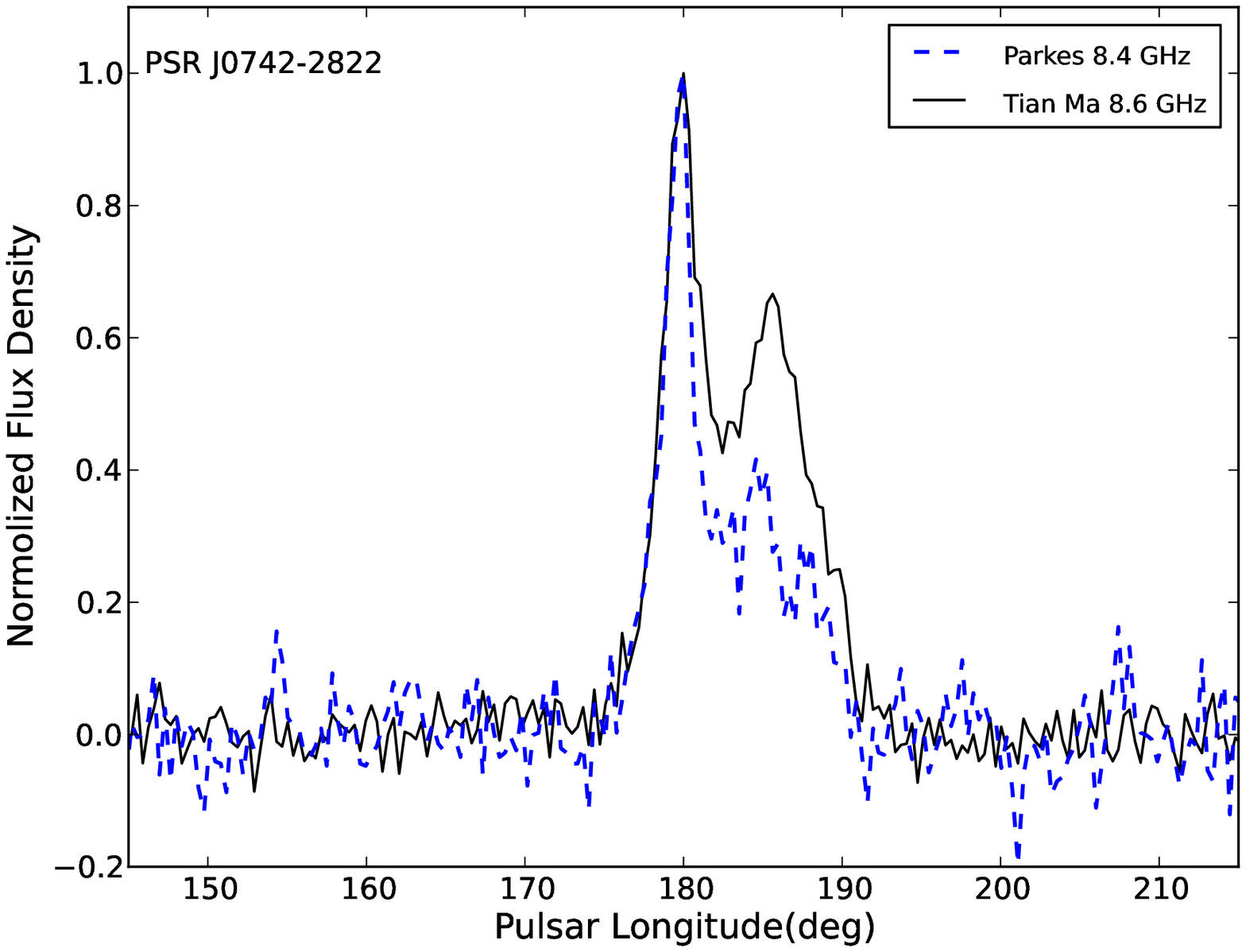}}&
\end{tabular}
\end{center}
\end{figure}

%\begin{figure}[h]
%\caption{Comparison of Effelsberg (1024 bins) at 4.85~GHz and TMRT (256 bins) at 8.6~GHz integrated profiles for PSR J1705-1906.}
%\label{fg:ipmp}
%\begin{center}
%\begin{tabular}{cc}
%\resizebox{0.53\hsize}{!}{\includegraphics[angle=0]{J1705-1906_4_8.eps}}&
%\end{tabular}
%\end{center}
%\end{figure}

\begin{figure}[h]
\begin{tabular}{cc}
\resizebox{0.53\hsize}{!}{\includegraphics[angle=0]{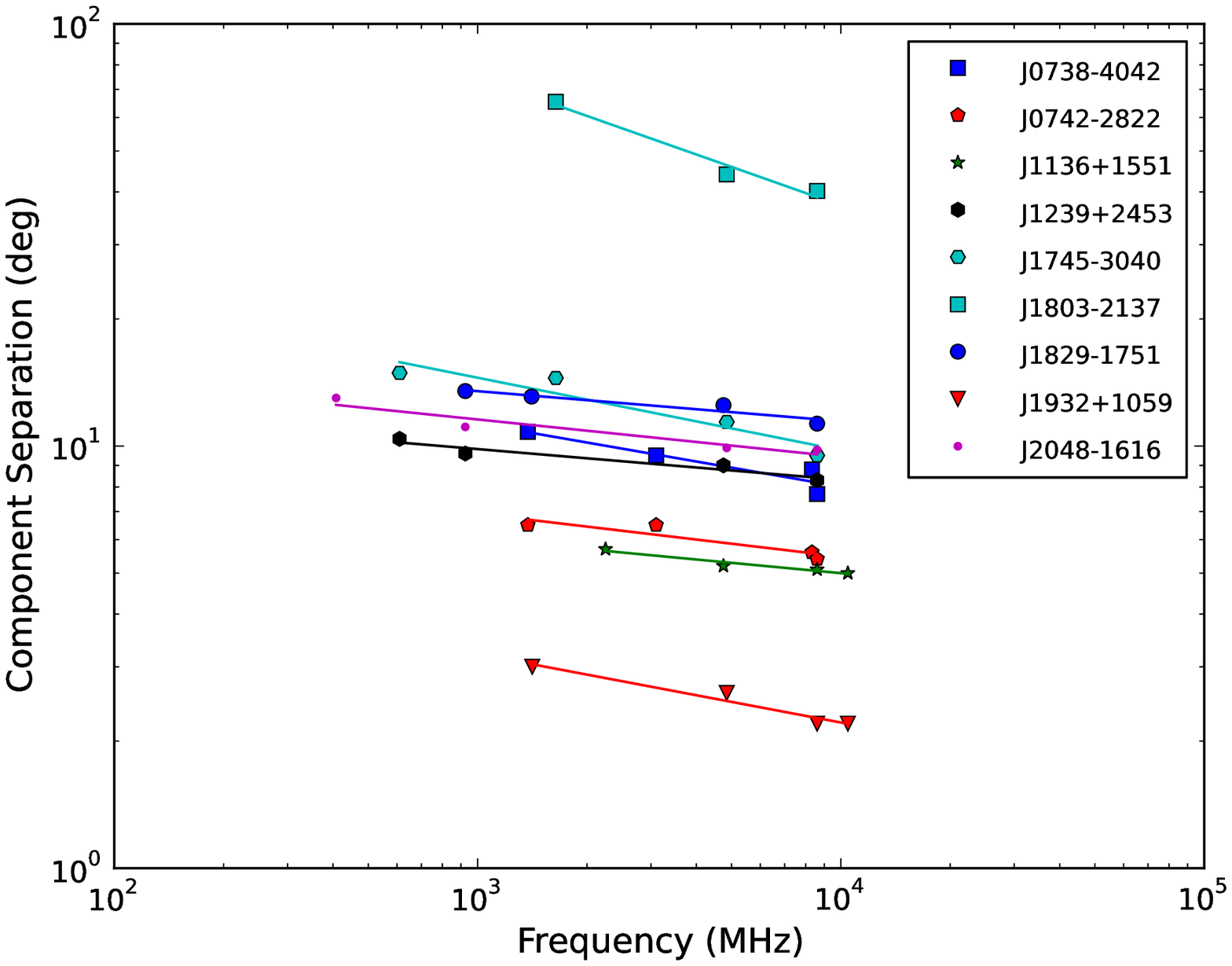}}&
\resizebox{0.53\hsize}{!}{\includegraphics[angle=0]{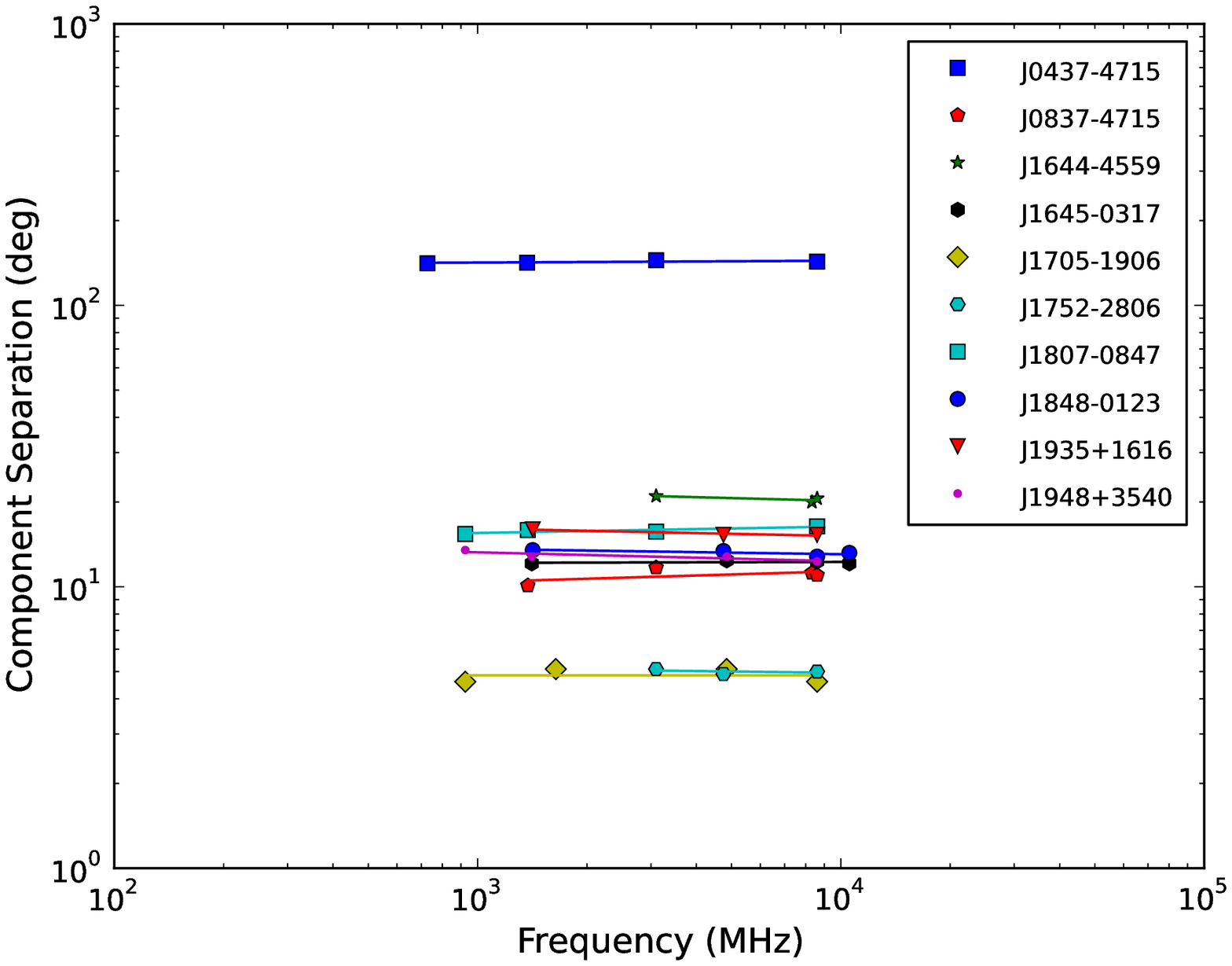}}\\
\end{tabular}
\caption{Separation of profile outermost components as a function of
  frequency for nine pulsars where the separation decreases with
  increasing frequency (left panel) and for ten pulsars with
  essentially constant component separation (right panel). The lines
  are the fitted power law for each pulsar.}\label{fg:comp-sep}
\end{figure}

\begin{figure}[h]
\begin{center}
\begin{tabular}{cc}
\resizebox{0.53\hsize}{!}{\includegraphics[angle=0]{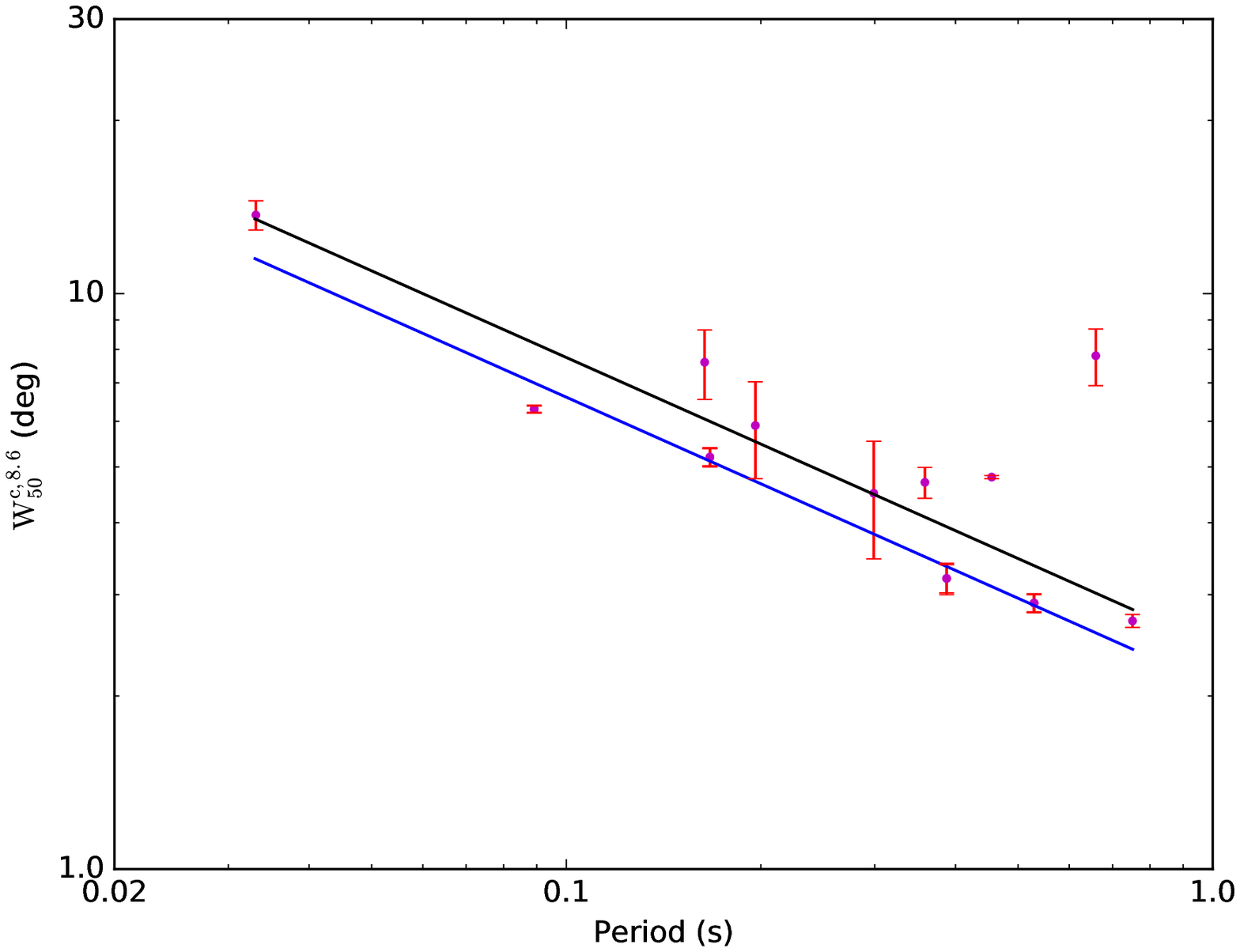}}\\
\end{tabular}
\end{center}
\caption{Observed core-component half-power widths at 8.6 GHz plotted against pulse
  period. Two lower-bound relations are shown, that from \citet{ran90}
  derived from 1.0 GHz observations (black line), and that for the 8.6 GHz data (blue line). }
\label{fg:wp}
\end{figure}

\end{document}